\documentclass{aa}
\usepackage{graphicx}
\usepackage{placeins}
\usepackage{txfonts}
\usepackage{hyperref}
\hypersetup{colorlinks,citecolor=blue}
\makeatletter
\renewcommand*\aa@pageof{, page \thepage{} of \pageref*{LastPage}}
\makeatother

\begin{document}

   \title{Multiples among B stars in the Scorpius-Centaurus association\thanks{Based on observation made with European Southern Observatory (ESO) telescopes at Paranal
    Observatory in Chile, under programme 1101.C0258 }}


   \author{R. Gratton \inst{1} 
          \and V. Squicciarini\inst{1,2}
          \and V. Nascimbeni \inst{1}
          \and M. Janson \inst{3}
          \and S. Reffert \inst{4}
          \and M. Meyer \inst{5}
          \and P. Delorme \inst{6}
          \and E. E. Mamajek \inst{7}
          \and M. Bonavita \inst{8}
          \and S. Desidera \inst{1}
          \and D. Mesa \inst{1}
          \and E. Rigliaco \inst{1}
          \and V. D'Orazi \inst{1,9}
          \and A. Vigan \inst{10}
          \and C. Lazzoni \inst{11,1}
          \and G. Chauvin \inst{12}
          \and M. Langlois \inst{13}
          }

   \institute{INAF - Osservatorio Astronomico di Padova,
              Vicolo dell'Osservatorio 5, I-35122 Padova, Italy\\
              \email{raffaele.gratton@inaf.it}  
        \and
        LESIA, Observatoire de Paris, Universit\'e PSL, CNRS, Sorbonne Universit\'e, Universit\'e Paris Cit\'e, 5 place Jules Janssen, 92195 Meudon, France
        \and
        Institutionen f\"or Astronomi, Stockholms Universitet; Alba Nova Universitets Centrum, SE-106 91 Stockholm, Sweden
        \and
        Landessternwarte, Zentrum f\"ur Astronomie der Universität Heidelberg; Königstuhl 12, 69117 Heidelberg, Germany
        \and
        Department of Astronomy, University of Michigan; 1085 S. University Ave, Ann Arbor MI 48109, USA
        \and
        Univ. Grenoble Alpes, CNRS, IPAG; F-38000 Grenoble, France
        \and
        Jet Propulsion Laboratory, California Institute of Technology; 4800 Oak Grove Drive, Pasadena CA 91109, USA.
        \and
        School of Physical Sciences, The Open University, Walton Hall, Milton Keynes MK7 6AA, UK
        \and
        Dipartimento di Fisica, Universit\`{a} di Roma Tor Vergata, Via della Ricerca Scientifica, 1 - 00133, Roma, Italy
        \and
        Aix Marseille Universit\'e, CNRS, LAM (Laboratoire d'Astrophysique de Marseille) UMR 7326, 13388 Marseille, France
        \and
        University of Exeter, Astrophysics Group, Physics Building, Stocker Road, Exeter EX4 4QL, UK
        \and
        Universit\'e C\^ote d’Azur, Observatoire de la C\^ote d’Azur, CNRS, Laboratoire Lagrange, France
        \and
        CRAL, UMR 5574, CNRS, Université de Lyon, ENS, 9 avenue Charles Andr\'e, 69561 Saint Genis Laval Cedex, France}

   \date{Received ; accepted }

 
  \abstract
   { The frequency, semi-major axis, and mass distribution of stellar companions likely depend on the mass of the primaries and on the environment where the stars form. These properties are very different for early- and late-type stars. However, data are largely incomplete, even for the closest environments to the Sun, preventing a cleaner view of the problem. }
   { This paper provides basic information about the properties of companions to B stars in the Scorpius-Centaurus association (age $\sim 15$ Myr); this is the closest association containing a large population of 181~B-stars. }
   { We gathered available data combining high contrast imaging samples from BEAST, SHINE, and previous surveys with evidence of companions from Gaia (both through direct detection and astrometry), from eclipsing binaries, and from spectroscopy. We evaluated the completeness of the binary search and estimated the mass and semi-major axis for all detected companions. These data provide a complete sample of stellar secondaries (extending well in the substellar regime) for separation $>3$~au, and they are highly informative as to closer companions. }
   { We found evidence for 200 companions around 181 stars. We did not find evidence for companions for only 43 ($23.8\pm 3.6$\%) of the targets, with the fraction being as low as $15.2\pm 4.1$\% for stars with $M_A>3.5$~M$_\odot$ while it is $31.5\pm 5.9$\% for lower-mass stars. This confirms earlier findings for a clear trend of a binary fraction with stellar mass. The median semi-major axis of the orbits of the companions is smaller for B than in A stars, confirming a turn-over previously found for OB stars. The mass distribution of the very wide ($a>1000$~au) and closer companions is different. Very few companions of massive stars $M_A>5.0$ M$_\odot$ have a mass below solar and even fewer are M stars with a semi-major axis $<1000$~au. However, the scarcity of low-mass companions extends throughout the whole sample. Period and mass ratio distributions are different for early B stars (up to B7 spectral type) and stars of a later spectral type: most early B stars are in compact systems with massive secondaries, while less massive stars are mainly in wider systems with a larger spread in mass ratios. We derived log-normal fits to the distribution of the semi-major axis and mass ratios for low and high-mass B stars; these relations suggest that it is not probable that the planets and brown dwarf (BD) companions to b~Cen and $\mu^2$~Sco are extreme cases in the distribution of stellar companions.}
   { We interpret our results as the formation of secondaries with a semi-major axis $<1000$~au (about 80\% of the total) by fragmentation of the disk of the primary and selective mass accretion on the secondaries. The formation of secondaries within the disk of primaries in close binaries has been proposed by many others before; it unifies the scenarios for formation of close binaries with that of substellar companions that also form within the primary disk, though on a different timescale. We also find that the observed trends with primary mass may be explained by a more prolonged phase of accretion episodes on the disk and by a more effective inward migration. Finally, in the Appendices we describe the detection of twelve new stellar companions from the BEAST survey and of a new BD companion at 9.599 arcsec from HIP~74752 using Gaia data, and we discuss the cases of possible BD and low-mass stellar companions to HIP~59173, HIP~62058, and HIP~64053.  }


\titlerunning{Multiples among B-stars in Sco-Cen}
\authorrunning{R. Gratton et al.}

   \maketitle
%

\section{Introduction}
\label{sec:Introduction}

A large fraction of the stars are not single \citep{Lada2006,Duchene2013} and this fraction increases with stellar mass: 30\%–40\% of the M stars \citep{Fischer1992, Delfosse2004, Janson2012},  50\%–60\% of solar-type stars \citep{Duquennoy1991, Raghavan2010, Duchene2013, Moe2017}), and more than 70\% of more massive stars \citep{Kouwenhoven2007b, Peter2012} are in multiple systems. The peak of their distribution with a period and semi-major axis (tens to hundreds of au: \citealt{Duquennoy1991, Raghavan2010, DeRosa2014}) is similar to the size of disks \citep{Najita2018}. These facts indicate that considering the formation of binaries is important when trying to understand how planets form. The mechanisms that lead to the formation of binaries are not well established (see e.g. \citealt{Tohline2002}). The favoured scenarios are turbulent fragmentation of clouds for a separation $>500$ au \citep{Offner2010, Offner2016} and disk fragmentation for a separation $<500$ au \citep{Kratter2010}. These values for the separation, however, apply to solar-type stars and they are possibly different for other ranges of mass. Disk fragmentation is expected to be more efficient around massive stars because of the larger value of the accretion rate from the natal cloud and hence the larger expected disk-to-star mass ratio during early phases of formation, when binaries are likely to form \citep{Machida2010, Kratter2016, Elbakyan2023}. In disk fragmentation, mass accretion on the secondary may be favoured with respect to accretion on the primary \citep{Clarke2012}; if the disk survives long enough, this would lead to a preference for equal mass binaries \citep{Kratter2010}. On the other hand, the disk may disperse before this condition is met, and hence the final mass ratio is not firmly established and may well be variable from case to case.

An accurate prediction of the outcome of binary formation from disk fragmentation is very complex because of the huge range of parameters involved and the complexity of the basic mechanisms that are often poorly understood \citep{Kratter2016, Meyer2018, Oliva2020}. Uncertainties concern the range of disk-to-star mass ratios and of the accretion of mass on the disk from the parental cloud, the threshold for the onset of disk instabilities, the migration of secondaries within the disk, the accretion rates on the stars, the loss of angular momentum related to magneto-hydrodynamical winds, and the role of ternary or higher multiplicity systems. The exploration of the wide range of parameters with detailed hydrodynamical models is extremely expensive in terms of computational time. In addition, the properties of binaries not only depend on the mass of the star, but also on the environment (see e.g. \citealt{Heggie1975, Duchene1999, Goodwin2010, Kaczmarek2011, Duchene2013}). \citet{Tokovinin2020} thus considered a parametric approach within a toy model in order to explore the impact of the many different parameters involved. This approach allowed for the role played by the different mechanisms to be outlined, but it cannot be used to make firm predictions, for example, on the distribution of mass ratios as a function of separation.

Within this context, we discuss in this paper the frequency, mass ratio, and separation in binaries around B stars in the Scorpius-Centaurus (Sco-Cen) association. The Sco-Cen association is at about 100-150 pc from the Sun \citep{deZeeuw1999} and it is well suited for this analysis for a number of reasons. Star formation is essentially complete in Sco-Cen. Binaries are young enough (age $<20$~Myr: \citealt{Pecaut2012}) and the association is so loose (density $<1$~star/pc$^3$) that the impact of the long-term evolution of binary systems related to the environment \citep{Heggie1975, Binney1987, Kaczmarek2011} likely does not strongly influence the properties of binaries even for a separation as large as a few thousands au\footnote{Since the density of Sco-Cen stars is comparable to that of the Solar neighbourhood, the formalism of \citet{Binney1987} indicates that encounters with field stars are actually more probable than those with other members of the association. This is due to the much larger spread in velocities. Of course, encounters are likely more probable in the birth place of individual stars, but this can be considered as a component of the binary formation mechanisms.}. Sco-Cen is close enough that high contrast imaging (HCI) provides a good view for the range of separation corresponding to the formation within the disk; its young age allows it to be complete even in the substellar regime, down to at least 0.01~M$_\odot$, for a separation between a few tens to a thousand au. Finally, Sco-Cen is a large association including hundreds of stars more massive than the Sun \citep{Mamajek2002} and an estimated total of $\sim 6,000$ stars \citep{Luhman2022} (even higher numbers $\sim 10,000$ objects are obtained including substellar objects: \citealt{Damiani2019, Luhman2022}), allowing large samples adequate for a statistical discussion. We considered  
all B stars in the Sco-Cen association listed by \citet{Rizzuto2011}. In addition to HCI, we used a wide range of other methods to detect and characterise companions.

The paper is organised as follows: Section~\ref{sec:Sample} describes the sample of B stars in Sco-Cen considered in this paper, discussing the adopted interstellar reddening and ages for the individual objects. Section~\ref{sec:detections} reviews the detections of companions around these stars, considering visual, eclipsing, spectroscopic, and astrometric binaries. In Section~\ref{sec:mass} we describe the methods used to characterise these companions in terms of mass and semi-major axis (not enough data exist to also derive the eccentricity distribution, which would however be important). We give an analysis of the completeness of our detections in Section~\ref{sec:completeness}. Section~\ref{sec:statistics} presents the results of our analysis.  Section~\ref{sec:discussion} presents a discussion by comparing the current results with those obtained from other surveys on binaries -- revealing trends over a large range of masses and in different environments -- and with a toy model of binary formation similar to that presented by \citet{Tokovinin2020}. We draw conclusions in Section~\ref{sec:conclusion}. In the Appendices we present the binary companions detected in the B-star Exoplanet Abundance STudy (BEAST) survey including an indication for a probable brown dwarf (BD) c close to HIP~59173 --, a reanalysis of data for some eclipsing binaries, the detection of a BD close to HIP~74752 and indication for the presence of two more close to HIP~62058 and HIP~64053 using Gaia data, and finally the tables containing the most relevant data for the whole sample. 

\section{Sample of B stars in Sco-Cen}
\label{sec:Sample}

\begin{figure*}[htb]
    \centering
    \includegraphics[width=18cm]{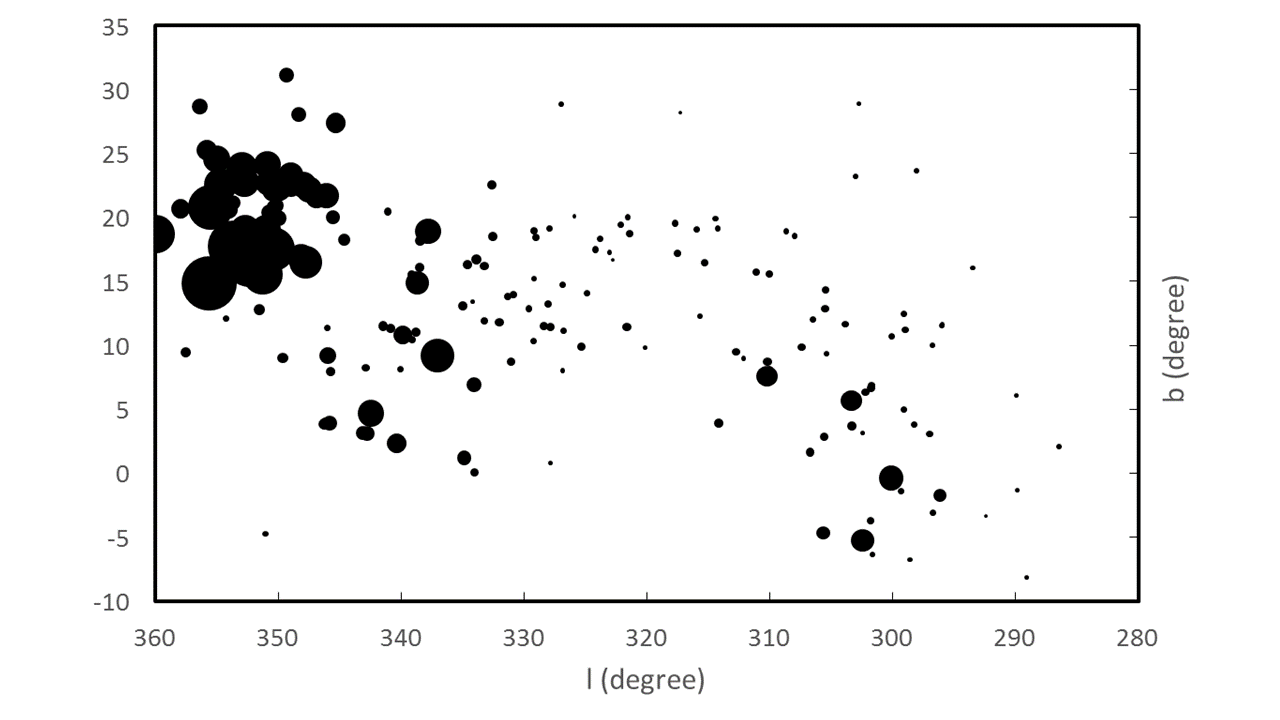}
    \caption{Map of the reddening assumed for the programme stars. The area of the blobs is proportional to the assumed value for the reddening E(B-V). The highest value is E(B-V)=0.994 mag for the B5III star HIP 80371 in the $\rho$~Ophiuchi cloud}
    \label{fig:reddening_map}
\end{figure*}

\begin{figure}[htb]
    \centering
\includegraphics[width=8.5cm]{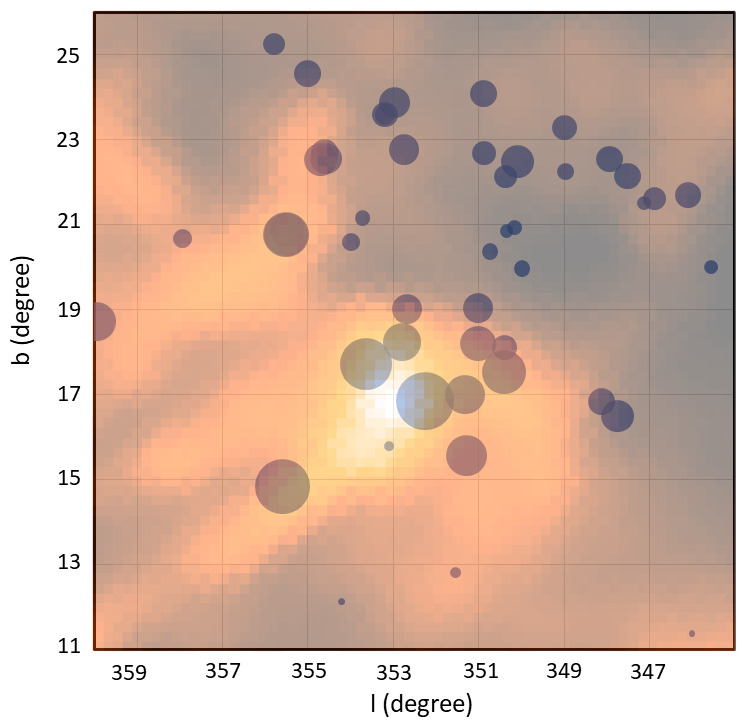}
    \caption{Comparison between the map of the reddening assumed for the programme stars in the region of the Upper Scorpius - Ophiuchus cloud (blobs) and the dust emission map from Planck mission  (heat colour map in transparency). The area of the blobs is proportional to the assumed value for the reddening E(B-V). The highest value is E(B-V)=0.994 mag for the B5III star HIP 80371 in the $\rho$~Ophiuchi cloud}
    \label{fig:reddening_map_ophiucus}
\end{figure}

\begin{figure}[htb]
    \centering
\includegraphics[width=8.5cm]{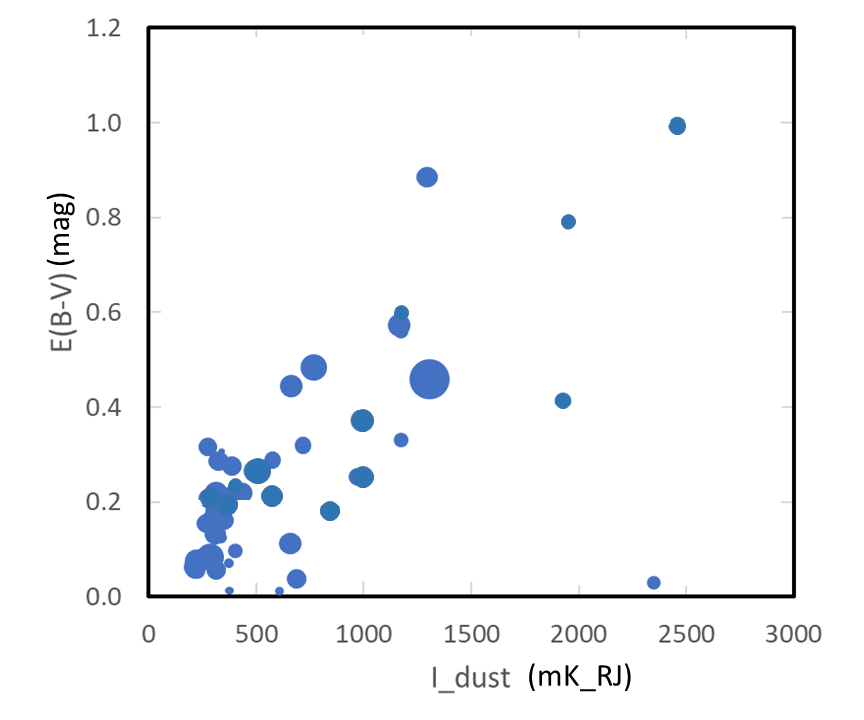}
\includegraphics[width=8.5cm]{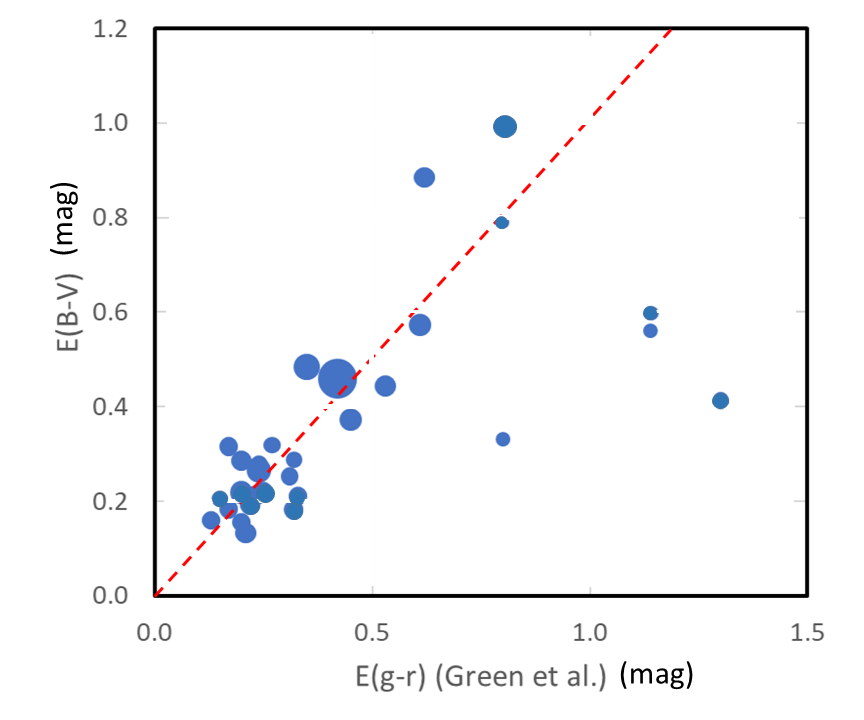}
    \caption{Comparison between the reddening assumed for the programme stars in the region of the Upper Scorpius - Ophiuchus cloud and other estimates. Upper panel: with the intensity of the emission as measured by the Planck mission. Lower panel: with the total reddening in the $g-r$ colour and in that direction obtained from the map by \citet{Green2019}; the dashed red line is the expected relation between $E(B-V)$ and $E(g-r)$ \citep{Schlafly2011}. Only objects with $E(B-V)>0.1$ are plotted here. In both panels the size of the points is related to the distance from the Sun (largest symbols are for the farthest stars). }
    \label{fig:reddening_idust}
\end{figure}

\subsection{B stars in Sco-Cen}
\label{sec:b-stars}

The sample of B-stars in the Sco-Cen association considered in this paper is mainly based on the list of members of the association by \cite{Rizzuto2011} that have a B-spectral type as listed in the SIMBAD database \citep{Wenger2000}. A few additional members listed in \cite{Janson2021a} (BEAST sample) were also considered. In total, our sample includes 181 stars. We kept throughout the analysis a few stars even though they have low membership probability both in \citet{Rizzuto2011} and from the online Banyan code \citet{Gagne2018}\footnote{See \url{https://www.exoplanetes.umontreal.ca/banyan/banyansigma.php}} because their proper motion might be influenced by far companions. These stars (HIP~52742, HIP~54767, HIP~57669, HIP~59196, HIP~65021, and HIP~76126) are located quite at the edge of the Sco-Cen association and might indeed not be members. HIP~59196 is a massive B2V star. It might be a member of the Argus association (age of $\sim$40-50 Myr: \citealt{Zuckerman2019}) according to the online Banyan code \citep{Gagne2018}. In this case the star would be evolved off the MS, a fact that could explain its position on the colour-magnitude diagram, though this might also be explained by the fact that the star is a Be. With such an older age, the mass would be about 7.6~M$_\odot$ rather than 10.5~M$_\odot$ as adopted in our analysis. In our analysis we have two very wide pairs that are likely physically linked with each other (HIP~63003 - HIP~63005, HIP~80062 - HIP~80063); we considered them as separate entries.

\subsection{Interstellar reddening}
\label{sec:reddening}

Magnitudes and colours were corrected for interstellar extinction towards the programme stars that was obtained by integrating the 3-d maps by \citet{Lallement2022}. We used the ratio between the absorption in the Gaia $G$\ band \citep{Gaia2022a} and in the 2MASS $K$\ band \citep{Skrutskie2006} from \citet{Wang2019}. However, some stars are in regions of high extinction, mainly in the $\rho$~Ophiuchi cloud but also in the Crux region. We found that for these stars the extinction must be much higher than expected from the 3-d maps by \citet{Lallement2022}. We then derived appropriate values for the reddening for these stars by forcing their $G-K$ colours, deblended for the presence of companions as described in Section~\ref{sec:massvisual}, to agree with those expected for stars having the same spectral type in the Table by \citet{Pecaut2013}\footnote{\url{https://www.pas.rochester.edu/ emamajek/EEM_dwarf_UBVIJHK_colors_Teff.txt}} for a main sequence star. This procedure cannot be applied to Be stars; in this case the extinction was obtained by forcing the absolute magnitude of the star to agree with that expected for main sequence stars of the same spectral type. Figure \ref{fig:reddening_map} shows the on-sky distribution of the reddening.

Figure~\ref{fig:reddening_map_ophiucus} shows a map that compares the reddening assumed for the programme stars in the region of the Upper Scorpius - $\rho$~Ophiuchi cloud and the dust emission from Planck mission \footnote{\url{https://pla.esac.esa.int/\#maps}} \citep{Planck2020}. The regions where there are stars for which we derived the largest reddening coincide with those with stronger dust emission in the Planck map. This is shown in the upper panel of Figure~\ref{fig:reddening_idust}, that compares the reddening values for the individual stars and the intensity of the dust emission; the size of the symbols is related to the distance of the stars. The discrepant point in this diagram, with a low value of $E(B-V)$ in comparison with a rather strong dust emission at this location is HIP~80815 (i~Sco); according to Gaia this star is at 125.9 pc, that is about 15 pc closer than the bulk of the stars in the Upper Scorpius and $\rho$~Ophiuchi associations. It is then reasonable to assume that this star is also in front of most of the dust seen in this direction.

We compared these reddening values with the 3-d maps by \citet{Lallement2022} and \citet{Green2019} (see lower panel of Figure~\ref{fig:reddening_idust}). We found that the main difference is about the distance of the Upper Scorpius - $\rho$~Ophiuchi absorption cloud: according to these 3-d maps, the cloud is at $\sim 230$~pc, that is much farther than the Upper Scorpius Association ($\sim 140$~pc). Hence, at the distance of the programme stars, these maps give a negligible reddening. However, the observed relation between colours and spectral type requires that a large fraction of the absorbing cloud should be closer than the Upper Scorpius Association. On the other hand, stars that are closer than average have a reddening estimated from the colour-spectral type relation that are smaller with respect to the expectation for that direction (if we consider the total galactic absorption). This shows that most of the galactic absorption in this direction is due to material at a distance comparable to the Upper Scorpius Association. This agrees very well with estimates  of $131\pm 3$~pc for the distance of the absorption clouds (Lynds 1688 and 1689) associated to the Ophiuchus complex \citep{Bontemps2001, Mamajek2008}.

\begin{figure}[htb]
    \centering
    \includegraphics[width=8.5cm]{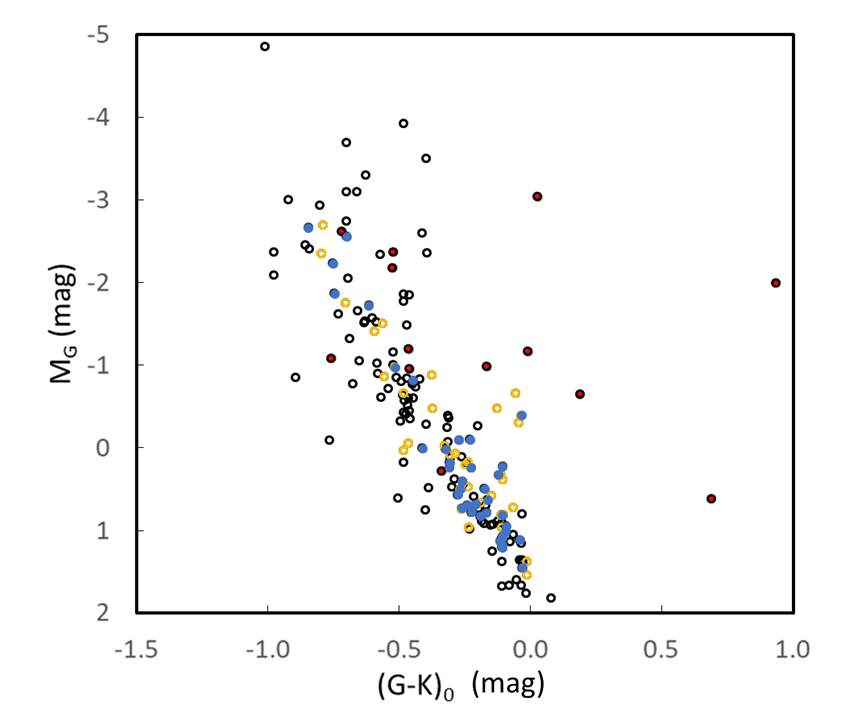}
    \caption{Dereddened $M_G-(G-K)_0$ colour-magnitude diagram for the primaries of the B-stars in Sco-Cen considered in this paper. Blue symbols are stars that have no indication of the presence of companions; black open circles are primaries in multiple systems, after deblending for the contribution of companions to the photometry; yellow open circles are stars that have indication for the presence of companions only from variation of the RV, from proper motion anomaly or GAIA RUWE parameters (see below). For these stars deblending of the companion contribution is not possible. Red filled circles are Be stars, with the disk contributing to the flux in the $K-$band. }
    \label{fig:cmd}
\end{figure}

Figure \ref{fig:cmd} shows the dereddened $M_G-(G-K)_0$ colour-magnitude diagram for the primaries of the B-stars in Sco-Cen considered in this paper. Whenever possible, deblending for the contribution of companions to the photometry was included, as explained in Section~\ref{sec:massvisual}. Of course, this correction was not needed for the single stars. In addition, it could not be applied to stars that have indication for the presence of companions only from variation of the radial velocities (RVs), from proper motion anomaly or the GAIA RUWE parameter (see next Section), because the nature of the companion is not well determined. The main sequence can be clearly seen, though there is still a quite significant scatter. Part of this scatter is due to the presence of Be stars, whose disks contribute to the flux in the $K-$band and make then the stars to appear redder than the main sequence. In addition, some spread of the main sequence is expected because of the difference in age - in fact the brightest and oldest stars are clearly evolved-off the main sequence. Finally, it is well known that rotation may also cause a broadening of the main sequence for B-stars due to both the Von Zeipel effect \citep{vonZeipel1924} and rotational induced mixing (see e.g. \citealt{Meynet1997, Heger2000, Brott2011}). However, part of the scatter is likely due to imperfections in the procedure adopted in this paper to correct for the contribution of the companions.

\subsection{Ages}
\label{sec:ages}

\begin{figure*}[htb]
    \centering
    \includegraphics[width=18cm]{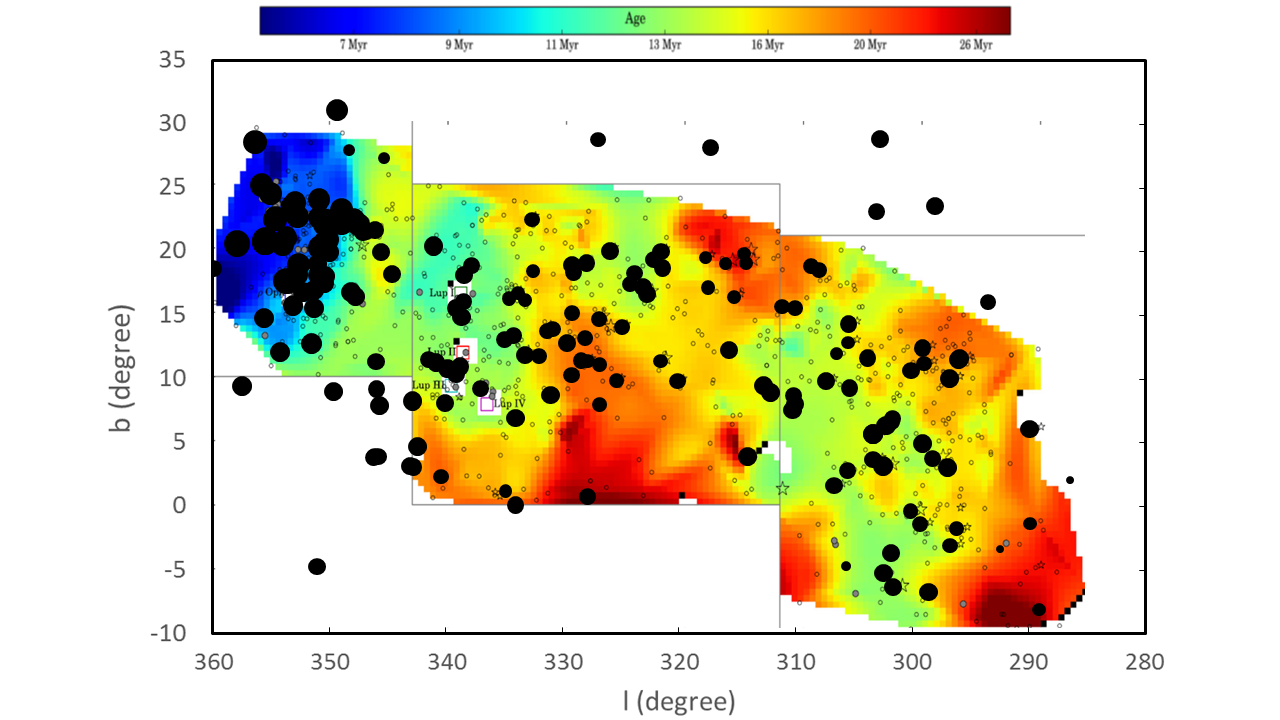}
    \caption{Map of the ages assumed for the programme stars and the age map by \citet{Pecaut2016} (coloured map in transparency; redder colours are oldest stars, blue are younger; the colour scale is above the plot). The area of the black blobs is inversely proportional to the age in our analysis. The smallest value is for HIP~54767 (84.5 Myr); this star is not actually member of Sco-Cen.}
    \label{fig:age_map}
\end{figure*}

We derived ages for the stars in the BEAST survey as described in \citet{Janson2021a}. We give preference to ages derived using common proper motion companions \citep{Squicciarini2021} and then to those obtained using the age map by \citet{Pecaut2016}. In the case of HIP~62434 we adopted an age of 12.0 Myr (see Appendix~\ref{sec:beast}). For objects not included in the BEAST survey, we adopted the age of the BEAST star projected closest to each of the remaining stars. Figure~\ref{fig:age_map} shows how ages of the star distribute on sky; this map is very similar to that obtained by \citet{Pecaut2016}. The median age is of 15.6 Myr, very close to the value usually adopted for the Sco-Cen association.

\subsection{Summary}
\label{sec:Summary}

The star with the brightest intrinsic G-magnitude in our sample is HIP~80112 ($\sigma$ Sco: $M_G=-4.85$, spectral type B1III+B1:V). The faintest one is HIP~82069 with $M_G=1.82$, that corresponds to star of 2.0 $M_\odot$ and a temperature of 9300 K according to the PARSEC isochrone \citep{Bressan2012} and to an A3.5 main sequence star with a mass of 1.86 $M_\odot$ and a temperature of 8600 K according to the table by \citet{Pecaut2013} for main sequence stars. The difference between the theoretical and empirical calibrations might be due both to the age of the Sco-Cen stars (much younger than average age for late-B early-A main sequence stars in the general field) and/or to the neglect of rotation in the stellar models we used. On this respect we notice that most of the fainter stars in the sample are fast rotators ($V \sin{i}>150$~km/s$^{-1}$: \citealt{Glebocki2005, Zorec2012, Solar2022}). On the contrary, brighter stars on average rotate slower than expected for early main sequence stars, because they are evolved off the main sequence and have then a larger radius. Most of them are indeed classified as luminosity class IV or even class III, and are $\beta$~Cephei pulsators (see e.g. \citealt{Sharma2022}).

\section{Companion detections}
\label{sec:detections}

Our study does not aim to determine all orbital parameters for the programme stars, that is in most cases beyond possibility due to the length of the orbits and scarcity of data; rather we focus on (even quite rough) determination of the masses of the components and on their semi-major axis distribution. For this reason, we did not try to find orbital solutions but rather we tried to be as complete as possible in the detection of companions over a wide range of separation (conscious that even so, a number of real companions likely went undetected). For this goal, we considered a variety of detection methods that covers a wide range in periods or semi-major axis and contrast or mass ratios. They include direct detections of the companions (visual binaries), eclipses, spectroscopy, and astrometry.

\subsection{Visual binaries}
\label{sec:visual}

Visual binaries can be detected using a variety of approaches, covering a range of different separations. Close binaries (separation of a few tenths of mas - that is of the order of a few au at the distance of Sco-Cen) have been discovered through interferometry. Binaries with separations between about 50 to 5000 mas (separation 5-1000 au) are best discovered through HCI or speckle interferometry. A complete survey of wider binaries (separation larger than 200 au) is provided by Gaia. Whenever available, both HCI and Gaia can be considered complete for stellar companions in their respective range of separation (see e.g. \citealt{Bonavita2022}), while interferometry reveals companions with a contrast up to about 4 magnitudes (in the $K$-band; see e.g. \citealt{Rizzuto2013}) that corresponds to a mass ratio $q=M_B/M_A>0.3$. Visual binaries can then provide a quite complete sample of binaries over a wide range of periods that covers the peak area of the distribution for solar-type \citep{Raghavan2010} and A-type stars \citep{DeRosa2014}.

\subsubsection{High contrast Imaging (HCI)}
\label{sec:highcontrast}

A large fraction of the stars in our sample (167 out of 181) has been observed in HCI at the ESO telescopes. 82 stars were observed with ADONIS at the ESO 3.6m telescope in the survey by \cite{Shatsky2002} and 72 in the survey by \cite{Kouwenhoven2005}. Many others were observed with NACO at VLT (see e.g. \citealt{Kouwenhoven2007a, Oudmaijer2010, Scholler2010, DeRosa2011}) and with NIRI at Gemini North (\citealt{Lafreniere2014}). A total of 127 stars have been observed with the SPHERE \citep{Beuzit2019} instrument at the ESO VLT telescope: 82 of them are in the BEAST survey \citep{Janson2021a}, 20 in the SHINE survey \citep{Desidera2021}, and 25 in other studies. Companions to 12 of the 14 stars not observed in HCI have been detected using other techniques at separation where HCI would be sensitive or shorter. The remaining two stars for which there is no information about companions are HIP~63007 and HIP~78183: they are considered as single stars throughout this paper.

Binary companions detected in the BEAST survey are described in the Appendix~\ref{sec:beast} to this paper.

Detection of stellar companions in the SHINE survey are described in \citet{Bonavita2022} and two additional BD companions to stars in our target list are described in \citet{Vigan2021}.

\subsubsection{Interferometry}
\label{sec:interferometry}

\cite{Rizzuto2013} performed a search for close companions to B-stars in Sco-Cen using the Sydney University Stellar Interferometer; three additional stars in our sample were observed by \citet{Hutter2021} using the Navy Precision Optical Interferometer and the Mark III Stellar Interferometer. Interferometric observations are then available for a total of 53 stars in our sample; 22 companions were detected around them at separations ranging from 7 to 130 mas. This corresponds to a range of projected separation between $\sim 1-18$ au. The limiting contrast of their observations (about 4 mag) implies a typical mass ratio $q=M_B/M_A>0.3$. 

Interferometry is available only for about 30\% of the star in the sample; they are bright stars with magnitude $G<4.7$, that corresponds to $M_G<-1$ (about 90\% of these bright stars have been observed). These stars have spectral type earlier than B4. The very high frequency of companions found may be a consequence of the high mass of the stars observed (masses larger than $\sim 5$ M$_\odot$ according to the table by \citealt{Pecaut2013}). In addition, we notice that most of the companions detected by \cite{Rizzuto2013} and \cite{Hutter2021} would have been discovered using other techniques too, including HCI \citep{Shatsky2002}, spectroscopic binaries  \citep{Pourbaix2004, Chini2012}, astrometric binaries \citep{Makarov2005}, proper motion anomaly \citep{Kervella2022} or a large value of the Gaia RUWE parameter, also indicative of binarity \citep{Belokurov2020}. Actually only two of the 22 companions detected using interferometry have not been also detected using alternative methods. They are the companions of HIP~81266 and HIP~86670.

\subsubsection{Gaia}
\label{sec:gaia}

Companions with projected separation larger than about 1 arcsec ($\sim 140$ au at the distance of the Sco-Cen association) have separate entries in the Gaia eDR3 \citep{Gaia2021} and DR3 \citep{Gaia2022a} catalogues; these data are available for all targets. We considered as companions objects with full (5-parameter) astrometric solution and with parallax and proper motion similar to that of the B-star. The contrast provided by Gaia allows detection of companions with mass ratios $q\sim 0.03$ - that is roughly the hydrogen burning limit for most targets in the survey - at separation larger than 5 arcsec - that is the typical limit of HCI surveys. This means that Gaia provides quite complete data about stellar companions with semi-major axis larger than about 700 au - and additional detections for closer ones also detected by HCI imaging. We limited our search to companions within 60 arcsec, that is about 8400 au. Within this limit, Gaia provided a total of 54 detections. Two of these detections (the companions of HIP~74752,  see Appendix~\ref{sec:HIP74752}, and HIP~77900, see \citealt{Petrus2020}) are actually BDs and the companion to HIP~77858 is very close to the hydrogen burning limit. Sixteen of the companions found by Gaia were also found in previous surveys using HCI and/or from previous visual binaries surveys.

There is a not negligible chance that the far companions detected this way might be stars in Sco-Cen projected close to the programme stars but unrelated to it. The typical surface density of stars in Sco-Cen (including substellar objects) is about eight stars per square degree, though it may be higher than this value in some region (e.g. Upper Scorpius). On average, we then expect to find $\sim 0.007$ unrelated Sco-Cen stars projected within 60 arcsec from any of the programme stars. The probability of finding at least one such contaminant in our sample of 181 stars is then about 71\%, and on average we expect to find 1.2 contaminants in the sample. On the other hand, the probability of finding a similar contaminant within a projected separation of 1000 au (that is, $<7$~arcsec) for a particular star is $\sim 10^{-5}$ and over the whole sample is 1.6\%. These numbers are low with respect to the observed number of detections and we neglect this small possible correction to our statistics.

\subsubsection{Visual binaries from the literature}
\label{sec:literature}

In order to be as complete as possible and refine the parameters for the multiple systems, we also inspected the Washington Double Star Catalogue \citep{Mason2001}, the Multiple Star Catalogue by \cite{Tokovinin2018}\footnote{ \url{http://www.ctio.noirlab.edu/ atokovin/stars/stars.php} }, and the catalogue of data from speckle interferometry by \cite{Mason2009, Scholler2010, Hartkopf2012}.

\subsection{Eclipsing binaries}
\label{sec:eclipsing}

Short period binaries may be discovered as eclipsing (EB) or spectroscopic binaries (SB). 

\begin{table}[htb]
  \caption[]{Stars with light curve (LC) analysis}
  \label{t:eb}
  \begin{tabular}{cccc}
  \hline
  G range &  Targets & Stars    & EB or           \\
          &          & with LC info & Reflecting\\
  \hline
$<4$      & 34 & 30 &  4 \\
$4-5$     & 37 & 29 &  2 \\
$5-6$     & 37 & 32 &  4 \\
$6-7$     & 43 & 37 &  0 \\
$7-8$     & 24 & 19 &  1 \\
$>8$      &  6 &  3 &  0 \\
  \hline
  \end{tabular}
\end{table}

We searched for EBs in the catalogues by \citet{Malkov2006} and \citet{Avvakumova2013} and added the stars found to be EBs from TESS light curves by \citet{IJspeert2021, Sharma2022}. We also inspected the short cadence TESS EBs catalogue \citep{Prsa2022}, but we found entries only for HIP 74950 and HIP 82514, both previously known EBs with adequate solutions \citep{Budding2015}. Additional stars in Upper Scorpius have been observed by K2 \citep{Rebull2018}, one of them being in common with TESS. In total, TESS short cadence or K2 light curves are available for 150 stars, that is 82.9\% of the programme stars; missing stars are in areas not covered by TESS or K2, still mostly in Upper Scorpius. None of the programme stars is in the Gaia EB catalogue \citep{Mowlavi2022}, in the Gaia DR2 variability catalogue \citep{Gaia2019}, and in the ASAS detached eclipsing variable catalogue \citep{Rowan2022} because they are too bright. Light variations for HIP~67464 ($\nu$ Cen: listed as EB in SIMBAD) and HIP~76297 ($\gamma$ Lup) are likely caused by the reflection effects of the light from the B-star on the companion, but there is no real eclipse as shown by the analysis of \citet{Jerzykiewicz2021}. To our knowledge, no additional transits were discovered by TESS or K2 around the programme stars, including HIP~65112 that is listed as an eclipsing binary in \citet{Malkov2006} and \citet{Avvakumova2013} but rather it is a pulsating variable \citep{Sharma2022}. Summarising, we found detections of nine EBs in our sample with two additional stars showing reflection light variations.

The EB  with the longest period in the sample is HIP~67669 (17.428 d, \citealt{Avvakumova2013}, where it is however noticed that it is not well clear that this object is really eclipsing), that should correspond to a semi-major axis of 0.21 au. We may assume that the TESS+K2 sample is complete up to this separation.  The EB with the lowest mass ratio is HIP~67669 ($q=0.16$) that is the only one with a secondary having a sub-solar mass ($M=0.67$ M$_\odot$, \citealt{Avvakumova2013}). However, TESS and K2 have the potentiality of discovering binaries with much lower mass ratio, down to the substellar regime, also among B-stars (see e.g. \citealt{Rizzuto2017}). So, the non-detection of such secondaries should be related to their rarity, if any.

A summary of the EB and reflection binaries discovered in our sample is given in Table \ref{t:eb}. If we limit to the sample of stars observed by TESS and K2 (where the search of EB with periods shorter that 17.4 d should be complete) the incidence of EBs is much higher for the brighter stars. While $11.0\pm 3.5$\% of the stars with $G<6$ (that should roughly correspond to $M_G$=0.6, that is expected for a B9.5 star) are EB or reflecting binaries, only one (that is $2.0\pm 2.0$\%) of the stars fainter than this limit is an EB. We further notice that some EB might have been missed among the 31 stars not observed by TESS or K2; indeed, if the fraction is the same than among the stars with TESS or K2 data ($7.3\pm 2.2$\%), we expected a couple of EBs among these stars.

\subsection{Spectroscopic binaries}
\label{sec:spectroscopic}

Since B-stars often have high rotational velocities and few lines suited for RV determinations, results cannot be of high precision and we expect that only systems with rather short periods and large RV amplitude can be discovered this way. We searched for entries corresponding to our stars in the S9 catalogue of spectroscopic binaries \citep{Pourbaix2004}, in the Multiple Star Catalogue by \cite{Tokovinin2018}, in \cite{Stock2021}, and in the list of RV measurements of B stars in the Sco-Cen association \citep{Jilinski2006}. No SB could be obtained from the Gaia catalogue because they were included in this last catalogue only if temperature is $<8300$ K \citep{Gaia2022}, that corresponds to a spectral type later than A3 according to \citet{Pecaut2013} tables. We added a few other known spectroscopic binaries \citep{Quiroga2010, Levato1987}; this makes a total of 69 stars classified as SB or EB. 

In addition, we considered stars that while not having appropriate orbital solution, have been tested for RV variations in the spectroscopic survey of bright stars by \citet{Chini2012} (92 stars in our sample, mostly among the brightest stars in the sample), \citet{Levato1987} (54 stars), \cite{Stock2021} (70 stars), and in Gaia DR3 (\citealt{Katz2022}: 37 stars, 9 of them being also in the \citealt{Chini2012} sample). For the RVs listed by Stock, we considered nightly averages and assume that the internal errors are the largest between the internal errors for individual observations and the nightly RV scatter (in both cases divided by the square root of the number of observations). We then considered as RV variables those stars whose $\chi^2>2$ with respect to a constant value; this method is similar to the usual analysis of the variance considered in this context (see e.g. \citealt{Conti1977, Levato1987}), but takes into consideration that the internal errors from this heterogeneous collection of data is highly variable. In the case of Gaia, we considered as RV variables those stars with a probability to be constant $<0.05$, which happens for 14 stars. In most cases (11 out of 14) robust RV amplitudes are quite large ($>9$ km/s, with four cases $>50$ km/s, likely associated to compact systems with large secondaries. These are HIP~78702, HIP~79739, HIP~80815, and HIP~83508\footnote{All these objects do not have TESS photometry; the two first have been observed by K2 and inspection of the light curve does not indicate they are EB. For the two last, we cannot exclude that they are undetected EBs or reflecting binaries.}), but in three cases (HIP~62434, HIP~76395, and HIP~81474) they are tiny ($<5$ km/s) and could only be found because their specific internal errors are small. We also considered six stars with RVs from HARPS spectra \citep{Trifonov2020}; in this case we find significant variations only for stars that were already known to be SB. Finally, we searched for our stars in the AMBRE project catalogue \citep{Worley2012} based on spectra acquired with FEROS at ESO La Silla, but found data - and no RVs - only for HIP~75264 ($\epsilon$~Lup), that is a known SB2 \citep{Thackeray1970, Pablo2019}.

Combining all these data, we found that information about eclipses and/or RV variations are available for 155 of the programme stars; in addition to the 69 EB or SB, RV has been found to be variable for 18 more objects, though it is not clear if in all cases the variability is due to a Keplerian motion. A total of 68 stars do not show RV variations; we assumed their RV to be constant. The longest period found for an SB in our sample is about 11 yr, corresponding to a semi-major axis of about 30 au; however, the vast majority of the objects have shorter periods, corresponding to separation $<2$ au and RV semi-amplitude $>30$ km/s. The actual fraction of short period binaries is not well known, because it is not clear how many stars were really tested for RV variations; for a summary, see Table \ref{t:sb}. It should be noticed that while 43 out of 68 stars with $G<5$ (that is, 63\%) have been found to be SB\footnote{If we include also stars whose RV is variable, 53 out of 68 stars with $G<5$ are possible SB, that is 78\%.}, the fraction of known SB is much lower for fainter stars. This is likely influenced by a higher fraction of SB among the most massive stars: for comparison, we notice that \citet{Chini2012} found that for the B stars the radial velocity variability fraction decreases from 61\% for B0 to 15\% for B9 stars. However, the lower number of SB known among late B stars also reflects incompleteness at faint magnitudes; in fact, while all the stars with $G<5$ have been tested for RV variation, this happens only for 60\% of the stars with $G>7$. 

\begin{table}[htb]
  \caption[]{Stars with RV analysis}
  \label{t:sb}
  \begin{tabular}{cccccc}
  \hline
  G range &  Targets & Stars with   & SB or EB  &  RV  & RV       \\
          &          & RV info &     &  Var & constant \\
  \hline
$<4$      & 34 & 34 & 23 & 5 &  6 \\
$4-5$     & 37 & 34 & 20 & 5 &  9 \\
$5-6$     & 37 & 31 & 13 & 2 & 16 \\
$6-7$     & 43 & 38 &  8 & 3 & 27 \\
$7-8$     & 24 & 16 &  5 & 3 &  8 \\
$>8$      &  6 &  2 &  0 & 0 &  2 \\
  \hline
  \end{tabular}
\end{table}

The period range where binaries are detected as SB overlaps with detections of companions by some alternative techniques but not with others. For instance, only one of the stars known as an SB in our sample has a marginally significant proper motion anomaly according to \cite{Kervella2022} (see next subsection). On the other hand, a Gaia \citep{Gaia2022a} RUWE parameter larger than 1.4, which is indicative of binarity \citep{Belokurov2020} for stars with $G>4$ (see subsection~\ref{sec:ruwe}), has been found for 15 out of 34 of the known SB. On the other hand, a high RUWE was obtained for 9 stars with $G>4$ not classified as SB, though in three cases there is indication of RV variation. Hence, 62\% of the stars with RUWE$>1.4$ and $G>4$ are classified as SB and for 12\% more there is indication of RV variation.

We have RV information for a total of 47 of the 53 stars observed in interferometry by \citet{Rizzuto2013}, so there is a considerable overlap of the target samples because in both cases the main focus is on bright stars. Nine of the SB have also been detected in interferometry by \citet{Rizzuto2013}; three of them are systems of higher multiplicity, and the object detected in interferometry is not that responsible for the RV variations, the period of the SB being too short. The SB recovered in interferometry are those with periods leading to semi-major axis in the range 1-15 au (separation between 7 and 100 mas) and with mass ratios $q>0.3$. We note that three additional SB with similar periods were not detected in interferometry, though the target was observed by \citet{Rizzuto2013}; likely, they all have low values of $q<0.3$, that is below the expected threshold for detection. On the other hand, there are nine binaries detected in interferometry that are not classified as SB, though some data are available about RV variations; four of them are listed as RV variables in \cite{Chini2012} or have variable RV in Gaia DR3 \citep{Gaia2022a}, but the remaining five are considered as RV constant. This implies that 53\% of the companions detected in interferometry are classified as SB and for an additional 24\% there is indication of RV variation.

The systems classified as RV constant but detected as binary through astrometry or interferometry might be seen quite face on and/or have rather long periods; in these cases the RVs variations are too small to be detected. In both cases, they are about 1/4 of the total.

In addition to the classical SB searches, based on variations of the RVs, \citet{Gullikson2013} and \citet{Gullikson2016a, Gullikson2016b} developed a method based on the detection of signatures of the secondary on the high resolution spectrum of the star. This method is well suited for high contrast systems including an early type primary and a late-type secondary seen at short separation. The method works better in the near infrared, where the contrast between the two components is minimised. 18 of the stars in our sample have been analysed with this technique, and 6 companions have been found. All of them were also discovered as SB.

\subsection{Astrometric binaries}
\label{sec:astrometric}

We inspected the catalogue of astrometric binaries by \cite{Makarov2005}. None of the programme stars is included either in the Two Body Orbit catalogue or in the Accelerating Star catalogue of Gaia DR3 \citep{Gaia2022}. This is because these catalogues do not include early-type stars ($M_G\leq 1$).

\subsection{Gaia-Hipparcos proper motion anomaly}
\label{sec:pma}

\cite{Kervella2022} compared the proper motions in the Hipparcos and Gaia catalogues with that determined using the positions in these catalogues; if this proper motion is considered the real proper motion of the stars, the residuals, called Proper Motion anomaly (PMa), are a measure of the acceleration at the Hipparcos and Gaia epochs. They are then indication of the presence of companions. This datum is available for 158 of the stars in our sample - all missing objects but four are bright stars with $G<3.3$\ for which the Gaia DR3 solution \citep{Gaia2022a} is not reliable. We found that the PMa determined by \cite{Kervella2022} has $SNR>4$ for 44 of the programme stars; in 22 cases the object responsible for the PMa has been also observed using direct imaging, interferometry, and RV. A comparison with detections using these other techniques shows that the PMa is expected to be significant for objects with separation between approximately 2 and 40 au, with a peak sensitivity in the range $5-12$~au. We then considered the remaining 22 objects with a significant PMa as binary detections using the PMa method in this period range. At the distance of Sco-Cen, the PMa may be sensitive to stars with an even low value of the mass ratio $q\geq 0.01$.

\subsection{Gaia {\it RUWE} parameter}
\label{sec:ruwe}

The renormalised unit weight error (RUWE)\ parameter is an indication of the goodness of the 5-parameters solution found by Gaia \citep{Lindegren2018}. \citet{Belokurov2020} showed that a high value of this parameter is an indication of binarity, at least for stars that are not too bright ($G>4$) and saturated in the Gaia scans; the threshold value is usually set at RUWE$>1.4$. This method is sensitive to systems with periods from a few months to a decade \citep{Penoyre2021}. The RUWE parameter is available for 144 out of the 147 stars with $G>4$; 31 of them have RUWE$>1.4$. 

Following \citet{Belokurov2020}, we expect that at the distance of the Sco-Cen association (100-150 pc) binaries with a value of RUWE$>1.4$ have an astrometric signal $\delta \theta>0.3$ mas (that is a projected shift of $>0.04$ au). Still following the relations given by \cite{Belokurov2020}, this corresponds to binaries with semi-major axis in the range 0.15-10 au (see Appendix~\ref{sec:bds}) for a more detailed explanation). We note that RUWE is mostly sensitive to binaries with intermediate values of the mass ratio $q\sim 0.5$. 

Binaries signalled by a high RUWE value are closer than those with significant PMa. There should however be a region of overlap, for binaries with semi-major axis in the range 2-10 au; these are companions that could be detected through interferometry (provided they have a mass ratio $q>0.3$), but are difficult to be discovered with other techniques. There are indeed ten stars with high RUWE and significant PMa; none of them is a SB; one of them has variable RV and 6 of them are labelled as constant in RV searches, suggesting a separation of at least a few au or a quite low mass ratio $q$. Only one of the companions was found in direct imaging campaigns (HIP~73624) suggesting that the remaining objects are closer than about 15 au. All these targets are too faint to be included in the interferometric surveys by \cite{Rizzuto2013} and \cite{Hutter2021}. Hence, almost half of the binaries discovered through PMa but not other techniques also have a high RUWE value indicative of binarity. 

\section{Mass and semi-major axis determination for the individual components}
\label{sec:mass}

We derived estimates of the masses and of the semi-major axis for all the 200 companions found. We describe the methods in this Section.

\begin{figure}[htb]
    \centering
    \includegraphics[width=8.5cm]{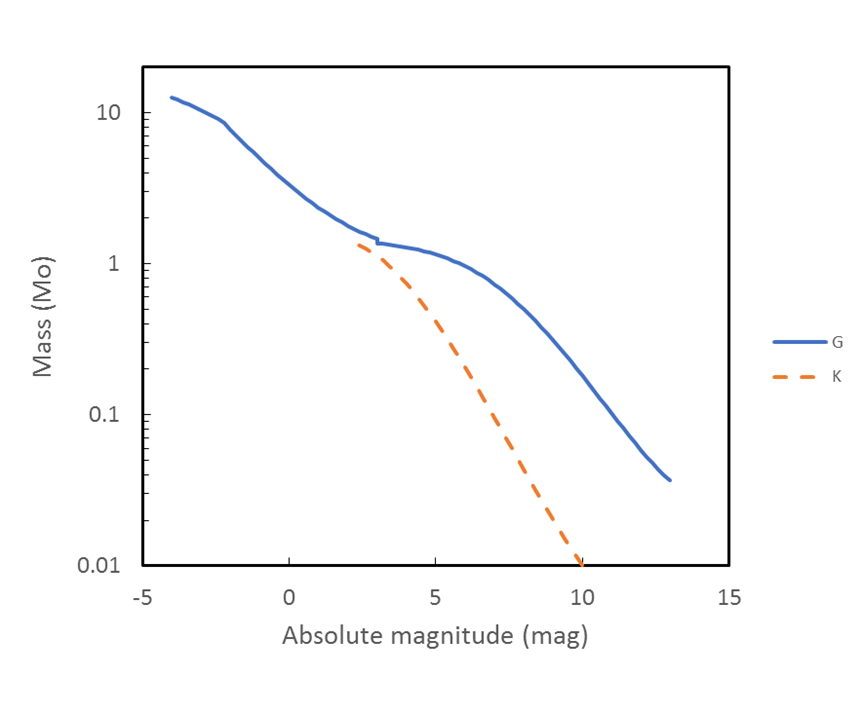}
    \caption{Relation between absolute magnitudes in the Gaia $G$ \citep{Gaia2022a} and 2-MASS $K$-band \citep{Skrutskie2006} and masses adopted for visual binaries}
    \label{fig:calibration}
\end{figure}

\subsection{Visual binaries}
\label{sec:massvisual}

The 124 visual companions among B-stars in Sco-Cen typically have long periods and, in general, there are not accurate orbital solutions available. When photometry is available for the individual components, masses could be obtained from the absolute magnitudes in the Gaia $G$ \citep{Gaia2022a} and 2-MASS $K$-band \citep{Skrutskie2006} and relations between masses and absolute magnitudes in the relevant bands. For the brightest stars ($M_G<-2$) that are evolved off the MS, these were obtained using the solar metallicity PARSEC isochrone \citep{Bressan2012} with an age of 15 Myr (appropriate for most of the Sco-Cen stars: \citealt{Pecaut2012}). For stars with $-2<M_G<3$ that are very close to the zero age main sequence we used the table by \citet{Pecaut2013}. For fainter stars that are still in the pre-main sequence phase we used the isochrones by \citet{Baraffe2015} (and the appropriate ages). The finally adopted calibrations are shown in Figure~\ref{fig:calibration}. Whenever the observed magnitudes refer to blended images and the mass difference between the components is expected to be less than a factor of three, corrections were considered to split the luminosity among the various components according to the measured contrast. This correction was applied for separation $<$0.3 arcsec for Gaia and $<2$ arcsec for 2MASS.

For these systems we assumed that the semi-major axis (in au) is equal to the projected separation divided by the parallax, that corresponds to the eccentricity distribution considered by \citet{Ambartsumian1937} of $f(e)=2 e$ (see \citealt{Brandeker2006}). This last paper indicates that this assumption underestimates the semi-major axis by about 25\% in the case of circular orbits.

We notice here that the ages we adopted for two stars with BD companions (HIP~78530 and HIP~78968, both in Upper Scorpius) are substantially younger than considered in the original analysis \citep{Vigan2021, Kouwenhoven2007a}. This results in lower masses of 19 and 22~M$_{\rm Jupiter}$ for HIP~78530B and HIP~78968B, respectively. While both objects are still BDs, as they were classified in the original papers, they are now considered to be closer to the Deuterium burning limit. The mass ratio values are $q=0.0083$ and 0.0104, respectively.

\subsection{Close binaries}
\label{sec:massclose}

\begin{table*}[htb]
  \caption[]{Parameters for eclipsing and reflecting binaries in the Sco-Cen association. }
  \label{t:eb2}
  \begin{tabular}{ccccccccl}
  \hline
HIP    &  $G$  &  per & a     & M$_A$  & M$_B$   & R$_A$  &   R$_B$    &  Ref. \\
       & mag & d     &  au & M$_\odot$ & M$_\odot$ & R$_\odot$ & R$_\odot$ &       \\
  \hline
\multicolumn{9}{c}{Eclipsing binaries}\\
63210 &  5.137 & 0.6496 & 0.026 & 3.32 & 2.37 & 2.09 & 1.67 & \citet{Harmanec2010}\\
64425 &  4.522 & 0.642  & 0.023 & 3.73 & 2.67 & 2.2  & 1.7  & \citet{Budding2010}\\
67669 &  4.513 & 17.428 & 0.212 & 4.20 & 0.76 &      &      & \citet{Avvakumova2013}\\
73266 &  7.256 &  0.586 & 0.019 & 2.32 &      &      &      & \citet{Sharma2022}\\ 
73807 &  4.602 & 15.5   & 0.245 & 4.70 & 3.48 &      &      & Appendix~\ref{sec:HIP73807} \\
74950 &  5.568 &  1.85  & 0.055 & 4.16 & 2.64 & 2.42 & 1.79 & \citet{Budding2015} \\
76600 &  3.646 &  3.448 & 0.082 & 6.10 & 2.74 &      &      & Appendix~\ref{sec:beast} \\
78168 &  5.810 &  9.20  & 0.152 & 5.58 & 2.62 & 2.73 & 1.69 & \citet{David2019, Maxted2018} \\
82514 &  3.070 &  1.446 & 0.051 & 8.30 & 4.60 & 3.9  & 4.6  & \citet{Budding2015,IJspeert2021} \\ 
\multicolumn{9}{c}{Reflecting binaries}\\
67464 &  3.289  & 2.625 & 0.077 & 8.70 & 1.04 &      &      & \citet{Jerzykiewicz2021} \\
76297 & 3.000  & 2.808 &  0.078 & 8.17 & 1.08 &      &      & \citet{Jerzykiewicz2021} \\
  \hline
  \end{tabular}
\end{table*}

Whenever possible (11 objects), masses and semi-major axis for EBs and reflecting variables were obtained from detailed studies \citep{Harmanec2010, Budding2010, Budding2015, Maxted2018, David2019, Jerzykiewicz2021} or from our reanalysis of existing data for the grazing eclipsing binaries HIP~73807 ($\pi$ Lup: see Appendix~\ref{sec:HIP73807}) and HIP~76600 (see Appendix~\ref{sec:beast}). We have not enough data about the secondary star in HIP~73266 \citep{Sharma2022}. Main data adopted in this paper are collected in Table \ref{t:eb2}.

Leaving aside the EBs, 15 of the remaining 28 SBs are SB2. The mass ratio for SB2 can be obtained from the ratio of the semi-amplitude of the two radial velocity curves. Masses for the individual components can then be derived from the observed total absolute $G$ magnitude and mass ratios assuming that the two components are normal main sequence stars obeying the mass - absolute $G$ magnitude relation used for the visual binaries. Once masses and periods are known, the semi-major axis can be obtained using the third Kepler law. In addition, for two stars with high mass ratio analysis of the individual components relevant data were available from \citet{Gullikson2016b} and \citet{Stelzer2006}; and for three additional stars we used the analysis made in the Multiple Star Catalogue by \citet{Tokovinin2018}.

For the remaining eight SB with orbit determination we can only use statistical arguments about the inclination. We assumed the median value of $i=60$ degree in order not to bias the sample.

\begin{figure}[htb]
    \centering
    \includegraphics[width=8.5cm]{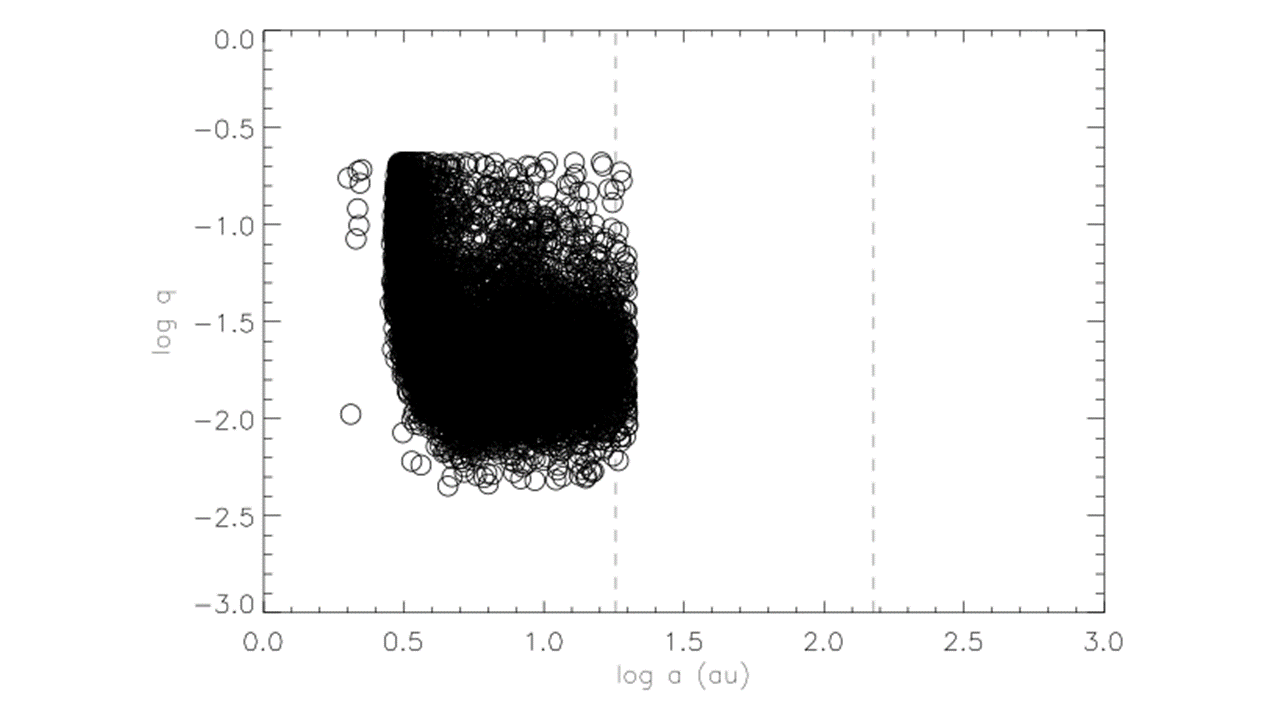}
    \includegraphics[width=8.5cm]{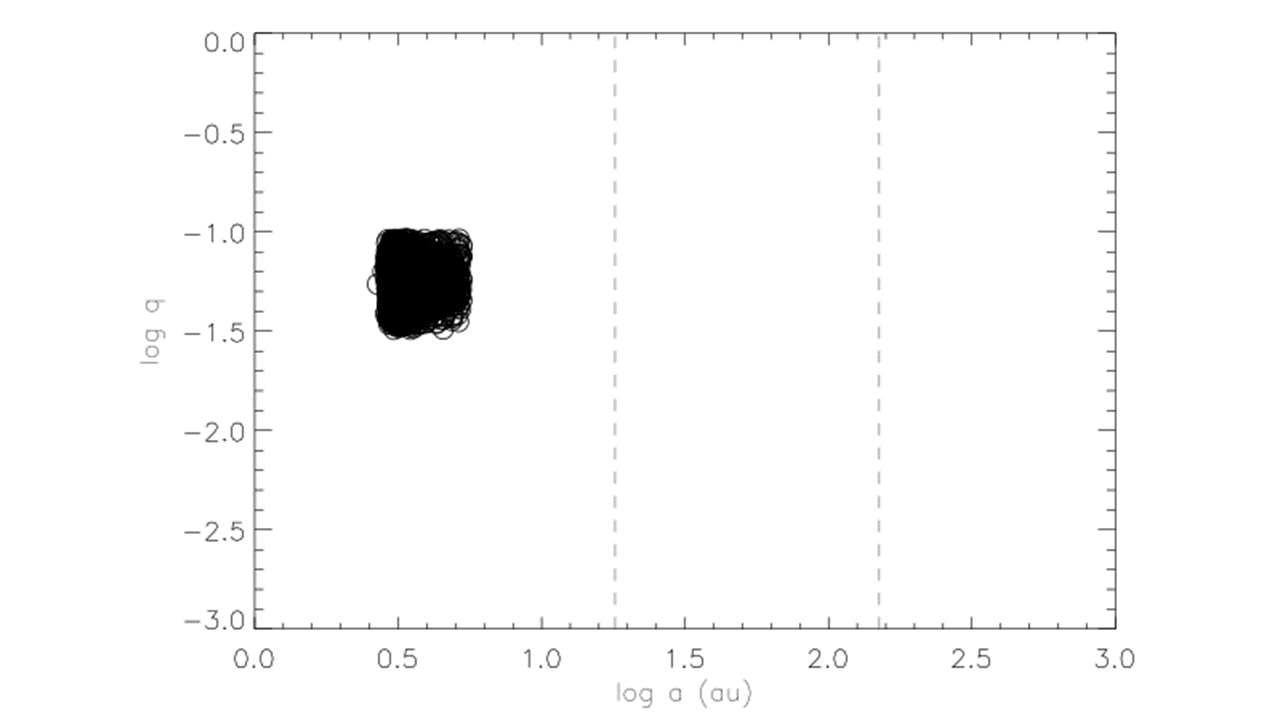}
    \includegraphics[width=8.5cm]{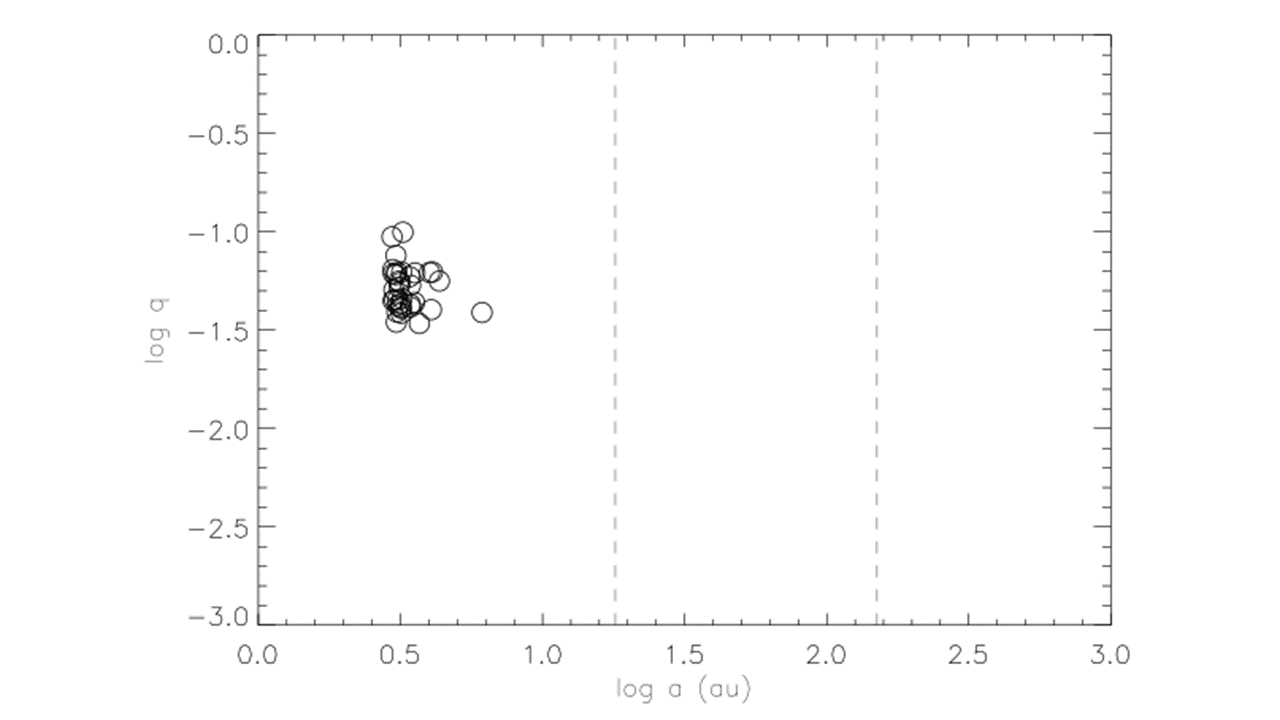}
    \caption{Values of the semi-major axis $a$ and mass ratio $q$ compatible with the observed value of PMa, RUWE, and scatter in RVs for a close companion to HIP~60855. The upper panel shows the values obtained only using the PMa; the middle panel those obtained also considering the RUWE, and the bottom panel the solutions compatible with all data. Dashed lines mark the semi-major axis corresponding to a projected separation of 0.12 and 1 arcsec, the approximate limit for detection using high contrast imaging (HCI) and Gaia, respectively. This particular companion is not expected to be detectable as a visual binary }
    \label{fig:dynamic_method}
\end{figure}

\subsection{Binaries only detected through PMa, RUWE, and RVs}
\label{sec:massdynamics}

\begin{figure}[htb]
    \centering
    \includegraphics[width=8.5cm]{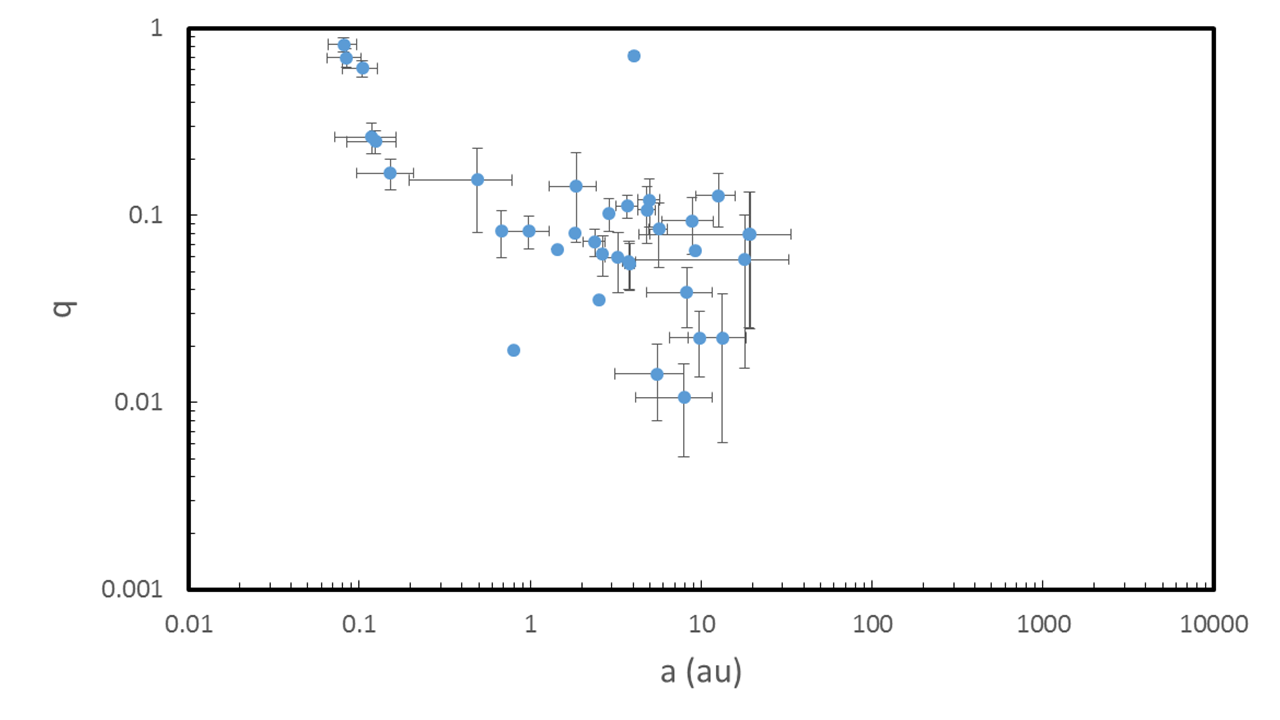}
    \caption{Relation between semi-major axis and mass ratio for the companions of the Sco-Cen stars discovered through Gaia PMa, RUWE, or variation in RVs that are not visual binaries or do not have orbit determination}
    \label{fig:dynamic_mass}
\end{figure}

Indication for binarity for 37 objects comes from PMa, RUWE, and RVs, but the secondary was not observed as a separate object and no period was determined. For these objects, we looked for solutions that are compatible with the observed values of the RUWE, of the PMa, and whenever available with the scatter in RV and the non-detection in HCI. This was done exploring the semi-major-axis - mass ratio plane using a Monte Carlo code. For simplicity, we adopted circular orbits\footnote{We also run the case of eccentric orbits, with uniform priors between 0 and 1 on eccentricity, 0 and 180 degrees in the ascending node angle $\Omega$, and 0 and 360 degrees in the periastron angle $\omega$. On average, we obtained differences of the values of $\Delta\log{a}=0.04\pm 0.03$ dex, r.m.s=0.15 dex, and $\Delta\log{q}=-0.02\pm 0.04$ dex, r.m.s=0.20 dex, in the sense circular - eccentric, for semi-major axis $a$ and mass ratios $q$, respectively. The r.m.s. of the differences obtained with circular and eccentric orbits is similar to the scatter of the values obtained for acceptable solutions for both assumptions (0.15 dex for both $\log{a}$ and $\log{q}$). We conclude that the assumption of circular rather than eccentric orbits does not affect significantly the derivation of $a$ and $q$, likely because not enough data are available.} but we left the inclination and phase to assume a random value. The adopted final values are the mean of those for solutions compatible with observations within the errors, and the uncertainty is the standard deviation of this population. An example of the derivation of $a$ and $q$ using this approach is shown in Figure~\ref{fig:dynamic_method}. Relevant data for all the stars for which we applied this method are given in Table~\ref{t:mass_dyn}; data without error bars are highly uncertain. We notice that the probability that the RV is constant is high ($>0.5$) for two stars with RVs from Gaia (HIP~62058 and HIP~65965), in spite of the fact that the ratio between the amplitude and internal error is also quite high. For these stars, we did not consider the RVs in the analysis. 

Figure~\ref{fig:dynamic_mass} shows the position of the companions discussed in this Section in the semi-major axis and mass ratio plane. These objects typically have quite low values of the mass ratio ($q\sim 0.1$) and semi-major axis in the range 0.1-20 au. Three of the stars (HIP~59173, HIP~62058 and HIP~64053) might have substellar companions. They are discussed in Appendices~\ref{sec:beast} and \ref{sec:bds}.

\begin{table*}[htb]
  \caption[]{Mass determination for additional stars with significant RUWE, PMa, or scatter in RVs.}
  \label{t:mass_dyn}
  \begin{tabular}{lccccccccccc}
  \hline
HIP     &       HD      &       RUWE    &       N$_{\rm obs}$   &       RV err     &       RV amp  &       Source  &       SNR(PMa)        &       M$_B$   &       err(M$_B$)      &       a       &       err(a)  \\
        &               &               &               &       km/s    &       km/s    &               &               &       M$_\odot$       &       M$_\odot$       &       au      &       au      \\
 \hline
54767   &       95783   &       0.932   &       5       &       0.18    &       1.47    &       Stock   &       0.08    &       0.335   &       0.06    &       ~0.231  &       0.084   \\
58901   &       104900  &       2.750   &               &               &               &       Chini   &       35.28   &       0.307   &       0.088   &       ~4.987  &       0.700   \\
59173   &       105382  &       1.440   &       5       &       1.78    &       1.35    &       Stock   &       6.40    &       0.062   &       0.027   &       5.533   &       2.414   \\
59747   &       106490  &       4.658   &       6       &       9.6     &       2.47    &       Stock   &       1.79    &       0.226   &               &       ~0.798  &               \\
60710   &       108257  &       1.489   &       2       &       3.53    &       10.54   &       Chini   &       1.17    &       0.730   &       0.350   &       ~0.491  &       0.294   \\
60855   &       108541  &       1.971   &       17      &       3.17    &       9.09    &       Gaia    &       5.98    &       0.233   &       0.082   &       ~3.259  &       0.528   \\
62058   &       110506  &       0.874   &       33      &       2.70    &       8.28    &       Gaia    &       5.13    &       0.050   &       0.026   &       ~7.913  &       3.773   \\
62327   &       110956  &       1.651   &       3       &       1.34    &       1.63    &       Stock   &       1.07    &       0.340   &       0.068   &       ~0.982  &       0.312   \\
66454   &       118354  &       1.133   &       12      &       12.65   &       23.55   &       Stock   &       14.97   &       0.125   &       0.044   &       ~8.253  &       3.437   \\
68282   &       121790  &               &       3       &       0.07    &       1.45    &       Stock   &               &       0.557   &       0.158   &       ~0.679  &               \\
71353   &       127971  &       3.166   &       3       &       0.16    &       0.44    &       Stock   &       4.04    &       0.193   &       0.032   &       ~2.386  &       0.349   \\
75141   &       136298  &               &       8       &       3.20    &       7.08    &       Levato  &       12.21   &       0.224   &       0.085   &       ~9.788  &       3.289   \\
76126   &       138485  &       1.756   &       8       &       3.25    &       7.78    &       Levato  &       6.20    &       0.331   &       0.095   &       ~3.783  &       0.335   \\
78702   &       143956  &       1.123   &       30      &       18.90   &       54.52   &       Gaia    &       1.54    &       1.344   &       0.136   &       ~0.104  &       0.024   \\
79044   &       144591  &       1.016   &       13      &       9.62    &       2.73    &       Gaia    &       3.31    &       0.600   &       0.087   &       ~0.124  &       0.039   \\
79098   &       144844  &       2.693   &       8       &       3.39    &       8.65    &       Levato  &       2.36    &       0.526   &       0.263   &       ~1.859  &       0.561   \\
81316   &       149425  &       0.940   &       14      &       3.19    &       5.59    &       Gaia    &       2.17    &       0.200   &               &       ~1.433  &               \\
81472   &       149711  &       0.684   &       5       &       0.10    &       2.60    &       Stock   &       5.45    &       0.778   &       0.110   &       ~3.819  &       0.733   \\
81474   &       149914  &       0.984   &       6       &       12.91   &       5.9     &       Stock   &       1.46    &       0.134   &               &       ~2.525  &               \\
81914   &       150591  &       0.764   &               &               &               &               &       41.59   &       0.460   &       0146    &       12.601  &       3.311   \\
53701   &       95324   &       2.535   &               &               &               &       Chini   &       5.26    &       0.159   &       0.039   &       ~2.655  &       0.142   \\
62683   &       111597  &       2.076   &               &               &               &       Chini   &       28.99   &       0.353   &       0.132   &       ~5.653  &       0.683   \\
64053   &       113902  &       1.365   &               &               &               &       Chini   &       5.82    &       0.058   &       0.042   &       13.322  &       4.892   \\
74479   &       134837  &       1.242   &               &               &               &       Chini   &       17.43   &       0.246   &       0.168   &       19.049  &       14.729  \\
75915   &       137919  &       1.113   &               &               &               &               &       11.82   &       0.171   &       0.126   &       17.951  &       14.682  \\
79399   &       145483  &       2.406   &               &               &               &       Chini   &       15.68   &       0.180   &       0.050   &       ~3.825  &       0.277   \\
80815   &       148605  &       1.263   &       13      &       35.96   &       98.66   &       Gaia    &       1.25    &       2.971   &       0.275   &       ~0.081  &       0.015   \\
83508   &       154021  &       0.844   &       19      &       31.51   &       70.00   &       Gaia    &       1.99    &       1.670   &       0.189   &       ~0.084  &       0.019   \\
66651   &       118697  &       1.071   &               &               &               &       Chini   &       30.82   &       0.184   &       0.062   &       8.857   &       2.967   \\
67973   &       121190  &       1.044   &       15      &       6.16    &       19.40   &       Gaia    &       1.00    &       0.801   &       0.145   &       ~0.118  &       0.046   \\
71724   &       128819  &       1.124   &       12      &       4.75    &       12.72   &       Gaia    &       24.60   &       0.372   &       0.259   &       19.640  &       17.260  \\
78754   &       143927  &       2.929   &               &               &               &               &       7.31    &       0.281   &       0.055   &       ~2.884  &       0.103   \\
79031   &       144661  &       1.531   &       7       &       2.32    &       6.28    &       Levato  &       2.27    &       0.257   &               &       1.820   &               \\
79599   &       145964  &       0.882   &       10      &       4.40    &       11.55   &       Gaia    &       1.10    &       0.407   &       0.077   &       ~0.152  &       0.056   \\
65965   &       117484  &       0.948   &       13      &       4.33    &       11.95   &       Gaia    &       4.25    &       0.134   &                &       ~9.201  &               \\
80063   &       147103  &       2.318   &               &               &               &               &       27.85   &       0.311   &       0.104   &       4.827   &       ~0.585  \\
80371   &       147701  &       0.986   &               &               &               &               &       19.73   &       0.267   &       0.184   &       19.370  &       14.330  \\
  \hline
  \end{tabular}
  \\
  Note: in the case of RVs from the compilation by Stock, N$_{\rm obs}$ is the number of nightly averages
\end{table*}

\subsection{Summary}
\label{sec:masssummary}

\begin{table}
\caption{Summary of companion detections using the various techniques.}
\begin{tabular}{lccc}
\hline
Method                      & Sample & Multiples & Comp. \\
\hline
Overall                     &   181  &    138    &   200      \\
            &        &           &            \\
Visual binaries             &        &           &            \\
- HCI                       &   167  &     56    &    65      \\
- Gaia separate entries     &   181  &     49    &    54      \\
- Interferometry            &   ~52  &     22    &    22      \\
            &        &           &            \\
Eclipsing binaries          &   150  &     11    &    11      \\
            &        &           &            \\
Spectroscopic binaries      &   155  &     87    &    87      \\
            &        &           &            \\
Astrometric binaries        &        &           &            \\
- RUWE                      &   144  &     31    &    31      \\
- PMa                       &   158  &     44    &    44      \\
\hline
\end{tabular}
\label{t:summary}
\end{table}

Table~\ref{t:summary} gives a summary of the companion detections obtained with the various techniques. Several companions were detected using multiple techniques, so the sum of the detections with different methods is much larger than the actual number of detected companions. Tables in Appendix~\ref{sec:longtables} gives details for the individual stars.

\section{Companion search completeness}
\label{sec:completeness}

\begin{figure*}[htb]
    \centering
    \includegraphics[width=18.0cm]{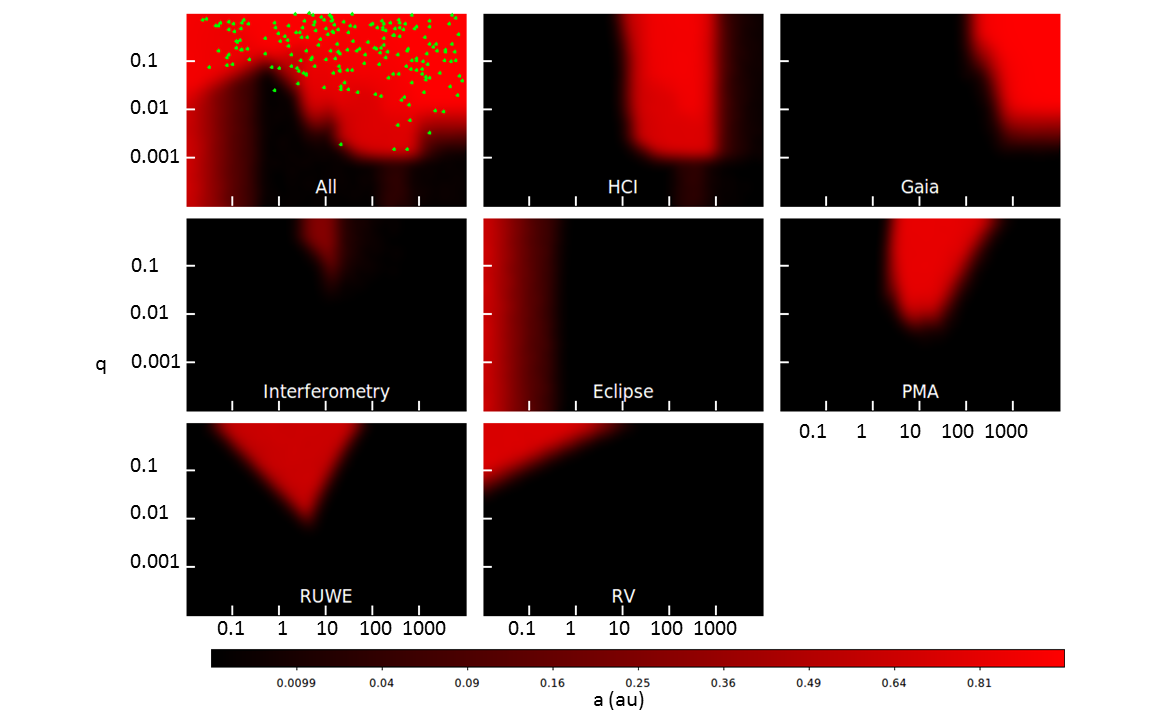}
    \caption{Completeness map of the search of companions in the semi-major axis $a$ - mass ratio $q$ plane. The upper left panel is the result obtained using all the techniques considered in this paper; the remaining panels are results for the individual techniques:  HCI = high contrast imaging; Gaia = separate entry in the Gaia catalogue; Interferometry; Eclipsing binaries; PMa = Proper Motion Anomaly; RUWE = Gaia goodness of fit RUWE parameter; RV = spectroscopic binaries. Different level of completeness are shown as different colours; the colour scale used is shown on bottom of the figure. Green points in the upper left image are the actual detections }
    \label{fig:completeness}
\end{figure*}

We found a total of 200 companions for which data about separation and mass were available. While extensive, this list is still likely incomplete. We prepared a Monte Carlo procedure in order to estimate the completeness of the search for companions around the programme stars. For simplicity, we considered circular orbits\footnote{Assuming circular orbits simplifies the issue, because we do not need to consider the effect of eccentricity and longitude of periastron, while the argument of the ascending node is however irrelevant in this context. On the other hand, the average relation between projected separation and semi-major axis depends on the actual distribution of eccentricities \citep{Brandeker2006}. Over a large sample, the difference is however at most 25\%, so the effect on our results is marginal.}. For each of the stars in the sample (with its own distance, reddening, and primary mass), we made an extraction of 10,000 companions with random values of the semi-major axis $a$, mass ratio $q$, inclination $i$ and phase. We considered uniform distribution in the logarithm for $a$ and $q$, (with $0.01<a<10000$ and $0.0001<q<1$), uniform distribution of phase between 0 and 1, and an isotropic distribution of inclinations. For each of these companions we estimated the relevant parameters: magnitudes in the $G$, $J$, and $K$ bands, position along the orbit at the observing epoch, appropriate value of the PMa and of the Gaia RUWE parameters, RV variation, and if the companion is transiting on the primary. In particular, the PMa and RUWE parameters were derived simulating a sequence of 70 Gaia visits uniformly spaced in time over the 34 months considered by Gaia DR3, each providing a position error of 0.3 mas along each of the coordinates. We then considered if the relevant observation is available for each of the targets, and compared the predicted signals with the detection limits for the various techniques. These limits were obtained as follows:

{\bf HCI:} we considered two different classes of observations: those obtained with ADONIS at the ESO 3.6 meter, and higher quality observations obtained with high contrast imagers equipped with coronagraphs on 8m telescopes (NACO and SPHERE at VLT, and GPI at Gemini). In the first case we used the limiting contrast given by \citet{Kouwenhoven2007a}; in the second one the curve shown in Figure~\ref{fig:contrast_limit} in Appendix~\ref{sec:beast}.

{\bf Separate entries in the Gaia DR3 catalogue:} we considered detectable those objects with $G<20$ at separation $>5$~arcsec and $G>10$ at 1 arcsec; we interpolated between these two values for separation between 1 and 5 arcsec for intermediate separations. This curve well reproduce the sensitivity limit found by \citet{Brandeker2019}

{\bf Interferometry:} The limiting contrast is as given by \citet{Rizzuto2013}.

{\bf Eclipsing binaries:} We assumed that all transiting sources with period $P<28$~days could be detected by either TESS or K2.

{\bf Proper Motion Anomaly:} We assumed that the companion is detected if the $SNR(PMa)>4$.

{\bf Gaia goodness of fit RUWE parameter:} We assumed that the companion is detected if RUWE$>1.4$.

{\bf Spectroscopic binaries:} after examination of available data bases, we considered detectable all those companions causing an r.m.s. of the velocities $>5.6$~km/s on a typical time interval of 3 years; this is twice the median internal error in the RVs.

We derived completeness by comparing the number of detected companions with that of simulated ones. To obtain maps in the semi-major axis $a$ - mass ratio $q$ plane (rather than simply clouds of points), we smoothed the maps of both simulated and detected objects with a bi-dimensional Gaussian with $\sigma=0.1$~dex in the logarithm. The various panels of  Figure~\ref{fig:completeness} show the overall completeness obtained combining the different techniques (that is, at least with one of these techniques), as well as those obtained for each individual technique. While none of the techniques alone cover the whole parameter space, we notice that their combination makes the search fairly complete for stellar companions ($q>0.02$) with semi-major axes larger $>3$~au. Within this range, the median completeness is 97\% and the minimum value (at the short separation, low mass ratio) is 44\%. The completeness is lower at shorter separation, but still quite good.  In fact, median completeness is 87\% for the stellar companions with $a<3$~au, and the fraction rises to 91\% for companions more massive than the Sun. However, only 47\% of the companions less massive than the Sun and with $a<3$~au are detected.

\section{Multiple stars statistics}
\label{sec:statistics}

\subsection{Binary fraction}
\label{sec:fraction}


\begin{figure*}[htb]
    \centering
    \includegraphics[width=18.0cm]{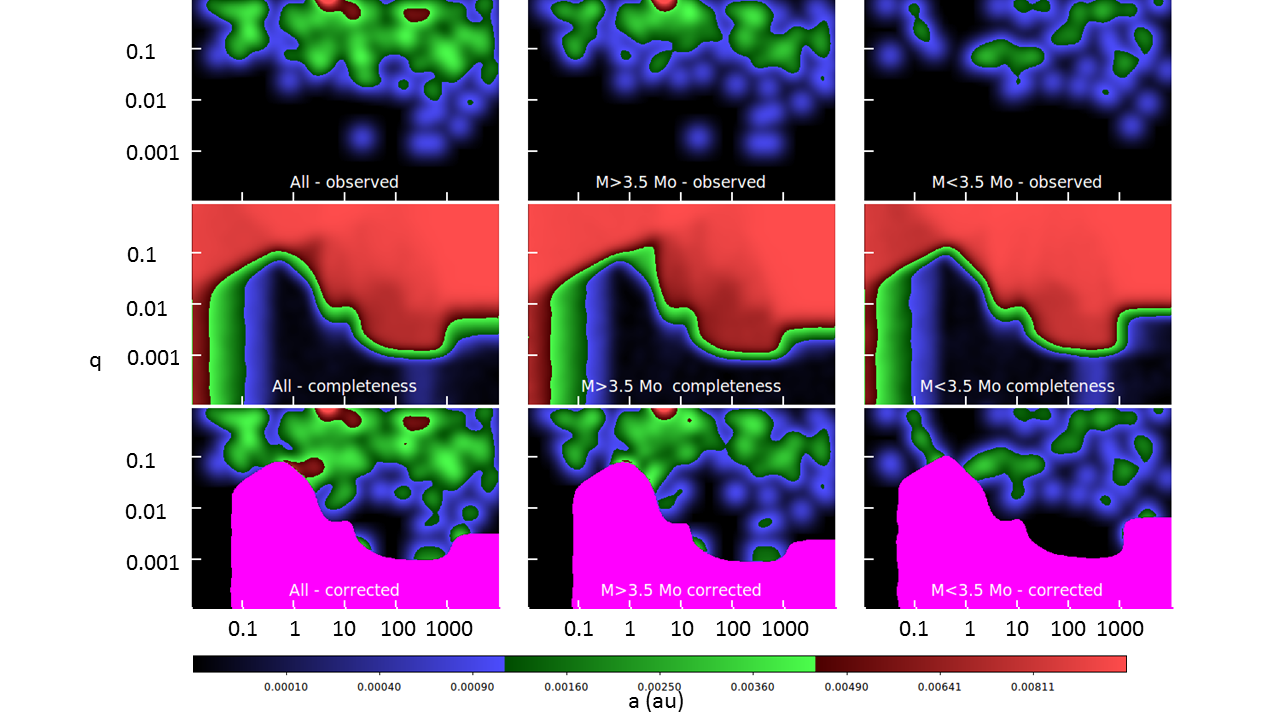}
    \caption{Smoothed distribution of the companions in the semi-major axis $a$ - mass ratio $q$ plane. Top row: observed distribution. Middle row: detection completeness maps. Bottom row: observed distribution corrected for completeness. Left column gives results for the whole sample; the central column gives results from stars more massive than 3.5~M$_\odot$; the right column for less massive stars. The magenta area in the lower row marks the region with completeness $<0.2$, not used in the analysis}
    \label{fig:a_vs_q}
\end{figure*}

Once data from the various techniques for our sample of 181 B-stars in Sco-Cen are combined, we found that there is no indication of binarity - that is, they are bona fide single stars - for only 43 stars, that is $23.8\pm 3.6$\% of the sample. Only 14 out of 92 stars ($15.2\pm 4.1$\%) with $M_A>3.5$~M$_\odot$ are bona fide single stars; for stars with mass lower than this limit, this ratio is 29 out of 89 stars ($32.6\pm 6.1$\%). According to the detections considered in this paper we found a total of 200 companions; 91 of the systems are binary, 34 are ternary, 11 have four components, and 2 five. On average, we detected 1.10 companions per star. Fifteen of these companions are substellar ($M<0.072$~M$_\odot$); two among these are planets ($M<0.013$~M$_\odot$). We note that these are lower limits; multiplicity may be higher because companions may be too small to be detected, may be themselves undetected multiple stars, and because we lack information from RVs, high precision photometric series, or interferometry for a significant fraction of the stars - in most cases the fainter ones. 

In order to provide data useful for discussing the origin of the systems, we looked for their distribution in semi-major axis and mass ratio plane. We created smoothed distributions in this plane as done for estimating completeness. Relevant data are given in Figure \ref{fig:a_vs_q}, where we considered separately the whole sample of the programme stars, those with $M>3.5$~M$_\odot$, and lower-mass stars. Here, the mass ratio is always the ratio between the mass of a companion and the total mass of the stars and of other companions that are closer to the primary than the companion considered; we neglect consideration of the fact that there are hierarchical multiples where the companion is itself a multiple star. The upper row of Figure \ref{fig:a_vs_q} displays the original distributions, the intermediate panel the completeness map appropriate for the mass bin considered, and the lower panel the distribution maps obtained after correcting for completeness. This correction was only done when completeness was higher than 0.2, else the corrected distribution was arbitrarily set at zero.



We may do a few considerations on the distribution of the companions in Figure \ref{fig:a_vs_q}. First, we notice that there is a scarcely populated region around 1 au. Companions in this region are mainly detected using RV variations.

The second point worth mentioning is the lack of low mass companions ($q<0.07$, that typically means stars with $M<0.2 \div 0.3$ M$_\odot$) at very short separation ($<0.2$ au). We notice that TESS and K2 (available for 151 out of 181 stars) would have likely detected low mass transiting planets with radii down to well below 1 R$_J \sim 0.1$ R$_\odot$. This corresponds to a mass of less than 0.001 M$_\odot$ (that is $q\sim 0.0003$) at the age of Sco-Cen \citep{Baraffe1998}.  By itself, the lack of detection of transiting hot Jupiters in a sample of 151 stars would not be highly meaningful. However, it contrasts with the detection of 11 EBs and reflecting binaries in our sample, whose companions are all more massive than 0.76 M$_\odot$ (the companion of HIP~67669, the longest period object among the EBs and the only one with a mass $<1$~M$_\odot$). So, companions at low separation are common among B-stars in Sco-Cen, mainly around those with mass $>3.5$ M$_\odot$, but they typically have a mass larger than the solar mass. This result agrees with the scarcity of transiting low mass companions detected around B-stars; so far, the hottest star hosting a transiting BD companion is HIP~33609 (\citealt{Vowell2023}: $M=2.38\pm 0.10$~M$_\odot$), a member of the Melange 6 moving group (age of $150\pm 25$~Myr) that is classified as an A0V star in SIMBAD. The period of this system is $39.471814\pm 0.000014$~d, that is very long for an eclipsing binary and the orbit has quite high eccentricity ($e=0.56\pm 0.03$). We note that this quite exceptional companion has semi-major axis $a=0.3$~au and mass ratio $q=0.017$.

For binaries at separation $>3$ au the search of stellar companions should be fairly complete, thanks to the contributions by Gaia (direct detections, PMa and RUWE), interferometry and moreover HCI. This last is the only one sensitive also to values of $q<0.01$ (that is substellar objects), save for BDs possibly discovered at large separation by Gaia. By construction, the BEAST sample did not include known visual binaries at the epoch of sample definition, but the consideration of stars not included in that sample (most of them with alternative though less deep HCI) allowed us to correct for this bias. The stellar companions seem quite uniformly distributed in semi-major axis, though this is an artefact of neglecting the mass of the star; see discussion in the next sub-section. However, most of them have a mass ratio $q>0.1$ and there is a scarcity of companions with mass ratios $q<0.01$ that is not due to selection effects in this range of separation, at least for semi-major axis $<500$ au. At very wide separation ($>1000$ au), there is a shift of the companions to lower masses, with a median value of $q=0.10$.

We conclude that our search should be fairly complete for separation larger than $\sim 3$~au and mass ratios $q>0.1$, and highly informative at shorter separation and lower masses. 

\subsection{Semi-major axis distribution}
\label{sec:semimajor}

\begin{figure}[htb]
    \centering
    \includegraphics[width=8.5cm]{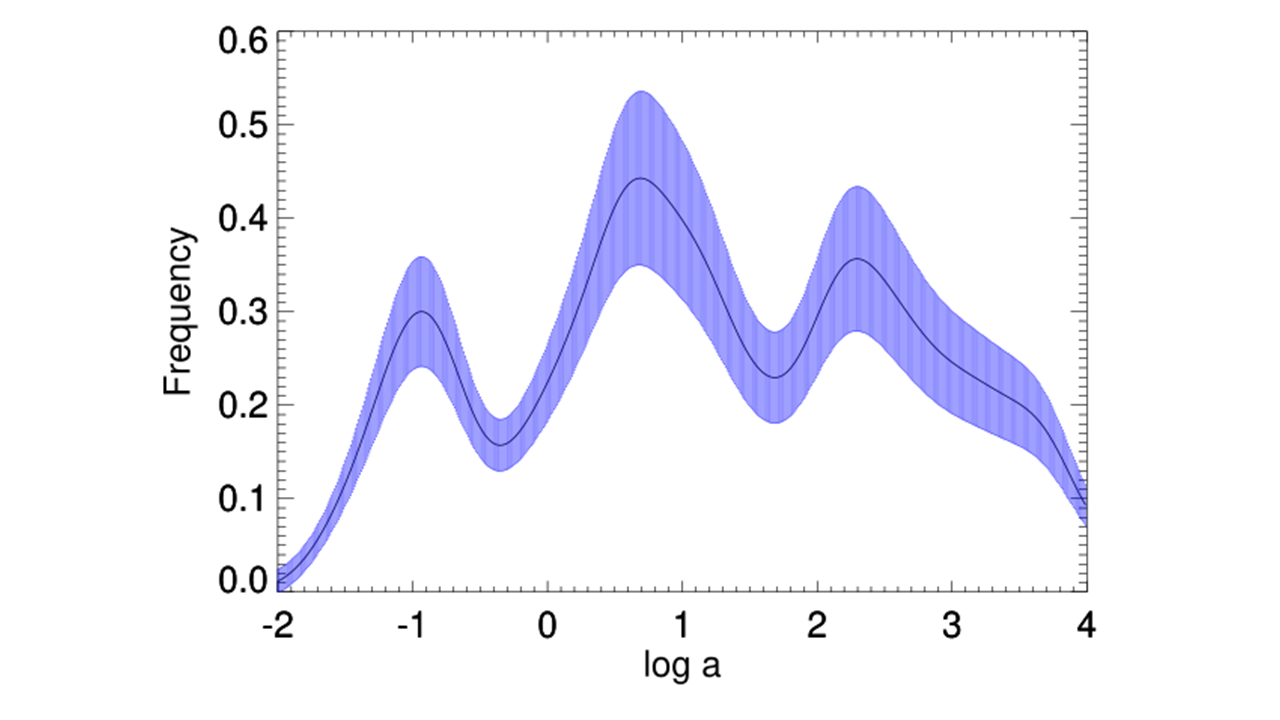}
    \includegraphics[width=8.5cm]{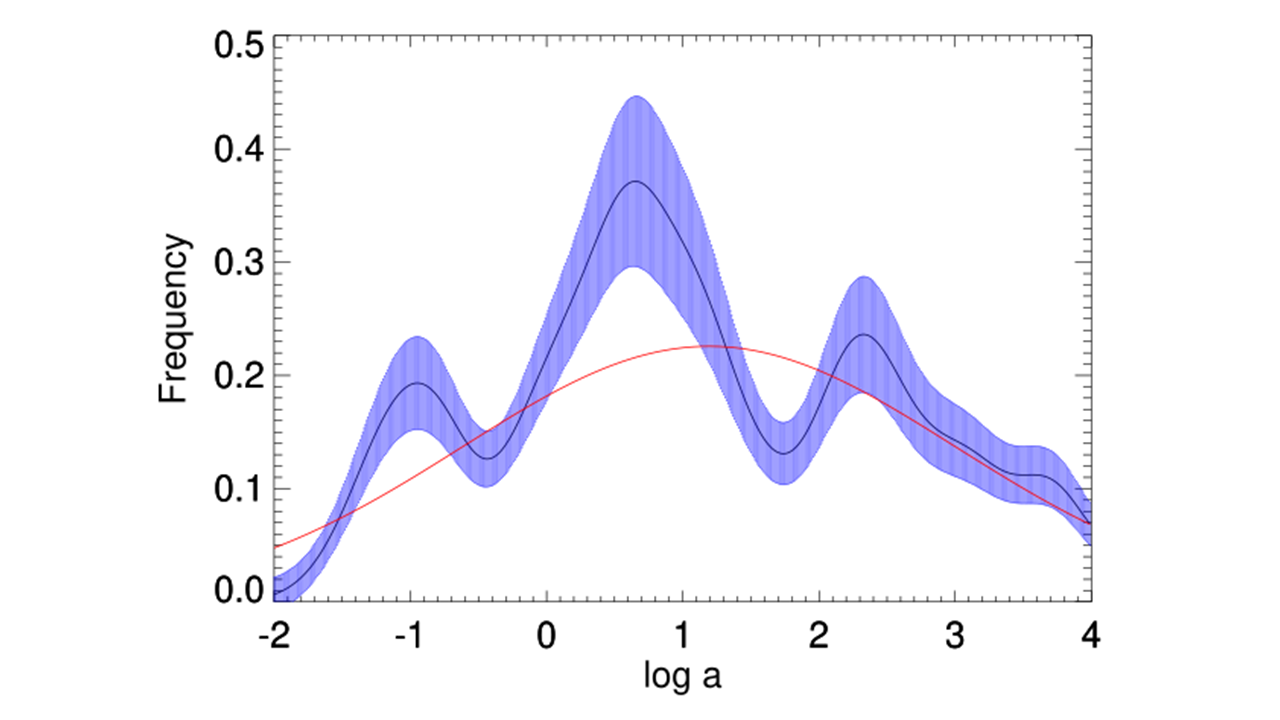}
    \includegraphics[width=8.5cm]{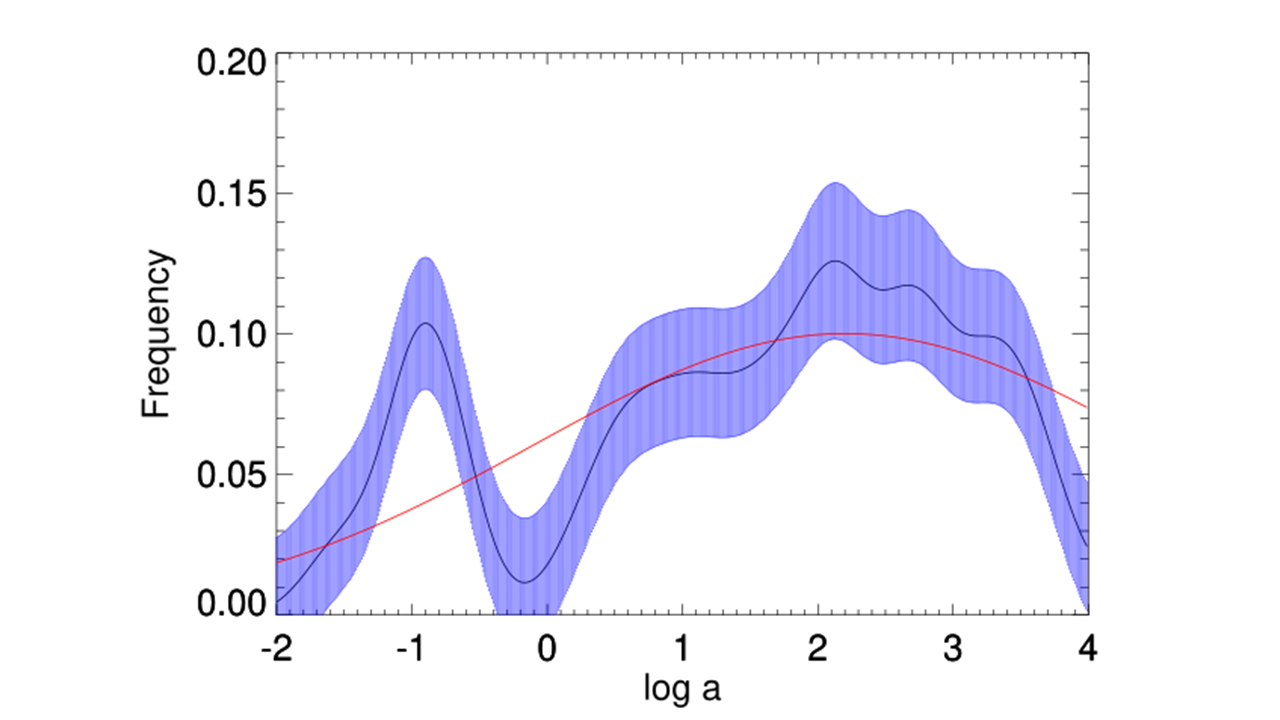}
    \caption{Distribution of companions with $\log{q}>-1.08$ as a function of the logarithm of the semi-major axis $a$ in au. The upper panel is for all B-stars, middle panel is for primaries with $M>3.5$~M$_\odot$, an the lower panel is for primaries with $M<3.5$~M$_\odot$. The shaded area corresponds to 1-$\sigma$ uncertainty. The red lines are fits with the log-normal curves given in the text}
    \label{fig:a_dist}
\end{figure}

Figure \ref{fig:a_dist} shows the distribution of detected companions to the B-stars in the Sco-Cen association with $\log{q}>-1.08$ as a function of the logarithm of the semi-major axis. We adopted this cut in order to avoid biases. We show the distributions for all stars, as well those for primaries with $M>3.5$~M$_\odot$ and those with masses below this limit. These two last distributions appear clearly different. The probability that they are extracted from the same population is very low; a two-side Kolmogorov-Smirnov test yields a probability of 0.002.

The distribution for the brighter and most massive primaries has a median value of only $7.8^{+8.8}_{-3.6}$~au and shows three distinct peaks: the first one (at $\sim 0.1$ au, that is $\sim 50$R$_\odot$) includes about 18\% of the companions. This peak is responsible of the large number of massive EB observed by TESS or Kepler2. The second peak is at a few au and includes roughly half of the companions. The third peak is at a few hundreds au and includes about a quarter of the companions. A log-normal fit to this distribution is:
\begin{equation}
\xi(\log{a/au})=0.226~\exp[-0.5~(\log{a/au}-1.19)^2/1.81^2],
\end{equation}
but it appears as a poor representation of the observed distribution.

The distribution for the less massive B-stars ($M<3.5$~M$_\odot$) has a much larger median value of $62^{+64}_{-51}$~au. It still has the compact binaries component with again about 18\% of the companions, that is very well separated from an extended distribution of companions in the range from 1 to a few thousands au by a distinct gap. The semi-major axis distribution can be described by a log-normal as:
\begin{equation}
\xi(\log{a/au})=0.100~\exp[-0.5~(\log{a/au}-2.20)^2/2.29^2],
\end{equation}

\subsection{A gap in the period distribution?}
\label{sec:gap}

We notice a gap between the short period and the other binaries at about 0.5-1 au (apparent separations of about 3-4 mas), corresponding to periods of about 40 d for the massive stars, and about 100 d for the less massive ones. We may detect companions with this semi-major axis through RV variations and the Gaia RUWE parameter. While this is the region where companion detection is less efficient, inspection of Figure~\ref{fig:a_vs_q} indicates that we should still be able to detect most companions with mass ratio $q>0.1$ in this semi-major axis range. However, pending a more careful search of similar companions through extensive RV surveys, we will leave open the possibility that this gap is an artefact of defects in the search for companions described in this paper. 

In general, detection of companions in this range of separation is difficult in the surveys based on RVs. Typically few companions are detected but large incompleteness are acknowledged (see for instance \citealt{Kobulnicky2014, Villasenor2021}). A relative lack of companions at about 0.5-1 au is possibly present in the analysis of solar-type stars by \citet{Raghavan2010} and among the O-stars by \citet{Sana2012}. The distribution of A-type stars by \citet{DeRosa2014} cannot be used for this purpose because they only considered wide binaries with separation $>10$~au. However they noticed that the trend for decreasing frequency of companions at small separation in their sample is inconsistent with the observed frequency of short period binaries \citep{Abt1965}.

\subsection{Primary and secondary masses: mass ratios}
\label{sec:massratios}

\begin{figure}[htb]
    \centering
    \includegraphics[width=8.5cm]{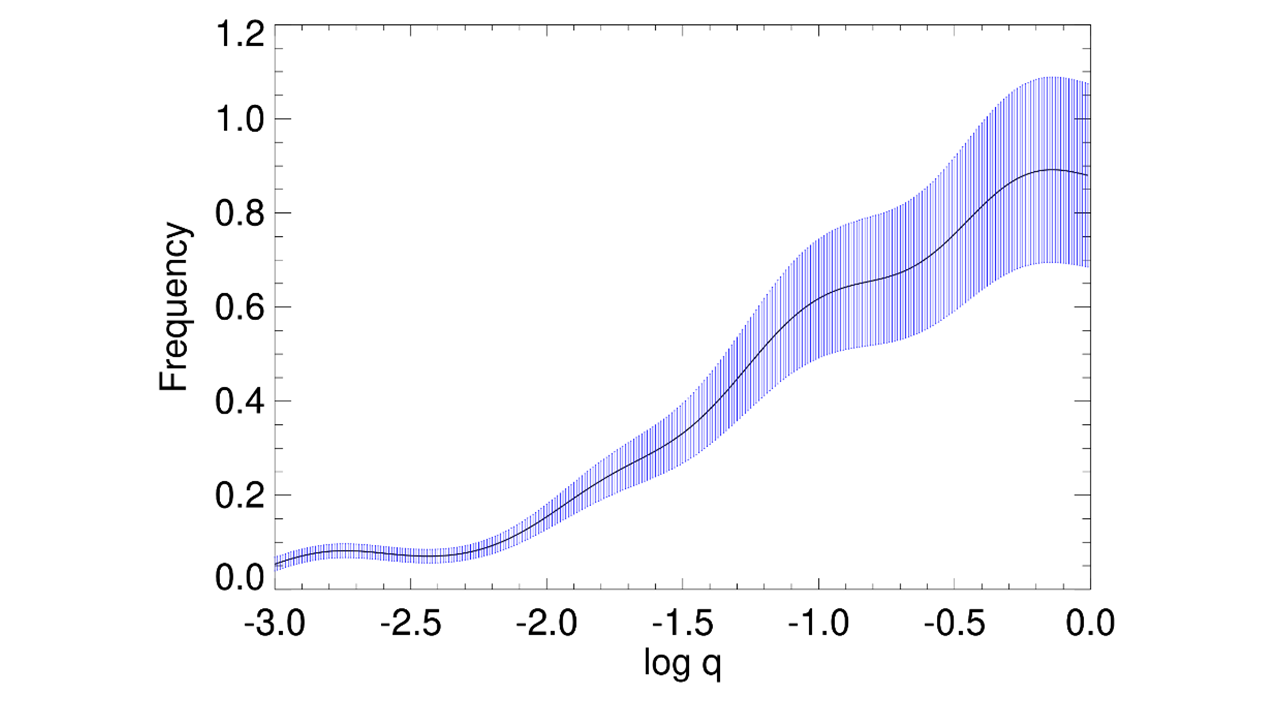}
    \includegraphics[width=8.5cm]{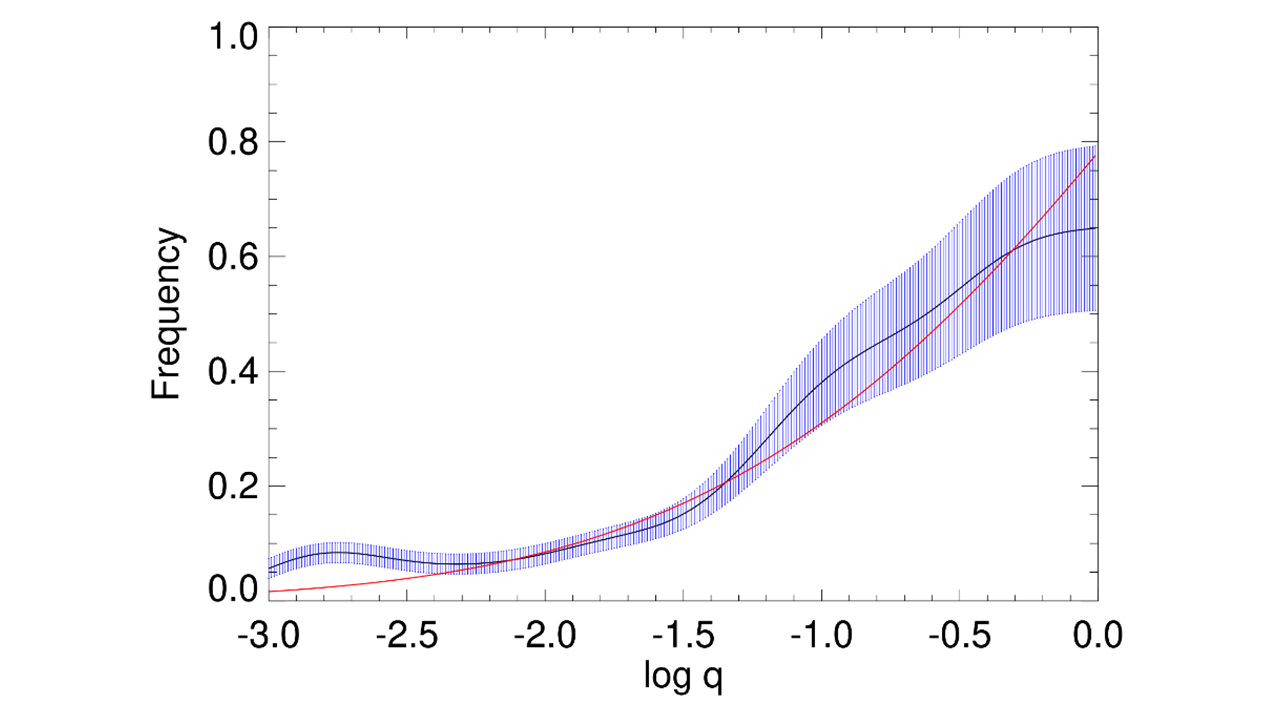}
    \includegraphics[width=8.5cm]{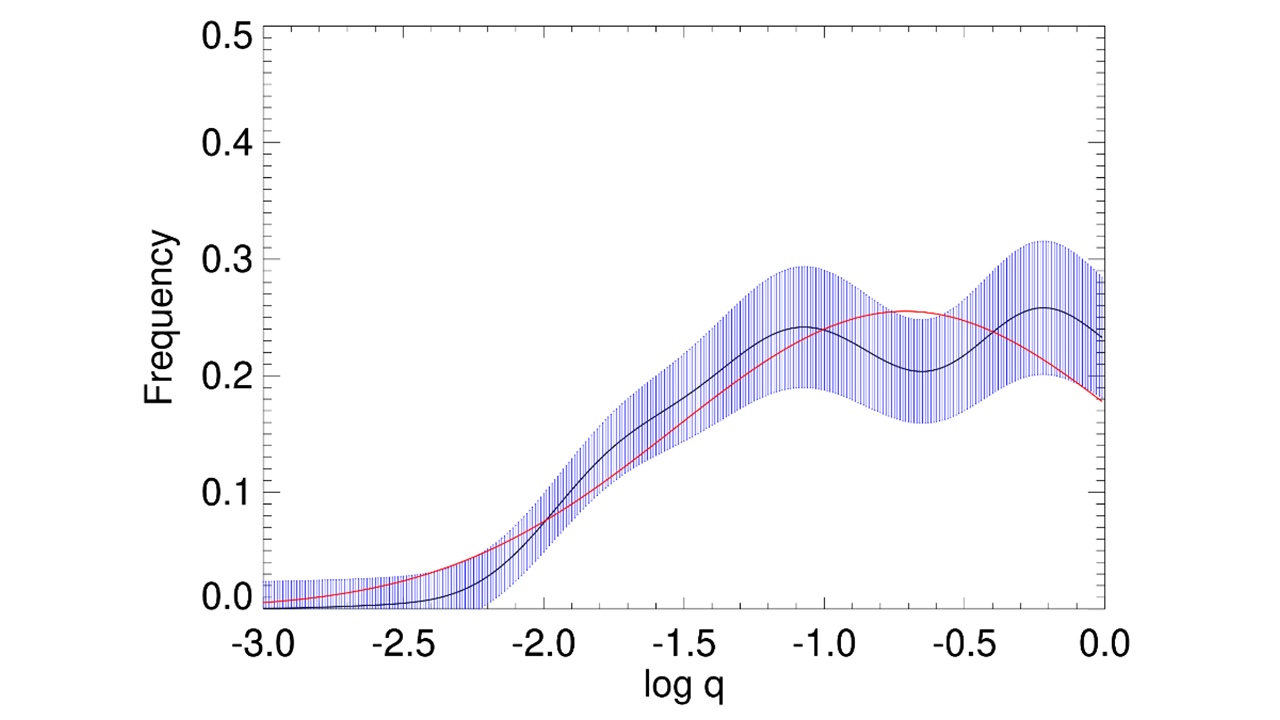}
    \caption{Distribution of companions with $0.5<\log{a}<3$ as a function of the logarithm of the mass ratio $q$. The upper panel is for all B-stars, middle panel is for primaries with $M>3.5$~M$_\odot$, an the lower panel is for primaries with $M<3.5$~M$_\odot$. The shaded area corresponds to 1-$\sigma$ uncertainty. Red lines are the fits with log-normal curves given in the text  }
    \label{fig:q_dist}
\end{figure}

Figure \ref{fig:q_dist} shows the distribution of the companions as a function of the mass ratio $q$, corrected for completeness effects. Since the distribution is different at wider separation and it is largely incomplete for low mass ratios at short separations, we considered only companions in the range 3-1000 au here, and separated the distribution of the companions of the most massive objects ($M>3.5$~M$_\odot$) from the less massive ones. The distribution of companions of very massive objects is quite narrow in terms of the mass ratio $q$ while that of the less massive is flatter, but in both cases very few companions have mass ratios $q<0.01$

The distribution with $q$\ of companions in the range 3-1000 au to massive stars ($M>3.5$~M$_\odot$) for $0.003<q<1$ is very well reproduced by a (truncated) log-normal law of the form:
\begin{equation}
\xi(\log{q})=1.654~\exp[-0.5~(\log{q}-2.02^2/1.65^2],
\end{equation}
while that for less massive stars ($M<3.5$~M$_\odot$) is given by:
\begin{equation}
\xi(\log{q})=0.255~\exp[-0.5~(\log{q}+0.71)^2/0.82^2].
\end{equation}
We notice that these two distributions give a fraction of substellar companions, that is with $q<0.014$ for the massive stars and $q<0.027$ for the less massive ones, of 8.6\% and 18.1\%, respectively. If we use eq. (1) to estimate the probability of the presence of companions with masses as the planet or BD around b~Cen \citep{Janson2021b} and $\mu^2$~Sco \citep{Squicciarini2022}, all having $q\sim 0.002$, or lower around massive stars, this is $\sim 1.9\,10^{-2}$. Taking into account the number of targets observed twice within the BEAST survey (so far 47: \citealt{Janson2021a}), the probability of extracting three (or more) such objects from a distribution as that observed for the stellar companions of massive B-stars is quite low ($\sim 6.1\,10^{-2}$).

On the other hand, for companions further than 1000 au, the distribution is reproduced by the following equation:
\begin{equation}
\xi(\log{q})=\exp[-0.5~(\log{q}+0.99)^2/0.57^2],
\end{equation}
which favours much lower mass companions.

\subsection{Primary and secondary masses: absolute values}
\label{sec:absolute}

\begin{figure}[htb]
    \centering
    \includegraphics[width=8.5cm]{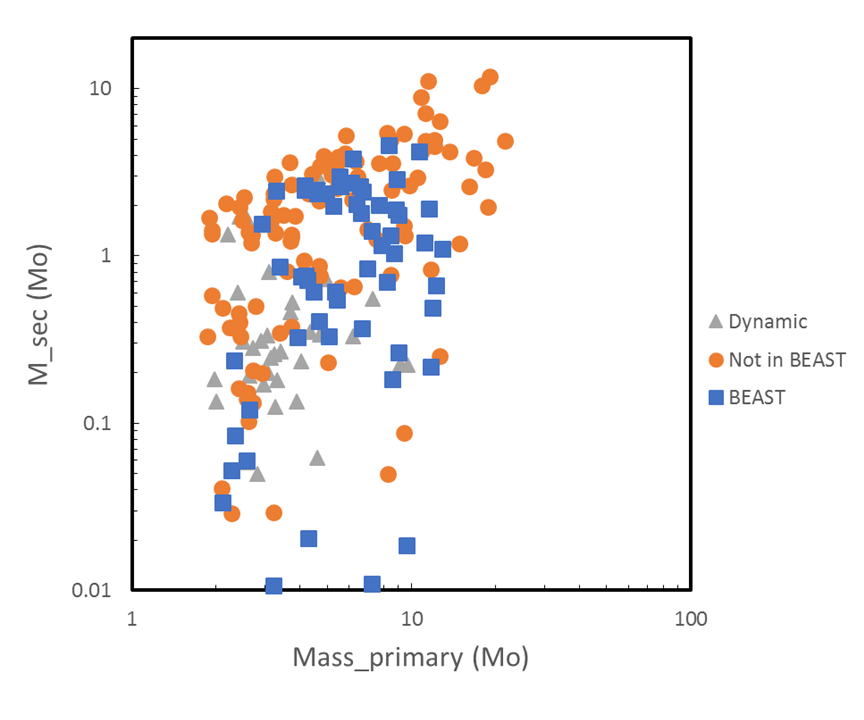}
    \includegraphics[width=8.5cm]{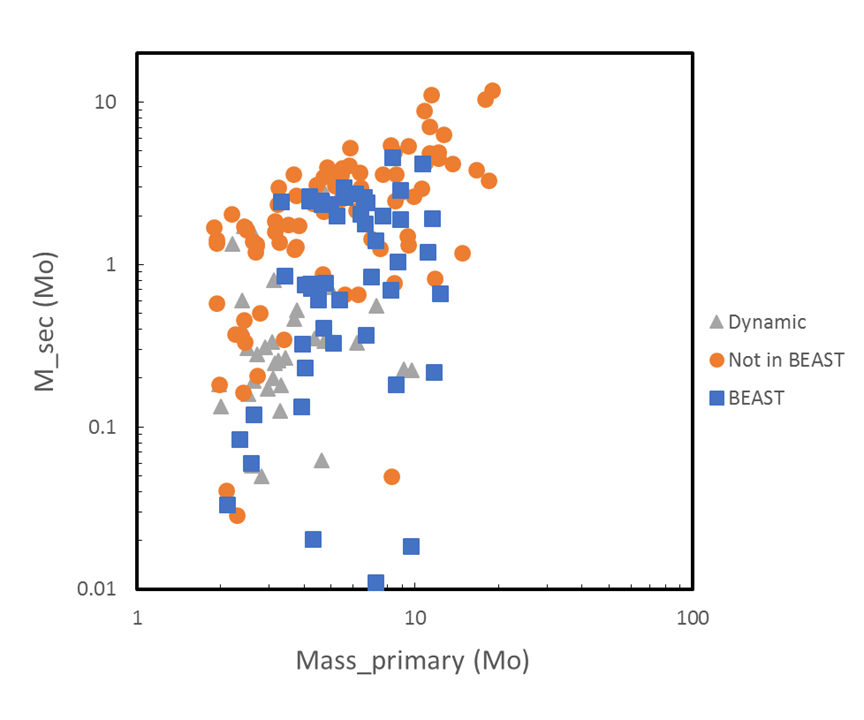}
    \caption{Relation between the mass of the companion $M_B$\ and the total mass of the system within within its orbit (mass of the primary). Upper panel: all companions. Lower panel: only companions within 1000 au from the star. Blue squares symbols are  companions detected through eclipses, RV curves, and imaging (including interferometry) on the BEAST sample; orange circles are stars not included in that sample. Grey triangles are companions only detected through PMa, Gaia RUWE parameter, and RVs. }
    \label{fig:ma_mb}
\end{figure}

\begin{table}[htb]
  \caption[]{Mass distribution of companions to massive stars $M_A>5$~M$_\odot.$}
  \label{t:mass_dist_massive}
  \begin{tabular}{lcc}
  \hline
Mass range & $a<1000$ au & $a>1000$ au \\
M$_\odot$       &    &   \\   
\hline
$M_B<0.072$     &  4 &  0 \\
$0.072<M_B<0.5$ &  7 &  6 \\
$0.5<M_B<1.0$   &  9 &  1 \\
$1.0<M_B<2.0$   & 11 &  5 \\ 
$M_B>2.0$       & 42 &  3 \\
  \hline
  \end{tabular}
\end{table}

Figure \ref{fig:ma_mb} shows the run of the mass of the companions as a function of the mass of the primaries (here, the total mass of the system within its orbit). This figure outlines a fact not obvious from the discussion of the mass ratios, that is the correlation existing between the mass of the companions and the mass of the primary. The correlation is even more clear if we eliminate very wide binaries (companions at separation $>1000$~au) that likely have a different origin from closer companions. The relation can actually be even stronger than shown in this plot, because we are neglecting the possibility that some of the companions are themselves multiple systems, and then have a higher mass than inferred from photometry (the source of the vast majority of the masses shown in this plot). Companions with masses $<1$~M$_\odot$ (that are the vast majority of the stars in the general field) are indeed rare as close companions to massive B-stars (see Table~\ref{t:mass_dist_massive}). Among the companions with separation $<1000$~au there are only seven M-stars (masses $0.072<M<0.5$~M$_\odot$) that are companions of primaries with a mass $>5$~M$_\odot$, while there are 53 companions in this range of separations more massive than the Sun. Even the companions with masses in the mass range $0.5<M<1$~M$_\odot$ are quite rare (only 9 companions found). Only 9\% of the companions to stars more massive than $>5$~M$_\odot$ are M-stars ($0.072<M<0.5$~M$_\odot$), while objects with this mass are some 77\% of the stars integrating for instance the \citet{Chabrier2003} initial mass function (IMF), 70\% using the IMF by \citet{Chabrier2005}, and 62\%, using the IMF by \citet{Kroupa2001}. We may also compare this distribution with that for the whole population of Sco-Cen members (see Figure 13 in \citealt{Luhman2022} or similar data in \citealt{Miret-Roig2022}). In this case, we may notice that a 15 Myr old star with a mass of 0.5 (0.072)~M$_\odot$ should have spectral type around M1.5 (M5.5) using the isochrones by \citet{Baraffe2015} and the temperature spectral type relation by \citet{Pecaut2013}. Using data by \citet{Luhman2022} we estimate that approximately $4750/6000\sim 79$\% of the stars of Sco-Cen are in the mass range $0.072<M<0.5$~M$_\odot$. This agrees with the expectation for the \citet{Chabrier2003, Chabrier2005} IMF's\footnote{We can notice that the observed fraction of B-stars in Sco-Cen of $181/6000\sim 3.0$\% also agrees very well with that expected with the \citet{Chabrier2003} IMF and is a bit lower than expected with the \citet{Chabrier2005} one (4.0\%). However, when considering the fraction of the total mass that is in systems including a B-star, we should also consider the multiplicity and the mass function of secondaries. For instance, summing up the mass of the primaries considered in this paper, we have a total of 841~M$_\odot$, but if we sum also all the companions this makes a total of 1157~M$_\odot$, that is larger by a factor of 1.38. This is important when for instance we try to estimate the total stellar mass of Sco-Cen from the mass of the B-stars.}. The very low fraction of low-mass star companions to B-stars contrasts with the fact that four substellar companions to these stars - objects that are much more difficult to be detected and for which incompleteness is likely much higher - have been found around such massive stars in this separation range \citep{Janson2019, Janson2021b, Squicciarini2022}. Since the population of such small objects is likely much larger, this suggests that they have a different channel of formation with respect the low mass stellar companions.

For the 34 companions further than 1000 au, detections are based on Gaia data and should be complete roughly down to the hydrogen burning limit. The mass distribution is very different from what obtained at shorter separations and it is reproduced by the equation:
\begin{equation}
\xi(\log{M_B/M_\odot})=\exp[-0.5 (\log{M_B/M_\odot}+0.31)^2/0.62^2)],
\end{equation}
where the masses are in M$_\odot$. This distribution is centred at 0.49~M$_\odot$. This is still shifted towards more massive stars with respect to the \citet{Chabrier2003} IMF that has a similar value for the $\sigma$ of the distribution but it is centred at 0.20~M$_\odot$. This mass distribution is however not very different from that obtained for Upper Scorpius by \citet{Miret-Roig2022}, if we consider the incompleteness at the planetary masses.


\section{Discussion}
\label{sec:discussion}

\subsection{Comparison with other samples}
\label{sec:other}

\begin{figure}[htb]
    \centering
    \includegraphics[width=8.5cm]{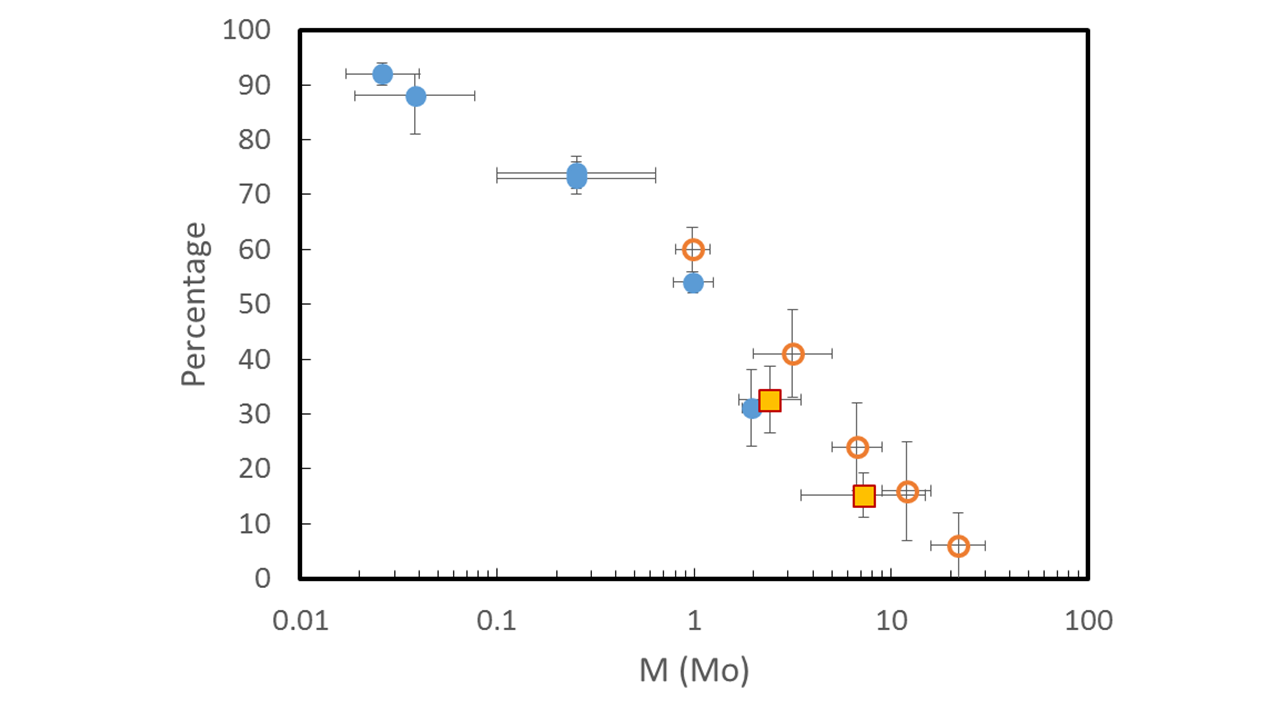}
    \includegraphics[width=8.5cm]{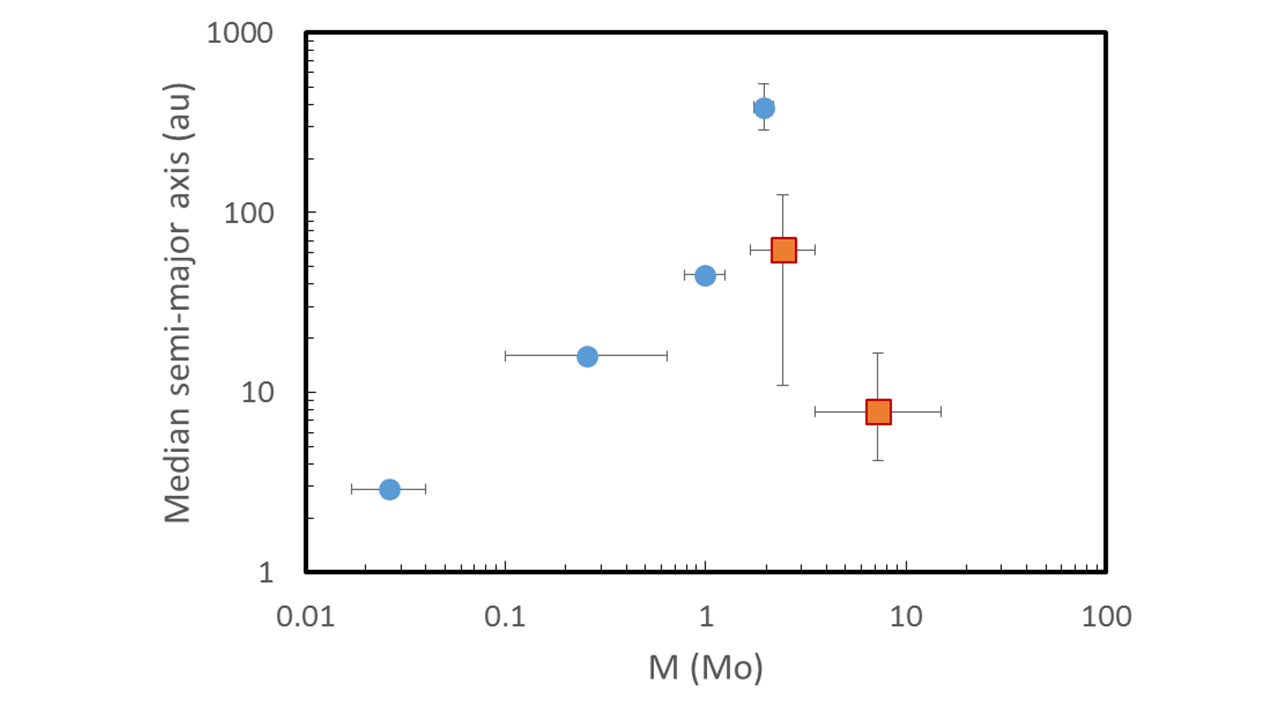}
    \caption{Runs of statistical properties of binaries with the stellar mass. Upper panel: run of the frequency of single stars from our data (orange squares with red edge), the samples in Table~\ref{t:frequency} (filled circles) and from \citet{Moe2017} (opens circles). Lower panel: run of the median semi-major axis. Horizontal error bars reproduce the mass range of the different samples  }
    \label{fig:frequency}
\end{figure}

\begin{table}[htb]
  \caption[]{Single star frequency and median semi-major axis with stellar mass.}
  \label{t:frequency}
  \begin{tabular}{lccc}
  \hline
Source &  Mass &  Single & Median $a$ \\
       &  M$_\odot$  &      \%        & au       \\   
\hline
\citet{Fontanive2018}  & 0.017-0.04 & $92\pm 2$    &    2.9  \\
\citet{Burgasser2006}  & 0.019-0.077 & $88^{+4}_{-7} $       \\
\citet{Delfosse2004}   & 0.10-0.64 &  $74\pm 3$    &       \\ 
\citet{Janson2012}     & 0.10-0.64 &  $73\pm 3$    & $\sim 16$ \\
\citet{Raghavan2010}   & 0.78-1.25 &  $54\pm 2$    &     45 \\
\citet{DeRosa2014}     & 1.75-2.18 & $31.1\pm 7.0$ &$387^{+132}_{-98}$\\
This paper             & 1.86-3.5  & $32.6\pm 6.1$ &$62^{+64}_{-51}$\\
This paper             & 3.5-15.0  & $15.2\pm 4.1$ &$7.8^{+8.8}_{-3.6}$\\
  \hline
  \end{tabular}
\end{table}

We may compare the low frequency of single stars and the value for the median semi-major axis for the companions we obtained for our sample of B-stars with data for other samples of stars from the literature. Combining our results for the Sco-Cen B stars with other samples in the literature (M-stars: \citealt{Delfosse2004, Janson2012}; solar-type stars: \citealt{Raghavan2010}; A-type stars: \citealt{DeRosa2014})  we get the values listed in Table~\ref{t:frequency}. The values given in this table are shown graphically in Figure~\ref{fig:frequency}. Our results extends the trends previously observed for a lower frequency of single stars with increasing stellar mass. Within the range covered by these different surveys ($0.2<M<10$~M$_\odot$), the fraction of single stars $f$ is well represented by a logarithmic trend with stellar mass: $f = -0.44 \log{M/M_\odot} + 0.48$. Of course, since $0<f<1$, this trend cannot be extended outside this range of validity. The frequency of single stars considered here is systematically lower than that given by \citet{Moe2017}. This difference can be explained as due to the fact that \citet{Moe2017} are only considering companions with $q>0.1$, while we are also considering lower mass companions that makes about 39.5\% of the total. Part of the difference might also be related to a larger number of wide companions (20\% of the companions have separation $>1000$~au). An excess of binaries in Sco-Cen T Tau stars with respect to stars of similar mass in the general field has also been noticed by \citet{Kohler2000}, as well as in many other low density star-forming environments \citep{Leinert1993, Ghez1993, Kohler2008}. However, higher density environments such as the Orion Nebula Cluster have a binary fraction similar to the general field \citep{Petr1998, Kohler2006, Reipurth2007}.

For what concerns the semi-major axis (and period) distributions, usually these are fit with log-normal laws. In general, it is found that the peak of these distributions steadily increase with mass up from M- to the A-stars (see e.g. \citealt{Janson2012, Duquennoy1991, Raghavan2010, DeRosa2014}). When considering the B stars in Sco-Cen, we found that log-normal fits are no longer adequate. We obtained an excess of companions to late-B stars at very large separations with respect to the \citet{DeRosa2014} distribution for A-type star that is quite well described by a log-normal law that peaks at 390 au: this is likely related to weakly bound objects that might be lost with further ageing of the systems (see also \citealt{Mathieu1994, Duchene2013}). However, for both late- and moreover for early-B stars, we obtained much lower values for the median separation. A similar low value of the position of the peak of the distribution with semi-major axis at about 10 au has been obtained by \citet{Sana2012, Kobulnicky2014}, and \citet{Moe2017} for their samples of OB-type stars - though these samples were aimed to determine the frequency of interacting binaries and are likely incomplete at large separations. This roughly agrees with the current result: the median semi-major axis increases with stellar mass only up to the A-stars (mass $\sim 2$~M$_\odot$) and then has a turnover and it decreases to a few tens au or less for massive stars. We may conclude that most systems around massive stars are compact - and a large fraction of them would likely interact in some phase of their evolution \citep{Sana2012, Kobulnicky2014, Moe2017}. These trends should be explained by binary formation scenarios (see e.g. \citealt{Moe2017}).

We found a strong correlation between the mass of the primaries and of the companions, and a scarcity of low mass stellar companions to massive stars. This is not entirely new since it was noticed already 15 years ago by \citet{Kouwenhoven2007b} in their analysis of binaries in the Sco-Cen association. These authors noticed that this correlation can be seen as an extension and widening of the BD desert observed around solar-type stars. Our new, much more complete data strongly supports this early conclusion. This result for the B-star in Sco-Cen likely reflects a major trend for the mass of the companions with the mass of the star, at least for relatively close binaries. Among the SBs studied by \citet{Chini2012}, 82\% of the spectra for O stars with $V\leq 10$ mag contain more or less separated multiple lines (SB2s) reflecting that the majority of systems contain pairs of similar mass. This is in agreement with results by \citet{Kobulnicky2007} who found that massive stars preferentially have massive companions. 

\subsection{Implications for binary star formation}
\label{sec:implications}



\begin{table*}[htb]
  \caption[]{Best parameters for the toy model.}
  \label{t:parameter}
  \begin{tabular}{lcccc}
  \hline
Parameter         & Prior Range &$M<3.5$M$_\odot$&$M>3.5$M$_\odot$&      All      \\
  \hline
$n_{\rm max}$      & [20, 60]    &   $38\pm 12$  &    $40\pm 12$  &   $38\pm 12$   \\
$\alpha$           & [-2.0, 0.0] &$-1.27\pm 0.17$&$-0.55\pm 0.23$&$-0.89\pm 0.16$ \\
$d_{\rm min}$ (au) & [30, 100]    &  $66\pm 21$  &   $68\pm 20$  &   $65\pm 20$   \\
$d_{\rm max}$ (au) & [200, 1200]  & $665\pm 286$ & $690\pm 281$ & $749\pm 256$  \\
$\mu$              & [0.0, 1.5]   & $0.82\pm 0.22$& $0.47\pm 0.15$& $0.57\pm 0.18$ \\
$\beta$            & [0.0, 1.0]   & $0.57\pm 0.29$& $0.44\pm 0.28$& $0.47\pm 0.29$ \\
  \hline
  \end{tabular}
\end{table*}

\begin{figure*}[htb]
    \centering
    \includegraphics[width=18cm]{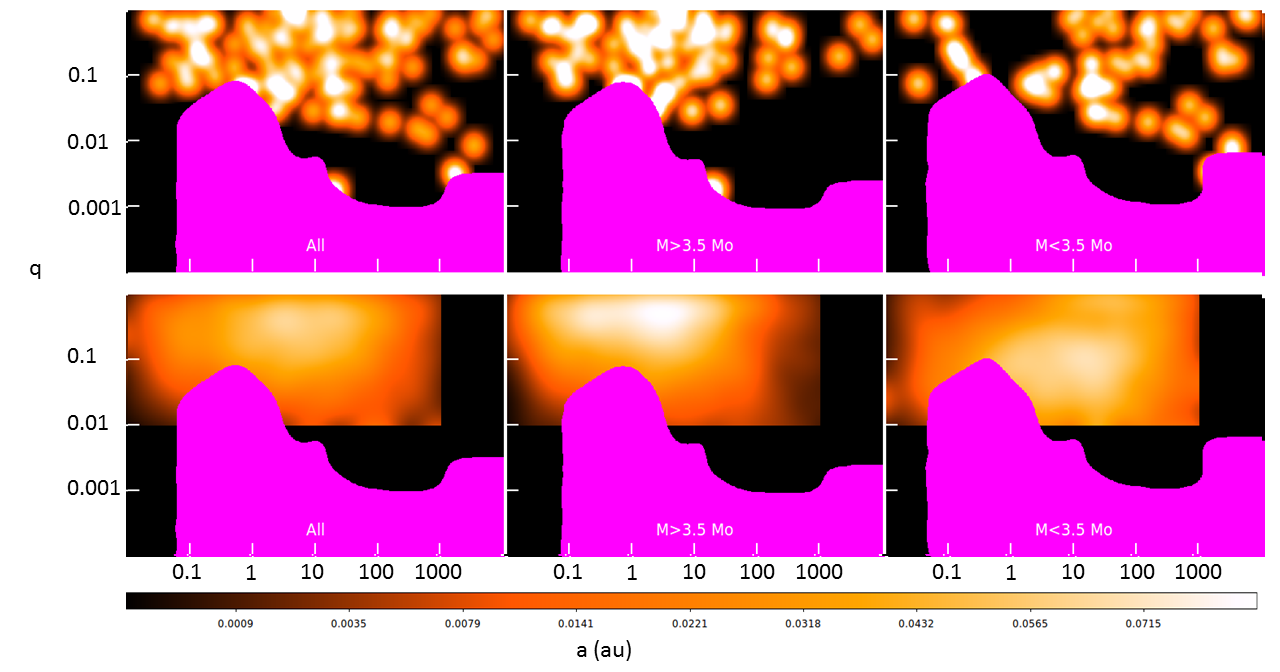}
  \caption{Comparison between observed (top row) and model (bottom row) maps of the smoothed distribution of companions in the separation (in au) vs mass ratio plane. We consider here the maps obtained considering only the closest companion, for consistency between the models and the observations. Left column is the whole sample; central column is for stars with a mass $>3.5$~M$_\odot$; right column is for stars with  a mass $<3.5$~M$_\odot$. The magenta area marks the region with completeness $<0.2$, not used in the analysis}
    \label{fig:simu}
\end{figure*}

We observed clear trends in the binary frequency, semi-major axis and mass ratio distributions as a function of stellar mass. This might be related to the same mechanism of binary formation or rather to the next evolution of binary systems. On this respect, \citet{Kaczmarek2011} found that the more massive a primary star, the lower the probability that the binary is destroyed by gravitational interactions; they then argued that the higher frequency of binaries in more massive stars is not due to differences in the formation process but can be entirely explained as a dynamical effect. However their statement is based on N-body simulations of the Orion nebula cluster, that has a high central density of $\rho=3.1 \times 10^3$~pc$^{-3}$. While the exact density appropriate for the formation environment of Sco-Cen B stars is not easy to assess, we may consider as typical the case of the L1688 and L1689+L1709 star-forming clouds in Ophiuchus (the youngest and densest region in Sco-Cen), that have density of a few tens pre-stellar cores per cubic parsec (see \citealt{Ladjelate2020}). This is two orders of magnitude below the density of the Orion nebula cluster. In addition, the trend for having more compact systems around the most massive stars observed in Sco-Cen (as well as in other OB associations: \citealt{Sana2012, Kobulnicky2014, Moe2017}) and the presence of a significant population of wide companions are unexpected if the only important effects causing the trends with stellar mass are due to interactions with other stars in the natal environment. 

We will then focus here on mechanisms related to the binary formation. The existing correlation between masses of primaries and companions for massive binaries with separation $<1000$~au (about 83\% of total) strongly supports a scenario where these companions form by disk fragmentation. This is because in this case we expect a threshold ratio between disk and star masses \citep{Kratter2016}; on the other hand, the different mass function for outer companions (separation $>1000$~au) rather supports the view where these companions form through a different channel (e.g. cloud fragmentation followed by capture within the star forming region). This is well within generally established scenarios \citep{Offner2010, Offner2016,Kratter2010}. However, we may go beyond this.

Figures \ref{fig:a_dist} and \ref{fig:q_dist} show that the distributions of companions with separation $a$ and mass ratio $q$ depend on the mass of the primary; companions around more massive stars are systematically closer and have a mass ratio closer to 1 than those around less massive objects. A full understanding of the complexity of binary formation requires very extensive hydro-dynamical computations (see e.g. \citealt{Kratter2016, Meyer2018, Oliva2020} and references therein), beyond the purposes of this paper. In order to understand what is the basic reason of these differences, we will rather compare the observed distributions with the expectations of a toy model for the binary formation, similar to that considered by \citet{Tokovinin2020}. This is a parametric approach, where the complex physics involved in the process of generating multiple stellar systems is described by simple dependencies and a Monte Carlo approach. The values of the parameters used in the models are only meaningful within the context of the model but the trends that can be obtained by comparing the observed distributions (e.g. of primary mass) with different models may give a physical understanding of the basic mechanisms involved.

As a first step in this comparison, we first constructed maps of the smoothed distribution of companions in the $a - q$\ plane; these maps were obtained replacing the point relative to each companion with a Gaussian distribution with a sigma equal to 0.2 dex. The upper panels of Figure~\ref{fig:simu} show the maps we obtained in this way considering only the closest companions, for three samples of stars: all the stars considered in this paper; only systems with primaries with a mass $>3.5$~M$_\odot$; and systems with primaries having  a mass $<3.5$~M$_\odot$. These maps show the same differences between high- and low-mass binary systems found in Figures~\ref{fig:a_dist} and \ref{fig:q_dist}.

We may then compare quantitatively these maps with analogous maps that can be obtained from the toy-model. The model by \citet{Tokovinin2020} considers formation of binaries by disk instabilities \citep{Kratter2010, Machida2010, Kratter2016}; in order to make these instabilities more likely, the model assumes that accretion of material from the interstellar matter onto the disk is not continuous, but it rather occurs in $n$ episodes (bursts). This is not an exotic assumption. In fact, accretion bursts were first introduced to explain the 'luminosity problem' \citep{Hartmann1996}; they offer an explanation for the luminosity burst events in regions of massive star formation, which have been found by \citet{Hunter2017, Caratti2017}, and \citet{Sugiyama2019}. \citet{Chen2020} reported the observation of disk substructures associated with an accretion burst event, thus providing a link between the two phenomena. In each episode, the \citet{Tokovinin2020} model assumes that there is a random possibility that a companion is generated as a low mass object at a separation in the range $d_{\rm min}-d_{\rm max}$; this may only occur if the mass ratio between the disk and the star is above a given threshold \citep{Kratter2016}. During the next episodes, the disk matter is accreted on both the primary and the secondary; we expect that in most cases accretion mainly occurs on the secondary \citep{Clarke2012}, so that the system tends to equal mass. In the model, this is considered by a parameter $\beta$ describing the exponent to the (instantaneous) mass ratio $q$ of the distribution of the accretion between companions and primary; $\beta$ should be in the range [0,1] and the lowest the value of $\beta$, the highest is the accretion on the secondary. Furthermore, at each accretion episode the interaction between the companion and the disk causes the companion to migrate on a new orbit \citep{Moe2018, Elbakyan2023}. This is a very complex phenomenon; in the model, it is described as an effect characterised by a factor $\gamma$, that may take a random value within a suitable range; a positive value of $\gamma$ means an outward migration and a negative value an inward migration. If the final position of the companion is within the radius of the star, it is assumed that a merging occurred \citep{Elbakyan2023}: the companion is destroyed and the mass of the star is increased correspondingly. We note that time does not enter explicitly in this model that rather considers a number of individual accretion episodes. Each of these episodes should be separated from the others by at least several dynamical times in order the disk instability to possibly take place. Considering Keplerian orbits with semi-major axis in the range 100-1000 au, that is periods of thousands of year for these massive stars, this actually means some tens thousands of year. Since the main accretion phase when the disk is likely to fragment lasts for a few $10^5$~yr \citep{Machida2010}, the number of accretion episodes is expected to range from very low values up to a few tens.

In general we adopted the same recipes described in \citet{Tokovinin2020}, but we modified them a little bit for our purposes. First, we only considered binaries and not systems of higher multiplicity. These are very important to obtain a realistic distribution of orbital eccentricities, but not so much to discuss other properties (mass and period distribution) and we refer to the closest companion alone. Second, we modified the mass accretion from the interstellar matter onto the disk at each episode to make it more realistic. Rather than a constant value, as considered by \citet{Tokovinin2020}, we assumed that this scales down with time, that is, the early episodes involve more mass. In practice, we assumed that the scaling runs with a power $\alpha$ of the episode. As assumed by \citet{Tokovinin2020}, the mass accreted in the various episodes has a random fluctuation drawn from a uniform distribution of $\pm 30$\% around the mean value expected for that episode. Third, we assumed that the total number of accretion episodes $n$ is not constant, but it may fluctuates randomly from a minimum value of 10 up to a maximum equal to $n_{\rm max}$. Fourth and most important, we modified the range of the parameter involved in the migration. \citet{Tokovinin2020} adopted a range for $\gamma$=[-3, 0] that only allows inward migration. With this assumption, all companions formed by disk fragmentation end up at short periods. However, the observed distribution with mass and separation for massive primaries is strongly different from that considered in the standard toy model by \citet{Tokovinin2020} for B-stars (compare fig. 6 in their paper with the distributions of our Figure~\ref{fig:a_vs_q}). According to that model, there are very few companions at wide separations and they should be very low mass objects. This is because according to their model, virtually all binaries that started their formation at separation of 10-1000 au will end up with periods $<100$ days. 
On the other hand, most of the stars in Sco-Cen have binaries in this range of separation. We then adopted for the migration parameter the range of values for $\gamma=[-3, 0]+\mu$. If $\mu$ is larger than 0, this distribution also allows outward migration. 

For each set of parameters $n_{\rm max}$, $\mu$, $\alpha$, $d_{\rm min}$, $d_{\rm max}$, and $\beta$, we run the model 1,000 times (that is simulating 1,000 systems) to define with a reasonable accuracy the distribution of companions. With these data we constructed maps of distributions in the $\log{a}-\log{q}$ plane, after the same smoothing applied to the observing data. Once normalised to the total populations, we may compare these model maps with the observed ones, and define a suitable goodness of fit parameter; in practice, we considered the mean quadratic residual between models and observation $r$. Since the model is only aimed at reproducing the properties of binaries generated by disk instability, the comparison is limited to semi-major axis $a<1000$~au. We consider here the maps obtained considering only the closest companion to each star, for consistency between the models and the observations. 

In order to explore the impact of the various parameters, we adopted a Monte Carlo approach: we computed 10,000 runs (each with 1,000 systems) with random values of the various parameters: $n_{\rm max}$, $\mu$, $\alpha$, $d_{\rm min}$, $d_{\rm max}$, and $\beta$ with uniform distributions within appropriate ranges (Column 2 of Table~\ref{t:parameter}). We then considered acceptable the best 100 (= 1\%) models (typically this means a value of $r<1.2 r_{\rm min}$, the total range of $r$ over all set of parameters covering about an order of magnitude). We then adopted as best value for each parameter the average obtained over the acceptable models and as error the standard deviation. As an example of this comparison, we show in Figure~\ref{fig:simu} the comparison between the maps given by the observation and the model providing the lowest value of $r$ for the whole sample, only for stars with $M_A>3.5$, and only for stars with $M_A<3.5$. We found that several parameters have little impact on the final value of $r$ and whatever value within the (wide) prior selection range looks adequate; this is for $n_{\rm max}$, $d_{\rm min}$, $d_{\rm max}$, and $\beta$. The only parameters that really matters are $\mu$ (the parameter involved in migration) and $\alpha$ (the exponent of the power law describing the time evolution of the mass involved in the accretion episodes), this last being indeed the parameter that makes most of the difference. We find that in order to reproduce the observed separation - mass ratio distributions, we need that the time evolution of the mass has a much shallower run with accretion episodes for massive stars ($M>3.5$M$_\odot$) in comparison to less massive ones $M<3.5$M$_\odot$. The difference is indeed very large, corresponding to roughly a factor of ten in the number of massive episodes of accretion (onto the disk). Overall, this is rather intuitive, the higher the mass of the star, the largest is the number of important accretion episodes, the higher the chance of companion formation, of migration, and of mass accretion on the secondaries. Models also suggests that inward migration (lower value of $\mu$) is more efficient around more massive stars. This might be a consequence of the larger mass of the disk in massive stars.

\section{Conclusions}
\label{sec:conclusion}

In this paper, we have considered the frequency and the distribution in mass and semi-major axis of the stellar companions to B stars in the young Sco-Cen association. This is a low-density environment, where the impact of gravitational interactions with other stars in the birth environment is likely not very important and the observed distributions more faithfully reproduce the trends due to the formation mechanism. To this purpose, we considered a list of 181 B stars in this association and assembled a lot of information regarding the presence of companions using a variety of methods. Companions have been imaged using a variety of techniques, including HCI from ground and the Gaia satellite, as well as other techniques (visual observations, interferometry, and speckle interferometry). Eclipsing binaries were discovered from ground-based observations; additional, rather complete data are available from the Kepler 2 and TESS missions. Spectroscopic binaries have been obtained using RVs (from ground and from Gaia) and the cross correlation of spectra. Astrometric binaries have been found using Hipparcos and Gaia data. We found information for a total of 200 companions and derived estimates of the masses, mass ratios, and semi-major axis for all of them. We compared these data with the expected detection limits of the various techniques and found that our search should almost be complete for a separation larger than $\sim 3$~au, and highly informative at a shorter separation. We derived completeness corrections considering the detection limits and number of objects observed with the various techniques; they are taken into account in our analysis.

We found that the vast majority of B stars have stellar companions and that single stars are quite rare ($23.8\pm 3.6$\%). The frequency of single stars is even lower among the most massive stars ($M_A>3.5$~M$_\odot$: $15.2\pm 4.1$\%), while it is somewhat higher among the less massive ones ($32.6\pm 6.1$\%). This result confirms earlier findings (see. e.g. the discussion in \citealt{Moe2017}).

The masses of the secondaries are correlated with those of the primaries, confirming an earlier finding by \citet{Kouwenhoven2007a} and the results obtained for other sets of stars \citep{Kobulnicky2007}. The mass distribution of the companion to the B stars in the Sco-Cen association is clearly different from that of field stars. This is more evident for stars with masses $M>5$ M$_\odot$, which rarely have M-star companions in this range of separation. However, the lack of low-mass companions extends at least down to $M\sim 2$ M$_\odot$. We found that the transition between a population of secondaries dominated by massive stars ($M>1$~M$_\odot$) and the usual population dominated by low-mass stars occurs at a separation of $\sim 1000$ au. We interpret this result as the formation of secondaries by fragmentation of the disk around the primary and selective mass accretion on the secondaries at shorter separation, and by cloud fragmentation at wider separations. We notice that while not original as this scenario has been proposed by many others before, considering disk fragmentation as the dominating scenario for formation of close binaries unifies it with that of substellar companions that also form within the primary disk.

We derived the distributions of the companions with a semi-major axis and mass ratio and found that there are systematic differences in both cases when comparing systems with primaries  with masses lower or higher than $\sim 3.5$~M$_\odot$. Systems around more massive stars are more compact and have a mass ratio closer to one. To explain these differences, we compared the observed distributions with predictions given by a toy model for the formation of binaries by disk fragmentation similar to that considered by \citet{Tokovinin2020}. We found that within that framework, the difference in the properties of binaries with primaries with a different mass is due to a different run of the mass of the accretion episodes with time: many more important accreting episodes should be considered for the most massive stars rather than for the low-mass ones. This gives many more opportunities to generate secondaries by disk instability (raising the binary frequency), a more pronounced inward migration, and a more appreciable growth of the companions up to a mass comparable with the mass of the primary.

\begin{acknowledgements}

R.G., S.D., V.D., D.M., E.R. acknowledge support 
from the PRIN-INAF 2019 'Planetary systems at young ages (PLATEA)' and ASI-INAF agreement n.2018-16-HH.0.

SPHERE is an instrument designed and built by a consortium consisting of IPAG (Grenoble, France), MPIA (Heidelberg, Germany), LAM (Marseille, France), LESIA (Paris, France), Laboratoire Lagrange (Nice, France), INAF-Osservatorio di Padova (Italy), Observatoire de Gen\`eve (Switzerland), ETH Zurich (Switzerland), NOVA (Netherlands), ONERA (France) and ASTRON (Netherlands), in collaboration with ESO. SPHERE was funded by ESO, with additional contributions from CNRS (France), MPIA (Germany), INAF (Italy), FINES (Switzerland) and NOVA (Netherlands). SPHERE also received funding from the European Commission Sixth and Seventh Framework Programmes as part of the Optical Infrared Coordination Network for Astronomy (OPTICON) under grant number RII3-Ct-2004-001566 for FP6 (2004-2008), grant number 226604 for FP7 (2009-2012) and grant number 312430 for FP7 (2013-2016).

This work has made use of the SPHERE Data Center, jointly operated by OSUG/IPAG (Grenoble), PYTHEAS/LAM/CeSAM (Marseille), OCA/Lagrange (Nice) and Observatoire de Paris/LESIA (Paris).

This research has made use of the SIMBAD database, operated at CDS, Strasbourg, France. 

This work has made use of data from the European Space Agency (ESA) mission {\it Gaia} (\url{https://www.cosmos.esa.int/gaia}), processed by the {\it Gaia} Data Processing and Analysis Consortium (DPAC, \url{https://www.cosmos.esa.int/web/gaia/dpac/consortium}). Funding for the DPAC has been provided by national institutions, in particular the institutions participating in the {\it Gaia} Multilateral Agreement.

This paper includes data collected by the Kepler and TESS missions and obtained from the MAST data archive at the Space Telescope Science Institute (STScI). Funding for the Kepler and TESS missions is provided by the NASA Science Mission Directorate. STScI is operated by the Association of Universities for Research in Astronomy, Inc., under NASA contract NAS 5–26555.

This research has made use of the Washington Double Star Catalog maintained at the U.S. Naval Observatory

This publication makes use of data products from the Two Micron All Sky Survey, which is a joint project of the University of Massachusetts and the Infrared Processing and Analysis Center/California Institute of Technology, funded by the National Aeronautics and Space Administration and the National Science Foundation.
\end{acknowledgements}

%
\bibliographystyle{aa} 
\bibliography{main} 
%

\begin{appendix}

\section{Binary companions detected in the BEAST survey}
\label{sec:beast}

In this Appendix we report about the companions detected in the B-star Exoplanet Abundance STudy (BEAST) survey \citep{Janson2021a}, that targeted 86 B-stars in the Sco-Cen association with SPHERE located on UT3 at the ESO Very Large Telescope \citep{Beuzit2019}. The stars were selected to be B-stars with high membership probability from \citet{Rizzuto2011} not included in the previous SHINE survey (\citealt{Vigan2021}: a total of 19 late B-stars were included in that survey) and not known to be visual binaries with separation $<6$~arcsec at the epoch of target selection. In addition, stars transiting at meridian within 3 degrees to zenith of Paranal were not included because of difficulties in their observation. The BEAST survey foresaw two observations per target, in order to use proper motion to confirm the physical link between the companions and the stars. It is not yet completed, lacking first epochs for four stars (HIP~78702, HIP~78933, HIP~79098, HIP~81474) and second epochs for 34 further targets. However, while crucial for detecting substellar objects, this incompleteness has not a significant impact for stellar ones. In fact, the four missing objects were already targeted in previous HCI that while not deep enough to detect substellar objects, were however fully adequate for stellar ones; and second epochs are only needed for the very faint substellar companions that have magnitudes similar to background M-stars, while it is very unlikely to find background objects as bright as the candidate stellar companions. We can then give here the results of the BEAST survey for stellar companions.

We acquired data with the typical observing procedure used for the BEAST survey \citep{Janson2021b}. Briefly, the high-contrast imager SPHERE, with the high-order AO system SAXO \citep{Fusco2006}, was used with the two infrared channels: the integral field spectrograph IFS \citep{Claudi2008} and the dual band imager IRDIS \citep{Dohlen2008, Vigan2010}. IFS and IRDIS were used in parallel mode; the observations were performed with SPHERE using the IRDIFS-EXT mode, that is using IFS in the $YH$ mode (wavelength range 0.95-1.65 micron, resolving power R$\approx$30) and IRDIS in $K1-K2$ mode (that is, 2.09 and 2.25 micron). IFS has a roughly square field of view (FoV) with a side of $\sim 1.76$~arcsec, while IRDIS has also a square FoV with a side of $\sim 11$~arcsec. We acquired the observations in pupil-stabilised mode with an Apodized Lyot Coronagraph with a focal mask having a diameter of 185 mas \citep{Boccaletti2008}. The observations were done in service mode; the same total integration time on source of 3072 s was adopted for all targets and individual detector integration time was adjusted depending on the brightness of the source in order to avoid saturation. The constraints set on atmospheric conditions allowed a uniform high quality of the observations, with a median seeing of 0.63 arcsec at zenith as measured by the DIMM. The median field rotation for validated observations was 34.4 degree. We also obtained on-sky calibrations for each scientific observation: they include a point spread function flux calibration, with the star offset with respect to the coronagraphic mask; centring calibrations, where we obtained satellite images symmetric with respect to the central star by imparting a bi-dimensional sinusoidal pattern to the deformable mirror; and sky calibrations that are important for background subtraction on IRDIS data at long wavelengths. 

\begin{figure}[htb]
    \centering
    \includegraphics[width=8.5cm]{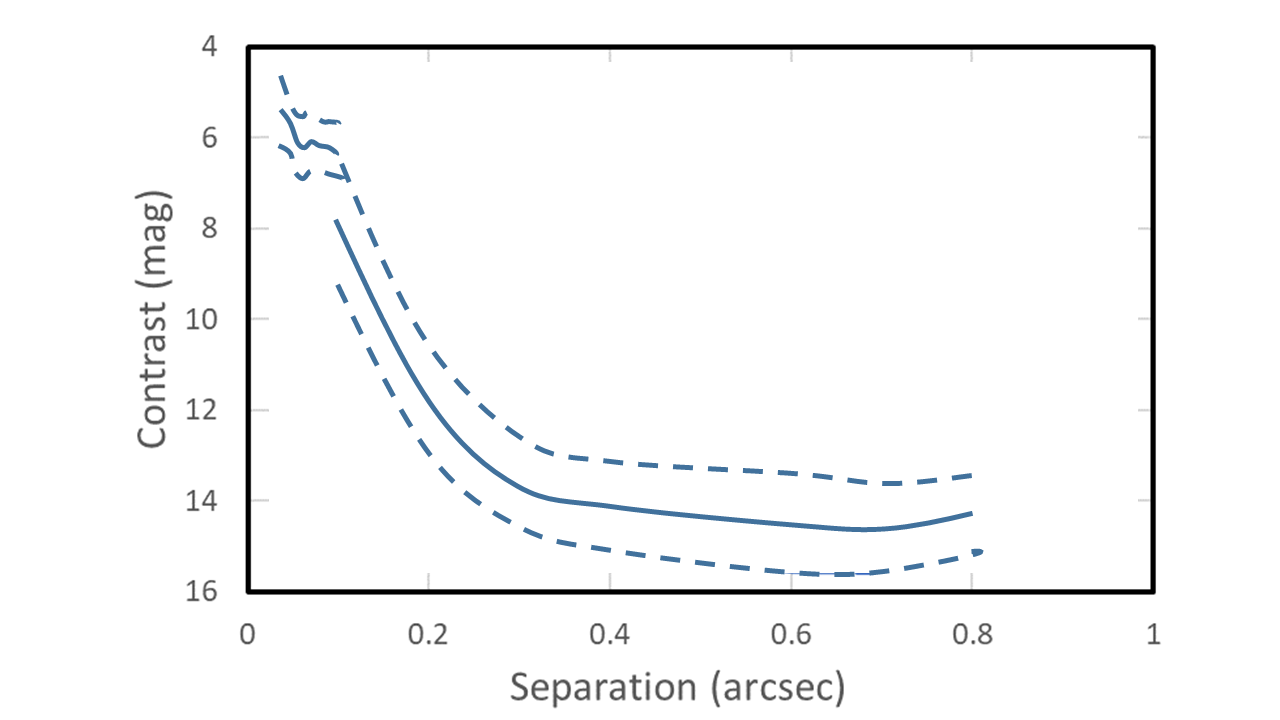}
    \includegraphics[width=8.5cm]{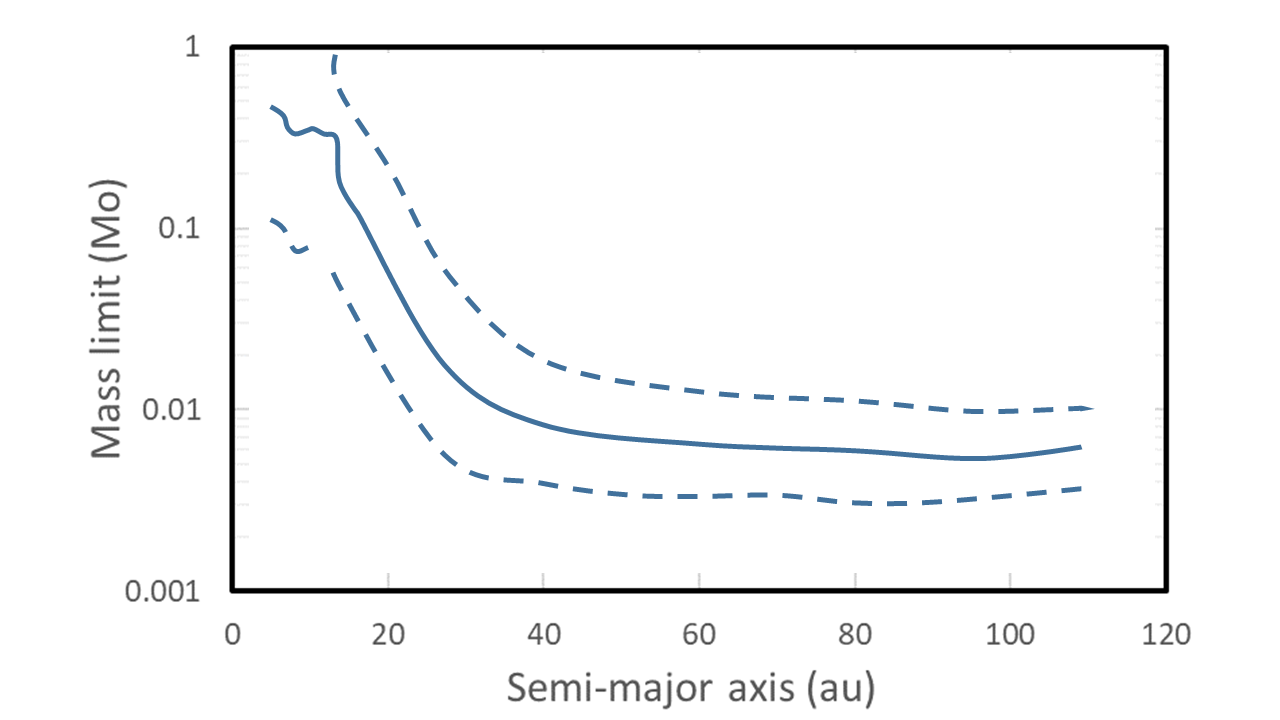}    
    \caption{Run of the limiting contrast with separation (upper panel) and of the limiting mass as a function of the semi-major axis (lower panel) obtained from the IFS BEAST observation. The solid line is the average value, the dashed lines mark the 1-sigma range in both directions. Limiting masses are obtained adopting 15~Myr old isochrones by \citet{Baraffe1998, Baraffe2015}. }
    \label{fig:contrast_limit}
\end{figure}

Data were reduced using the standard SPHERE pipeline (v. 15.0; \citealt{Pavlov2008}), and then by a suite of routines available in the SPHERE Data Center in Grenoble \citep{Delorme2017}. The final output of the data reduction procedure are four dimensional datacubes that include spatial (two dimensional), temporal and wavelength information. Data analysis was performed using the SPECAL routines \citep{Galicher2018} at the SPHERE Data Center as well as special routines based on simultaneous spectral and angular differential imaging based on the Principal Component Analysis method (PCA-ASDI: \citealt{Mesa2015}). Astrometrization was obtained as described in \citet{Maire2016}. The uniform quality of the observing material resulted in very similar limiting contrasts for all targets: the $5-\sigma$ limiting contrast at 0.5 arcsec is 14.3 mag with a standard deviation of 0.81 mag (see Figure~\ref{fig:contrast_limit}). This limiting contrast corresponds to a limiting mass of $<0.01$~M$_\odot$ at projected separation $>40$~au for a typical target.

In addition, very close bright companions (separation $<0.1$~arcsec) that are behind the coronagraphic mask in the science exposures could be detected using a method based on the flux calibration, where the star is offset with respect to the coronagraphic mask \citep{Bonavita2022}. Typically two such images are acquired, one before and one after the science sequence; given the time elapsed between the two exposures, the field rotation can be exploited to obtain a differential image that cancels static aberrations. With this procedure, the limiting contrast is typically about 6 mag at about 60~mas; this corresponds to a limiting mass of $\sim 0.4$~M$_\odot$ at a projected separation of $\sim 8$~au for a typical target.

While the BEAST survey is not yet completed, we already detected 17 companions around 15 stars. Given the selection criteria adopted for the survey, these companions are either of low mass or at very small separation. Five (around four stars) of these companions are substellar; they are discussed in dedicated papers together with one of the stellar companions that is in a triple system including a BD \citep{Janson2019, Janson2021b, Squicciarini2022, Viswanath2023}. Table~\ref{t:beast} summarises the main data for the remaining stellar companions. In the following we will discuss the individual objects.

\begin{table}[htb]
  \caption[]{Stellar companions detected on BEAST data.}
  \label{t:beast}
  \begin{tabular}{ccccccc}
  \hline
HIP &  JD  & Sep    &  PA  & d$J$  &  d$H$  &  d$K$\\
    & (+2400000) &  mas   &degree&  mag   & mag  & mag \\
  \hline
52742 & 58252.97 & 1092.0 &  9.7 &      &      & 6.73 \\
      & 58509.28 & 1091.7 &  9.7 & 7.70 & 7.40 & 6.74 \\
59173 & 58554.22 & 1276.6 & 129.9 &      &      & 5.30 \\
      & 58900.27 & 1271.8 & 130.1 &      &      & 5.30 \\
60009 & 58537.27 & 147.5 & 166.0 & 6.05 & 5.64 &     \\
      & 58887.32 & 149.7 & 171.5 & 5.97 & 5.57 &     \\
61257 & 58641.99 & 5512.7 & 324.49 &      &      & 6.05 \\  
      & 59655.23 & 5505.3 & 324.44 &      &      & 6.10 \\
62434 & 58574.18 &       &       &      &      &      \\
      & 59596.29 & 118.6 & 295.0 & 6.30 & 6.08 &      \\
      & 59674.20 & 106.0 & 190.6 & 5.86 & 5.87 &      \\
63005 & 58547.27 & 259.2 & 171.7 & 4.83 & 4.32 &      \\
      & 59637.29 & 284.7 & 180.5 & 4.69 & 4.28 &      \\
71860 & 59414.00 &  76.0 & 336.3 &      &      & 5.14 \\
      & 59673.28 &  67.7 & 258.7 &      &      &      \\
73624 & 58600.21 & 346.2 & 141.0 & 4.93 & 4.47 &      \\
74100 & 58915.34 & 549.4 & 276.9 & 4.25 & 3.83 &      \\
74449 & 59290.34 & 868.0 & 134.7 &      &      & 6.02 \\
      & 59773.04 & 871.9 & 134.5 & 7.46 & 6.91 & 5.90 \\
      & 60031.32 & 877.5 & 134.7 & 7.46 & 7.12 & 6.11 \\
76600 & 58582.28 &       &       &      &      &      \\  
      & 58691.99 &  60.2 & 289.9 & 1.49 & 1.56 &      \\
  \hline
  \end{tabular}
\end{table}

{\bf HIP~52742:} This Be star is likely not a member of Sco-Cen according to \citet{Rizzuto2011}; an age of $82.5\pm 21.8$~Myr is assigned by analysis of a single common proper motion companion star by \citet{Janson2021a}. With this age, the star should be slightly evolved off the main sequence and the appropriate mass using the PARSEC isochrones \citet{Bressan2012} should be 4.69~M$_\odot$. A candidate companion was detected on the IRDIS images (and at the very edge of the IFS image) at an apparent separation $1091.8\pm 1.4$~mas (projected separation of 176~au) and PA=$9.62\pm 0.13$~degree, with no detectable motion between the two observations (JD=528252.97 and JD=58509.31); we then considered it as a physical companion. The measured contrasts are d$J$=7.70, d$H$=7.40, d$K_1$=6.80, and d$K_2$=6.67 mag. Considering the distance modulus of the star and the 80 Myr old isochrones by \citet{Baraffe2015}, we derive a mass of $0.51\pm 0.03$~M$_\odot$ and a mass ratio of $q=0.109$.

\begin{figure}[htb]
    \centering
    \includegraphics[width=8.5cm]{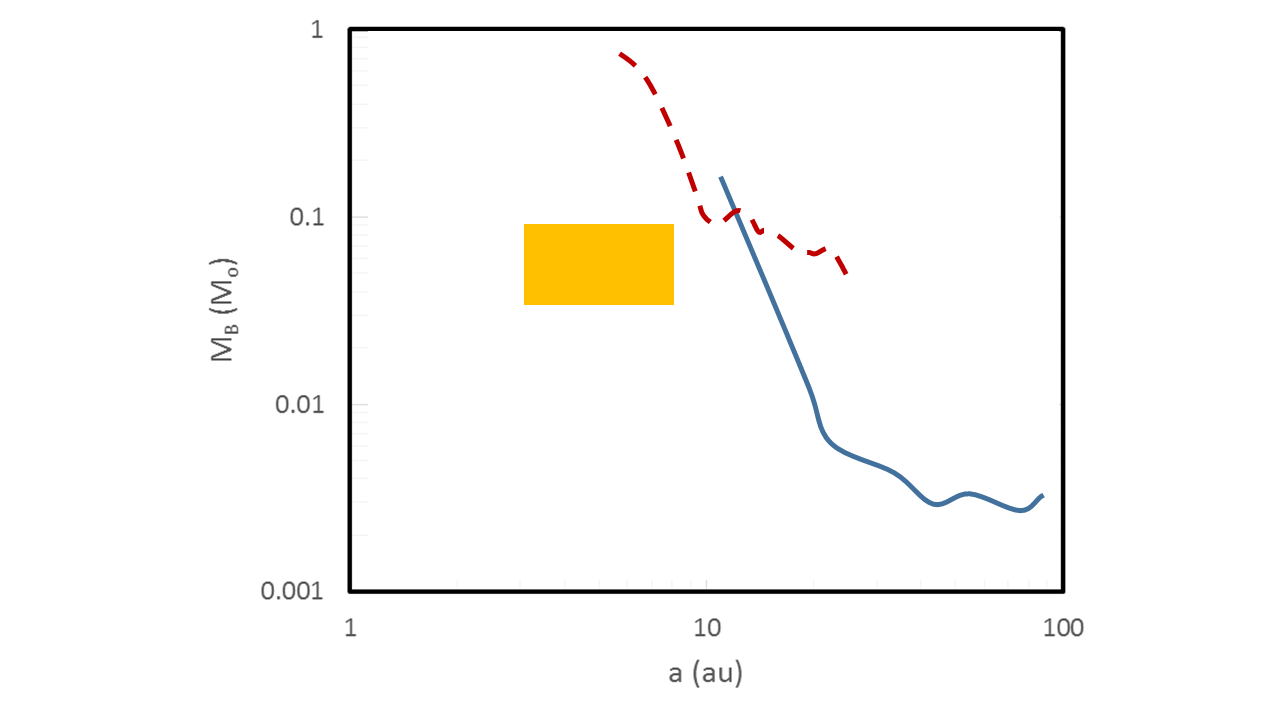}
    \caption{Comparison of the detection (upper) limit in mass as a function of separation for companions of HIP~59173 from coronagraphic (blue solid line) and not coronagraphic (red dashed line) obtained from data in the BEAST HCI and the semi-major axis and mass of the object compatible with the PMa, RUWE and RVs (orange area).}
    \label{fig:HIP59173_image}
\end{figure}

{\bf HIP~59173:} This is a B5V Be star with a high membership probability to Lower Centaurus Crux (LCC) \citep{Rizzuto2011}. The star is classified as an SB2 by \citet{Chini2012}; it has a quite large RUWE parameter (indicative of binarity) and a highly significant PMa \citep{Kervella2022} (S/N=6.4 and PA=$137\pm 9$~degrees). The star is then likely a quite close binary, though the companion was not detected in interferometry by \citet{Rizzuto2013}; this may indicate either a very short separation at the epoch of that observation or a low mass (or both). A candidate companion was found on the SPHERE IRDIS data with a separation of 1274~mas, that corresponds to 139~au at the distance of the star, and $PA\sim 130$~degre. Two epochs are available from BEAST data; the very small motion relative to the star between the two epochs supports a physical link with the primary. The PA agrees with that expected based on the PMa. However, the mass of this object (an M-star with a mass of 0.41~M$_\odot$) is an order of magnitude too small to explain the large observed PMa. The star should then be a triple. According to analysis of RUWE, PMa, and RVs the close companion is itself likely a BD or a low-mass star, with a mass of $0.062\pm 0.027$~M$_\odot$ and a semi-major axis of $5.5\pm 2.4$~au. Given the very small separation, this object is not expected to be detectable in the SPHERE high contrast images (see Figure~\ref{fig:HIP59173_image}). 

\begin{figure}[htb]
    \centering
    \includegraphics[width=8.5cm]{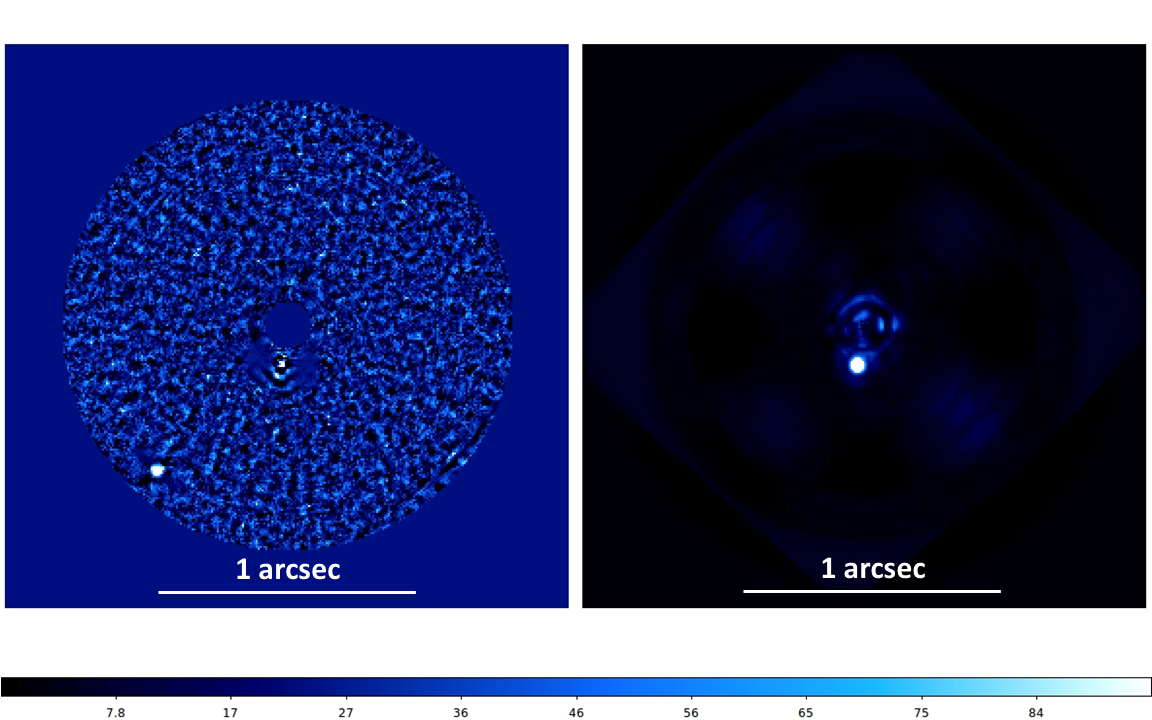}
    \caption{Images for HIP6009. Left: S/N map obtained on the SPHERE IFS data for HIP60009 analysed with PCA ASDI; right: The same data set after simple subtraction of a radial profile. The colour scale represents the S/N value on the left image, and an arbitrary relative intensity on the right}
    \label{fig:HIP60009_image}
\end{figure}

\begin{figure}[htb]
    \centering
    \includegraphics[width=8.5cm]{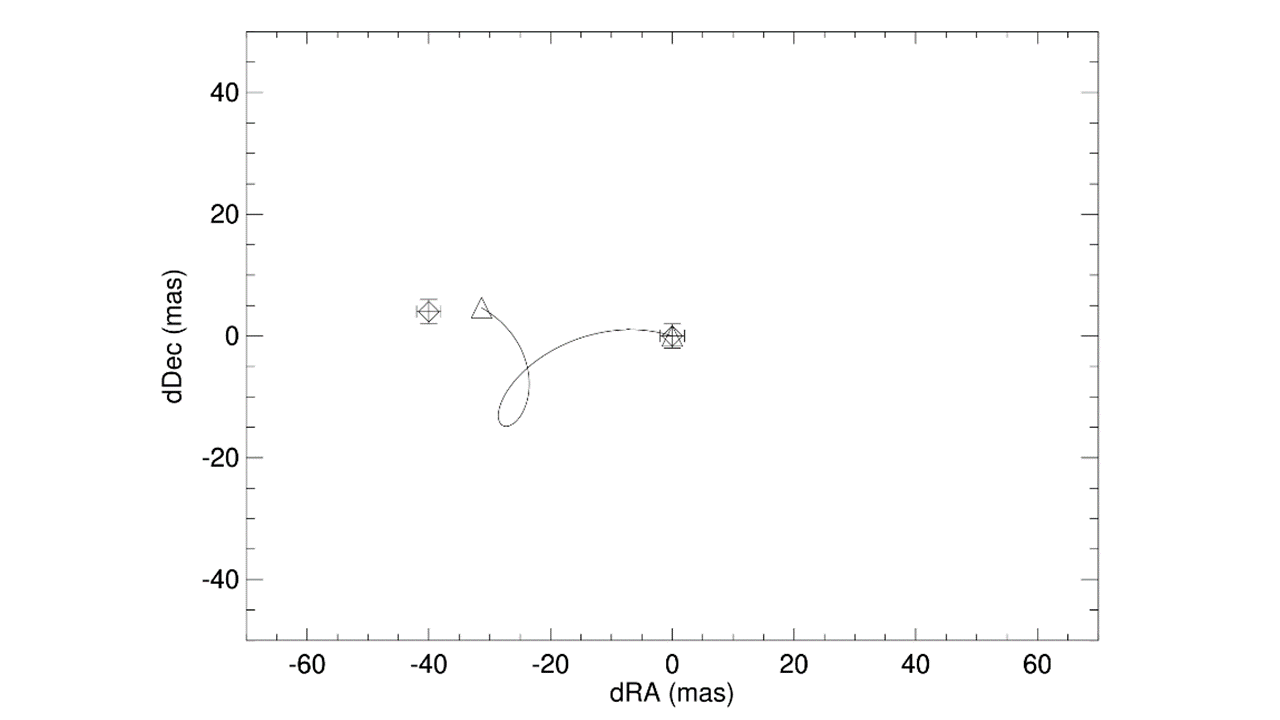}
    \caption{Relative position of the outer close companion candidate to HIP60009 (that at 0.75 arcsec from the star) at the two observing epochs (open circles). The triangle represents the position at the second epoch expected for a background object}
    \label{fig:HIP60009_motion}
\end{figure}

{\bf HIP~60009:} The star has a K-type stellar companion at a separation of 0.15~arcsec (projected separation $\sim 14$~au; expected period of $\sim 20$ yr) also detected by SPHERE (see Figure~\ref{fig:HIP60009_image}); this companion is likely responsible for the observed variation in the proper motion of the star. In addition there is an M2-type common proper motion companion at $\sim 11$ arcmin separation (projected separation of $\sim 60000$~au) revealed by Gaia. This companion is likely a pre-main sequence star because it is about 1.5 mag brighter than expected from the temperature. In the field of HIP~60009 there are two further groups of M-stars, also belonging to Sco-Cen: the first includes 3 stars at a separation of a few arcmin, and a second one 5 stars at a separation of $\sim 15$~arcmin W. These two groups are at a distance larger than HIP~60009 by about 15 pc.

The comparison between the position in the two BEAST epochs shows that the faint candidate companion at a separation of 0.75~arcsec is a background object (see Figure~\ref{fig:HIP60009_motion}).

{\bf HIP~61257}: The star has a spectral type of B9.5V and it is a high probability member of Lower Centaurus Crux (LCC) \citep{Rizzuto2011}. The star has a large population of more than 200 common proper motion stars so its age is well determined at $15.3\pm 0.3$~Myr \citep{Janson2021a}. The companion detected on the SPHERE IRDIS image has been also detected in Gaia DR3 at separation 5.519 arcsec, PA=324.46 degree, with an apparent magnitude of $G_B$=16.864. The contrast in the $K-$band is 6.08 mag. The companion should then have absolute magnitudes of $M_G=11.353$ and $M_K=7.19$. This is a very low-mass star with mass of $0.083\pm 0.01$~M$_\odot$, averaging the results obtained with the $G4$\ and $K$ bands and the \citet{Baraffe2015} 15~Myr isochrone. The mass ratio of $q=0.0354$ is then very extreme.

\begin{figure}[htb]
    \centering
    \includegraphics[width=8.5cm]{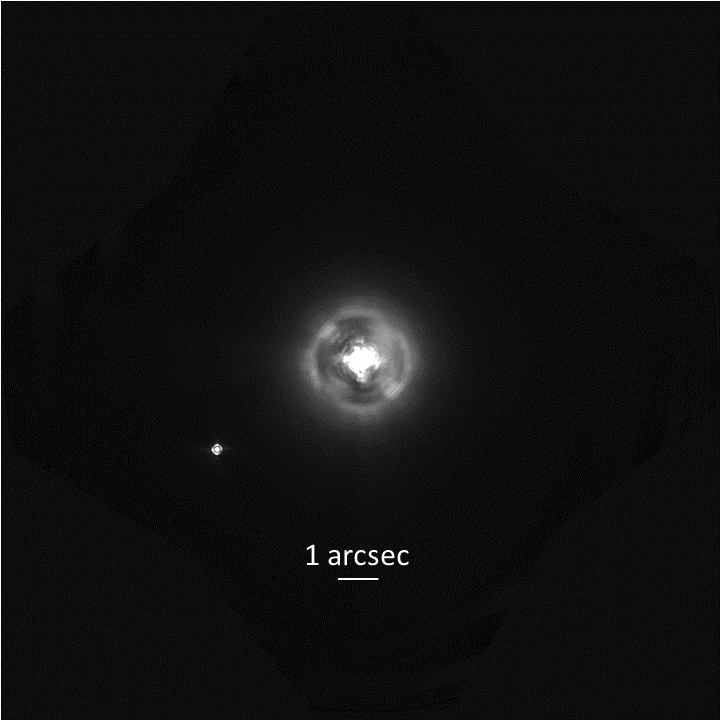}
    \caption{No-ADI IRDIS image of $\beta$~Cru. Component B is on the lower left (SE) of the star}
    \label{fig:HIP62434_image}
\end{figure}

\begin{figure}[htb]
    \centering
    \includegraphics[width=8.5cm]{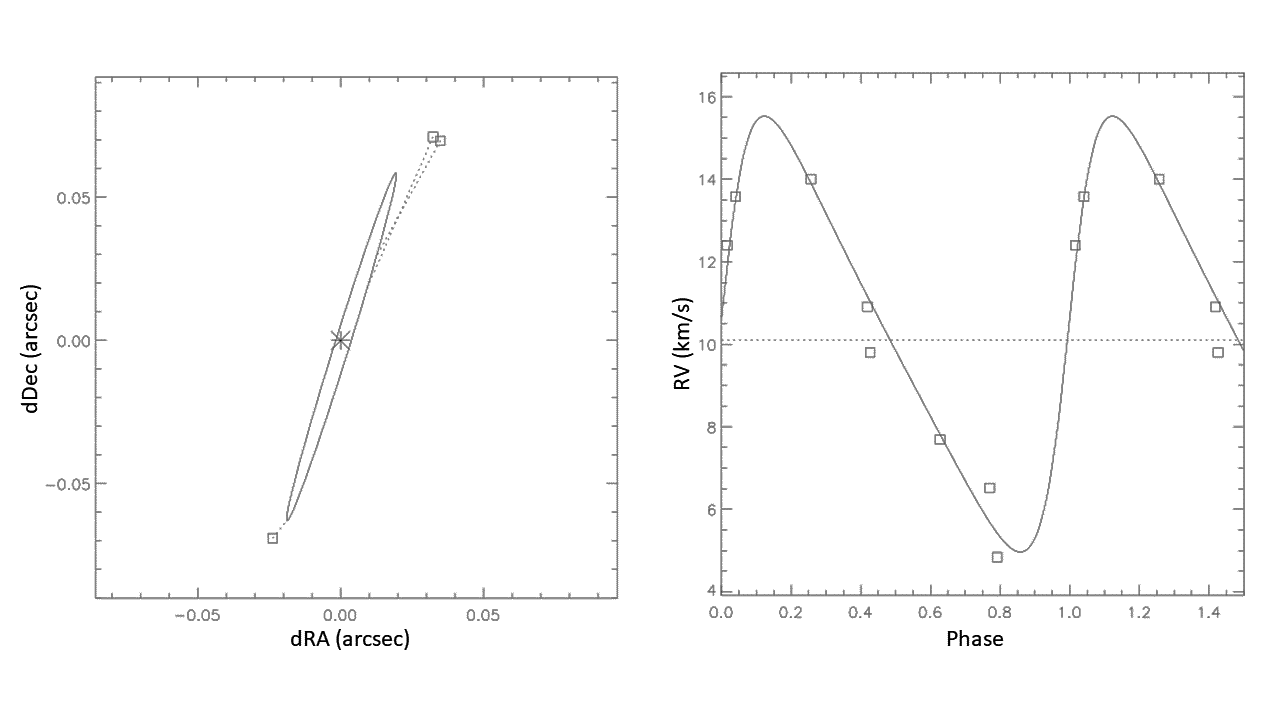}
    \caption{Orbital fit for $\beta$~Cru B. Left: astrometric orbit; Right: radial velocity orbit }
    \label{fig:HIP62434_orbit}
\end{figure}

{\bf HIP~62434 ($\beta$~Cru or Mimosa):} LCC member at 73\% \citep{Rizzuto2011}. We assumed a mass of 15~M$_\odot$; if this is correct, there is an upper limit to the age else the star should have already made a SN explosion. According to the PARSEC isochrones \cite{Bressan2012}, a 15.0~M$_\odot$ star lifetime is 13.9 Myr. From this argument we may adopt an age of $\sim 12$~Myr for $\beta$~Cru; the star will likely explode as a supernova in the next 1-2 Myr. In agreement with this, the asteroseismic analysis by \citet{Cotton2022} found that the star has an age between 9.7 and 12.8 Myr and a convective-core mass between 25\% and 32\% of its mass. This is in good agreement with the need for higher-than-standard core masses as derived from eclipsing binaries in this mass range. We find the radius of the star to fall in the range from 7.3 to 8.9~R$_\odot$. This is slightly larger than implied by the angular diameter as measured by intensity interferometry if using the new Hipparcos reduction for parallax, $6.6\pm 0.6$~R$_\odot$, but in good agreement if using the original Hipparcos parallax determination, $8.2\pm 0.5$~R$_\odot$. Both our mass and radius estimates are in agreement with earlier values based on multicolour photometry. The range for the radius, combined with the inclination and spectroscopic estimate of $V \sin{i} \simeq 16$~km~s$^{-1}$, leads to an equatorial rotation velocity of $\simeq 22$~km~s$^{-1}$ and a surface rotation period between 13 and 17 days.

We recovered on the BEAST image an already known stellar companion at 4.2 arcsec (see Figure~\ref{fig:HIP62434_image}). The projected separation is 451 au. The motion between the two epochs (that are separated by 3.01 yr apart) is -2.4~mas in RA and 0.65~mas in declination, that is at a rate of -0.8~mas/yr in RA and 0.2~mas/yr in declination. This is to be compared with the stellar proper motion of -42.97 mas/yr in RA and -16.18 mas/yr in declination. This companion is then physically bound to $\beta$~Cru. While the motion is very small, it is roughly radial and then possibly consistent with an orbital motion seen at high inclination. We may characterise the companion as a 0.78~M$_\odot$ star; the mass ratio of $q=0.047$\ is quite extreme for a binary system. This is clearly the same object responsible for X-ray emission found by \citet{Cohen2008} using data from the Chandra satellite. This companion is likely a post-T Tau object, as proposed by Cohen et al. We notice that in this multiple system we have simultaneously a high-mass star evolved off the main sequence and a low-mass star that is still on the pre-main sequence phase.

In addition, we discovered in our images a second closer companion (that we will call AB) projected at about 0.1 arcsec (about 9 au) from the star; while photometry is a bit uncertain, this companion should have a mass slightly higher than that of the Sun ($M=1.9\pm 0.07$ M$_\odot$). This closer companion is roughly at the right distance to produce periodic RV variations for the primary with a period of ~5 yrs deduced from the RV variations observed by \citet{Aerts1998}; the amplitude of the RV curve is $K=5.9\pm 0.8$~km~s$^{-1}$; this would require $q=0.134\pm 0.018$. This is within the errors of the value of $q=0.127\pm 0.007$ observed from photometry. It is then reasonable to identify AB as the responsible of the RV variations.

An attempt of use the ORBIT \citep{Tokovinin2016}  \footnote{\url{https://zenodo.org/record/61119\#.Xg83GxvSJ24} } programme to fit the observed astrometric orbit with the parameters of the spectroscopic orbit (for the distance given by Hipparcos and a total mass of 16.4~M$_\odot$ and forcing $K=5.28\pm 0.3$~km~s$^{-1}$, in order to reproduce the observed mass ratio), yields an inclination of $92.2\pm 0.7$~degree and $\Omega=342.6\pm 2.2$~degree. The fit is fairly good (total reduced $\chi^2=0.82$; see Figure~\ref{fig:HIP62434_orbit})

The system is then triple\footnote{\citet{Rizzuto2013} assumed the companions found by \citet{Cohen2008} and \citet{Aerts1998} are the same object but this cannot be because at a separation of 4 arcsec ($\sim $360 au) the period should be much longer than 5 yr}. In addition, we notice that the orbit of AB should be seen almost edge on and the PA of the wide companion (that we will call B) is very similar to that of AB at quadrature. This suggests that the two orbits are nearly coplanar. The orbit orientation is about 90 degrees distance from the stellar axis PA for the primary derived by \citet{Cotton2022}: $205\pm 8$~degree), that is the stellar axis is aligned quite close to the projected short axis of the orbit of AB and B. This also agrees with the whole system being quite close to coplanar. However, the stellar inclination ($i\sim 46$~degree) obtained combining polarimetric and spectroscopic observation by \citet{Cotton2022} is much less edge-on than the possible orbit derived from astrometry.

The star is also signalled as a visual binary by \citet{Rizzuto2013} with $sep=42.56$ arcsec, $PA=74.12$ degree, $dK=7.45$\ from 2MASS; however the corresponding Gaia source (6056720446549602304) is a background star (parallax of 1.21 mas). They do not detect any companion in their visual interferometric observations and set upper limits of about 2 mag in contrast for any companion. 

{\bf HIP~63005 ($\mu^2$~Cru):} This B5V star is a Be star and a far companion to HIP~63003 ($\mu^1$~Cru). The two stars have a large cohort of common proper motion companions suggesting an age of about 12.5~Myr \citep{Janson2021a}. It has a nearby candidate companion at about 0.259 arcsec. With a proper motion of -13.4 mas/yr in RA and -9.5 mas/yr in declination relative to the primary, the candidate companion is moving very differently from the values of 28.386 mas/yr in RA and 10.447 mas/yr in declination expected for a background object. We then conclude that the observed motion is the orbital motion of a secondary around the primary in a physical binary. The secondary is at projected separation of 29 au. We then expect a period of $\sim 73$~yr. This object is likely responsible for the large and significant PMa (S/N=15.96: \citealt{Kervella2022}). The PMa corresponds to a motion with a transverse velocity of 1.3~km~s$^{-1}$ directed towards a position angle PA=107 degree. Differentiating the positions measured at the two BEAST epochs, the projected orbital velocity of the companion on the sky plane is 16.4 mas/yr (=9.4~km~s$^{-1}$) directed towards PA=305 degree, roughly opposite to the PMa measured for the primary, as expected. Identifying the Hipparcos-Gaia motion with the secular motion of the star, we then roughly expect a mass ratio $q=1.3/9.4=0.14$. Given the uncertainties, this value agrees quite well with the value of $q=0.194$ determined from the mass we derived from the photometry of the two components ($M_A=3.8$~M$_\odot$ and $M_B=0.72$~M$_\odot$; see Table~\ref{t:mass}).

{\bf HIP~71860 ($\alpha$~Lup)}: this B1.5III star is among the brightest and most massive ones (mass of 11.6~M$_\odot$) in the BEAST sample; it is a $\beta$~Cep pulsator and it is evolved off the main sequence. It was observed in interferometry by \citet{Rizzuto2013} but no companion was detected, likely because the observation is quite shallow (contrast of about 3 mag); also, there was no detection in the speckle interferometry observation by \cite{Mason2009} (epoch 2001.5691). We detected a $dK=5.14$~mag close companion (separation of about 76 mas, that is a projected separation of 12.7 au); the object is too faint to be detected in these previous studies. Given the large mass of the primary, the mass of the secondary is however quite large, 2.25~M$_\odot$. The period is expected to be short ($\sim 10$~yr) so we expect a significant orbital motion even if the two epochs are quite close with each other (259~d). The radial velocity is considered to be constant \citep{Chini2012}; however, the RV variations expected due to this companions ($<6$~km~s$^{-1}$) are likely below the detection limit of those observations.

{\bf HIP~71865 (b~Cen:)} The detection of the planetary companion to HIP~71865 is described in \cite{Janson2021b}. The star is itself a close binary revealed in interferometry \citep{Rizzuto2013}. Here we adopted masses of 5.24 and 2.0~M$_\odot$, in good agreement with what assumed by \cite{Janson2021b}.  

{\bf HIP~73624:} This  B3V star has a rather high value of the RUWE$=1.472$ and a significant PMa anomaly ($SNR=7.04$) at PA=190.4~degree \citep{Kervella2022}, indicating it is a close binary. We detected a companion at a separation of 346 mas, that is a projected separation of 52 au, PA=141.0~degree and contrast $dH=4.47$~mag in a single epoch with BEAST acquired 3.32 yr after the Gaia DR3 epoch. We obtained a mass of 0.75~M$_\odot$ from the photometry. We expect a period of $\sim 24$~yr for this object (similar to the baseline between Gaia and Hipparcos, so that we might expect that PMa represents the component of the instantaneous velocity of the star at the Gaia epoch). We will then assume that this object is that responsible for the observed PMa. 

{\bf HIP~74100:} This star is labelled with variable radial velocity \citep{Levato1987, Chini2012}, though this result may be dubious given the very high rotational velocity of 370~km~s$^{-1}$ \citep{Zorec2012}; it has a low value of RUWE and significant PMa ($SNR=4.21$ at $PA=280\pm 14$~degree, \citealt{Kervella2022}). A companion is detected in a single epoch in the SPHERE data at a separation of 549.4 mas (projected separation of 79 au) and PA=276.9 degree, with a magnitude offset of $dH=3.83$~mag; the mass of such a bright object should be 0.86~M$_\odot$ using the 15~Myr isochrone by \citet{Baraffe2015}. In this case we expect a period of $\sim 350$~yr, much longer than the Hipparcos-Gaia baseline. For such a long period, we expect that the PMa represents the acceleration (at the Gaia epoch) and should then be aligned towards the secondary (see \citealt{Bonavita2022b}). The alignment is indeed well within the errors. In addition, the evolutionary mass cited above is within 25\% of that expected for a companion responsible of the PMa with a semi-major axis equal to the projected separation; this agreement is good in consideration of the uncertainties. In our analysis, we will then assume that the observed companion is that responsible for the PMa.

\begin{figure}[htb]
    \centering
    \includegraphics[width=8.5cm]{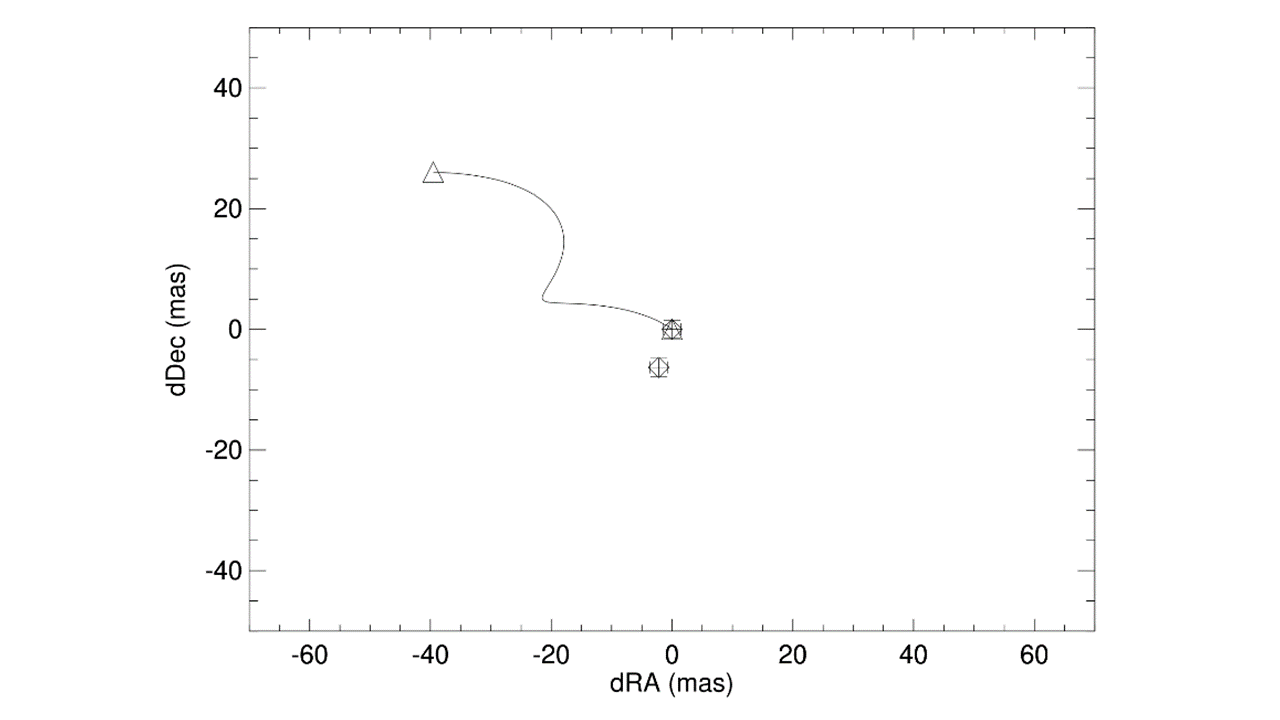}
    \caption{Relative position of close companion candidate to HIP74449 at the two observing epochs (open circles). The triangle represents the position at the second epoch expected for a background object}
    \label{fig:HIP7449_motion}
\end{figure}

{\bf HIP~74449 (e Lup):} this is a B3IV/V close spectroscopic binary member of Upper Centaurus Lupus (UCL).  In \citet{Janson2021a} an age of 15 Myr was adopted for this star; we confirm this age estimate using five common proper motion companions in Gaia eDR3 that gives an average age of $14.1\pm 0.7$ Myr. HIP~74449 is a compact (almost contact) binary \citep{Buscombe1962} composed of two massive components, and its total mass is of the order of 8-9 M$_\odot$. 

BEAST data show the presence of a close physical companion (HIP~74449B) projected at about 0.87 arcsec.  We may use the observations at the two epochs to check if the motion of this object between the two epochs is compatible with that expected for a background object. This comparison is done in Figure~\ref{fig:HIP7449_motion}. The relative motion is clearly very different from that expected for a background object. We then confirm that this candidate is physically linked to HIP~74449 and will call it HIP~74449B (the two components of the inner binary being then HIP~74449AA and HIP~74449AB). On the other hand, there is evidence for an orbital motion of HIP~74449B in almost exactly the radial direction. The projected separation of HIP~74449B from its primary is of about 130 au, suggesting a period of ~500 yr; the observed motion of about 9.5 mas in 2.03 yr is well compatible with an orbital motion. So we conclude that it is very likely a very low-mass star ($M=0.265\pm 0.057$~M$_\odot$). The mass ratio of $q=0.029\pm 0.006$\ is quite extreme for stellar binaries. We note that given the brightness of the central binary, this companion lies not much beyond the snow line for this system. Hence it seems very likely that this companion formed within the primary disk. We derived a spectral type of M2.5 from a best matching of the IFS spectrum with templates in the library by \citet{Leggett1996}. According to the table by \citet{Pecaut2013}, this corresponds to a temperature of 3470 K, in reasonable agreement with that expected for a 15 Myr old 0.26~M$_\odot$ star, for which we expect a temperature of 3320 K according to the isochrones by \citet{Baraffe2015}, that actually corresponds to a spectral type between M3 and M3.5.

Two additional companions are seen on the IRDIS images, at sep=2439 mas, PA=85.4 degree, and sep=3673 mas, PA=243.3 degree, in the 2021 observations (and contrast of about 10.7 and 11.8 mag, respectively). They were also recovered on the 2022 observations, but they clearly are background objects.

We notice that three of the Gaia wide common proper motion companions of HIP~74449 form a small group of stars at about 6 arcmin (projected separation of about 50 kau, but possibly some depth in space) from HIP~74449A.  This group includes a 0.5~M$_\odot$ star (2MASS J15122658-4425015) and a wide binary made of a low-mass star (2MASS J15122633-4424550: M=0.26~M$_\odot$) and a BD (2MASS J15122658-442501: 0.064~M$_\odot$). What is interesting is that this group of stars is aligned (with PA in the range 134-141 degree with respect to HIP~74449A) very close to the PA of HIP~74449B. This suggests that all these stars formed along the same original filament (that had this orientation on sky), and might still be physically bound to HIP~74449 (in this case the period can be as large as several Myr).

\begin{figure}[htb]
    \centering
    \includegraphics[width=8.5cm]{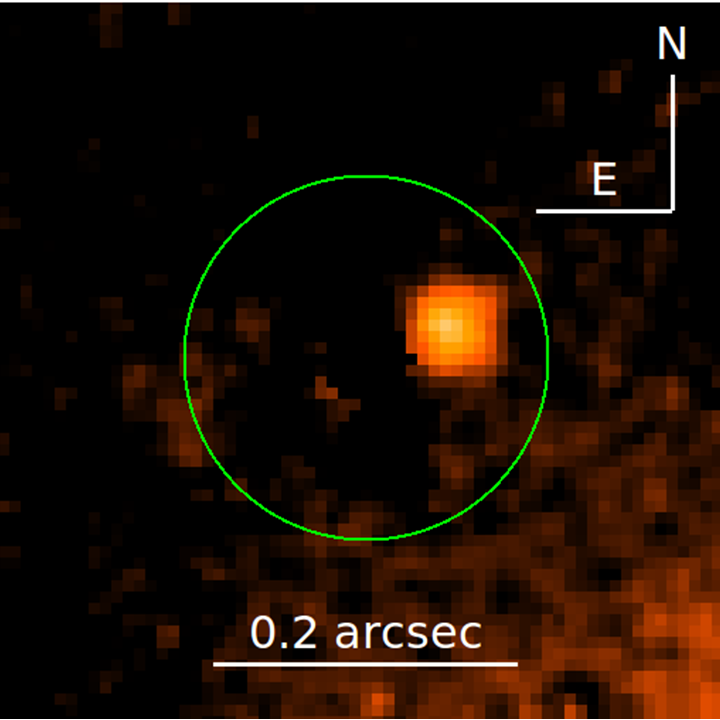}
    \caption{Analysis of the non-coronagraphic image of HIP76600 analysed using the same technique described in \citet{Bonavita2022}; colour scale is intensity arbitrary units}
    \label{fig:HIP76600_image}
\end{figure}

{\bf HIP~76600 ($\tau$ Lib):} HIP~76600 is a ternary system in Lower Centaurus-Lupus \citep{Jilinski2006} composed of a close spectroscopic binary and of a wide companion providing a clear astrometric signal. We recovered the wide companion in the SPHERE observations (see Figure~\ref{fig:HIP76600_image}). In addition, the star is also a heart-beat pulsator  (that is, an eccentric binary with tidal distortions exciting pulsations) with frequencies of 4.63 and 4.82 d$^{-1}$ \citep{Sharma2022}.

HIP~76600 is a $3.29066\pm 0.00001$ d double line spectroscopic binary \citep{Levato1987} with $K_1=75\pm 8$ km/s, $K_2=167\pm 12$ km/s, $v=-14\pm 4$  km/s, $a_1 \sin{i}=3.3\,10^6$ km, $a_2 \sin{i}=7.3\pm 10^6$ km. The mass ratio is then $q=0.45$. Using the table by \citet{Pecaut2013}, the primary (that should dominate the spectrum) is a 6.1~$M_\odot$ star and the secondary is a B9V star with a mass of 2.75~$M_\odot$. If this were true, then the semi-major axis of the binary should be 0.089~au. Given the projected separation along the line of sight from the spectroscopic orbit, we deduce that the inclination should be about $i=55$~degree (with a quite large error bar). Given the significant orbital eccentricity (e=$0.28\pm 0.05$) and orbital orientation ($\omega=114\pm 15$ degree) the system is close to eclipsing, justifying its peculiar light curve.

HIP~76600 is flagged as an astrometric binary in Hipparcos (see \citealt{Makarov2005}) and there is a discrepancy of about 10 mas/yr between Hipparcos and GAIA DR2 astrometry. The separation between the two components of the spectroscopic binary projected on the plane of sky is $\sim 0.05$~au that should be 0.4~mas at the distance of HIP~76600. The astrometric signal should be $<0.2$ mas, much less than the discrepancy between Hipparcos and GAIA DR2. We then expect that the astrometric signal is due to a third component.

There is a star (source 6209197287409304320) in the GAIA DR2 catalogue with parallax and proper motion compatible with that of HIP~76600 at 947 arcsec from it (that is, $\sim $0.5 pc projected distance). The parallax of this star is  $\pi=8.04\pm 0.10$ mas, $PM_{\rm RA}=-21.39$ mas/yr, $PM_{\rm DEC}=-27.77$ mas/yr, $G=13.991$, $b_p=2.668$, $T_{\rm eff}=3826$ K. This is of course too far to be responsible for the astrometric variation, but it can be used as a reference for the orbit of the wide companion.

No companion to HIP76600 was observed using speckle interferometry on 4m telescope \citep{Mason1999}. This of course implies an upper limit to separation and/or luminosity.

A companion with $sep=12.1$ mas, $PA=5.58$ degree, $dmag=2.85$ mag was detected on 14/07/2010 using interferometry with SUSI on a 15 m baseline (angular resolution of 7 mas: \citealt{Rizzuto2013}), in 25 optical wavelengths channels (from 550 to 800 nm).

We used the flux calibration to find the companion likely responsible of the astrometric perturbation. A companion was automatically detected using a procedure that differentiates the observations taken before and after the science exposure. We find $sep=60.2$ mas, $PA=289.9$ degree, that corresponds to a projected separation of 6.8~au. The contrast is 2.95 mag in $Y$, 2.76 mag in $J$, and 2.77 mag in $H$ band, in good agreement with the magnitude difference in the optical from interferometry. This implies $M_Y=1.31$, $M_J=1.30$, $M_H=1.49$ mag. This is an A1V-A2V star, with a mass of 2.1~$M_\odot$. 

The separation is much larger than that obtained in the interferometric observation. This suggests that this last was acquired when the companion was projected very close to the star with respect to the typical values appropriate for its orbit: that is, either the orbit is highly eccentric, or it is seen nearly edge-on, or both. If we assume that the projected separation at the SPHERE epoch is the semi-major axis of the orbit and use the masses obtained from luminosities, then the period would be 5.3~yr. The ratio between the time elapsed between the two observations and this period is 1.70 that agrees quite well with an observation taken in conjunction and the other one in quadrature. If this were true, the PA of the SPHERE observation should be close to that of the orbital plane and the inclination of the orbit of this star should be quite different from that of the spectroscopic binary.

The orbit of this object should be easy to find with Gaia, and possibly even before, combining Hipparcos and Gaia data. A preliminary estimate that is based on the difference in proper motion between Hipparcos and Gaia and a mean motion of the system given by nearby common proper motion star shows a good agreement, supporting the period considered here.

{\bf HIP~79098:} The detection of a BD companion to HIP~79098 is described in \cite{Janson2019}. The star has large RV variations \citep{Levato1987, Worley2012} and the Gaia solution gives a high value of the RUWE parameter indicating that the star is itself a binary; the S/N of the PMa is low suggesting that the binary is quite close. However, as discussed \cite{Janson2019}, literature results are quite contradictory and no clear consensus about the mass of the secondary and orbital period could be obtained. The analysis made in Section~\ref{sec:detections} indicates a mass of $0.53\pm 0.26$~M$_\odot$ and a semi-major axis of $a=1.86\pm 0.56$~au, well within the uncertainties described in \cite{Janson2019}.

{\bf HIP~81208:} The detection of a BD and a further low-mass stellar companion to HIP~81208 is described in \citet{Viswanath2023}. We do not find indications for additional companions.

{\bf HIP~82514:} The detections of a robust low-mass BD companion and a possible second one to HIP~82514 ($\mu^2$~Sco) are described in \cite{Squicciarini2022}. The star has a further far stellar companion with mass of 0.48~$M_\odot$ at 44 arcsec revealed by Gaia.

\FloatBarrier

\section{Reanalysis of HIP~73807}
\label{sec:HIP73807}

HIP~73807  ($\pi$~Lup) is a well-known visual binary discovered by John Herschel almost two centuries ago, with a projected semi-major axis of about 1.59 arcsec and a period of 517 yr \citep{Nitschelm2004}. The two components have very similar luminosity: the contrast is only 0.06 mag in Gaia. According to \cite{Nitschelm2004}, both components are spectroscopic binaries, one being an SB2 and the other an SB1, so the system is quadruple. \citet{Gullikson2016b} rather considered three components, with masses of $M_A=4.5$, $M_B=4.3-4.7$, and $M_C=3.5-3.9$~M$_\odot$. This object is included in the catalogue of OBA eclipsing binaries observed by TESS \citep{IJspeert2021} with a period of 15.50 d, depth=0.0099 and eclipse length of length 0.17709~d. The light curve indicates a grazing eclipse in a detached system; this is confirmed by \citet{Sharma2022}. Two nearly identical eclipses are visible in the TESS observation; this suggests that there is a single eclipse per cycle and that the orbit is eccentric, with a different separation between the two stars at the epochs of conjunction and opposition of the secondary; since the eclipse is grazing, the eccentricity may be even low. It is unclear which of the two visual components have the eclipses.

The analysis of the RV by \citet{Jilinski2006} yields two spectroscopic components with  velocities of A (-52.5 km/s) ad B (81.5 km/s)\footnote{These are not the amplitudes of the radial velocity curves, but instantaneous velocities.}. Assuming these are the velocities of the components of the eclipsing binary and adopting a centre of mass RV as given by the table of \citet{deBruijne2012} (4.5 km/s), the mass ratio is $q=M_B/M_A=0.74$. If we then use the relation between mass and absolute magnitude of Section~\ref{sec:visual}, the solution that yields the right absolute magnitude have masses of $M_A=4.70$ M$_\odot$ and $M_B=3.48$ M$_\odot$\footnote{Since the two visual components have very similar luminosity, this result is essentially independent of which of the two objects is eclipsing}. We considered the STEV isochrones \citep{Bressan2012} with an age of 21.7 Myr appropriate for this star (see Table~\ref{t:targets}). The expected radii of the stars in the eclipsing system are 3.33 and 2.49 R$_\odot$ and the contrast in the TESS band is about 0.82 mag. The secondary contributes about 47\% of the luminosity of the primary, and about 16\% of the total luminosity of the system. Since the observed luminosity dip at the eclipse is only $1.3$\%, only about 2.6\% of the primary or about 5.2\% of the secondary should be eclipsed at maximum depth. A simplified solution (circular motion) with a flux dilution factor of 0.5 due to the contribution by the other visual component of the system and a semi-major axis of $a=0.245$~au (consistent with the adopted masses and period) yields the correct eclipse depth for an inclination of $i=84.5$~degree. In this solution, the eclipse length (0.262 d if the primary is eclipsed, 0.271 d if it is the secondary) is longer than observed, indicating that at the epoch of the eclipse the separation is smaller (0.77 or 0.72 times, respectively) than the semi-major axis. This requires an at least moderate eccentricity ($e>0.32$\ or $e>0.35$), that is in any case consistent with the rather long period.

However, we cannot exclude that the eclipse is not on the SB2 component,
with arguments similar to those made above. In fact, we found that a very similar light curve can be obtained also if the secondary mass is small compared to that of the primary. High precision radial velocity curves for both components are required to identify the orbital solutions for the two components.

\FloatBarrier

\section{A BD close to HIP~74752 detected by Gaia}
\label{sec:HIP74752}

{\bf HIP~74752}: Gaia allowed detection of a very faint companion to HIP~74752, a star in the BEAST survey with spectral type B9.5 and a nearly 100\% membership probability to Sco-Cen. HIP~74752 has a mass of 2.26~M$_\odot$; the age estimated from the closest star in the BEAST survey is 14.8~Myr \citep{Janson2021a}. The companion is very remarkable because it is a BD with a mass of 0.053~M$_\odot$, and the mass ratio to the star (that has a mass of 2.265 M$_\odot$) is $q=0.023 \pm 0.001$. Given the large projected separation of 1256~au, the origin is unclear, but it is very interesting for a full discussion of BEAST companions.

This BD companion to HIP~74752 was not noticed so far. The star has been observed within the NICI survey \citep{Nielsen2013}, that had a FoV of $18\times 18$~arcsec, and it would have surely been noticed if within their FoV. The Gaia DR3 separation is 9.599 arcsec and the PA=20.812 degree and it could have been slightly out of their FoV (depending on the field rotation).

The relative motion of the BD with respect the star is $-0.07\pm 0.36$ mas/yr in RA and $-1.63\pm 0.63$~mas/yr in declination, that is significant at $\sim $2.6~$\sigma$; this corresponds to a motion with $dSep=1.6\pm 0.7$~mas/yr at PA=$182\pm 22$~degree. The motion is in the direction of the primary, within the (large) errors. The relative speed on the sky plane is of 0.8 km/s, fully compatible with an orbital motion (circular speed at this separation from the star is 1.3 km/s).

Gaia also indicates that there is an additional stellar companion to HIP~74752, at 26.15 arcsec almost exactly south of the star (PA=182.633~degree), with a magnitude $G=15.139$~mag. This object is very red ($b_p-r_p=3.097$~mag). Again, the difference in proper motion with the star is compatible with a physical link with the star. This object has also an entry in 2MASS (J15163704-4222386) with $K=11.152\pm 0.021$~mag. This object is a low-mass star (0.237~M$_\odot$). 

The system is then a triple. 

\FloatBarrier

\section{Possible brown dwarfs or low-mass stars close to HIP~62058 and HIP~64053}
\label{sec:bds}

\begin{figure}[htb]
    \centering
    \includegraphics[width=8.5cm]{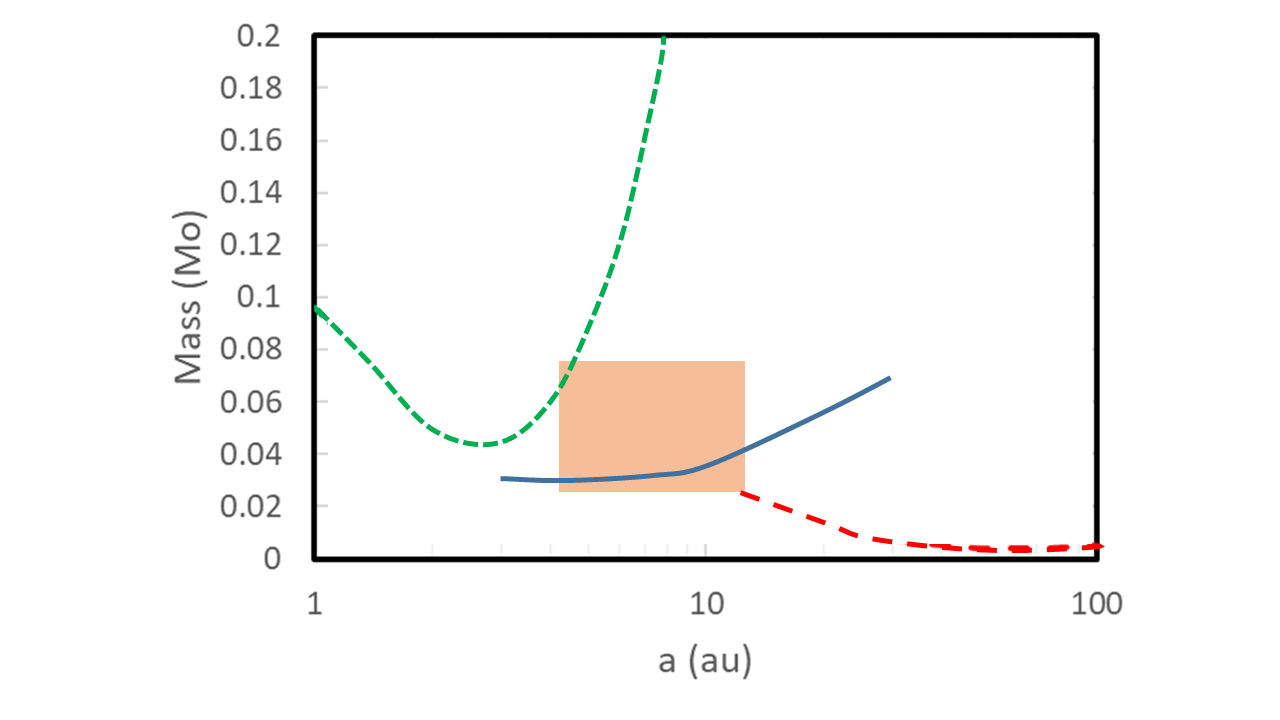}
    \includegraphics[width=8.5cm]{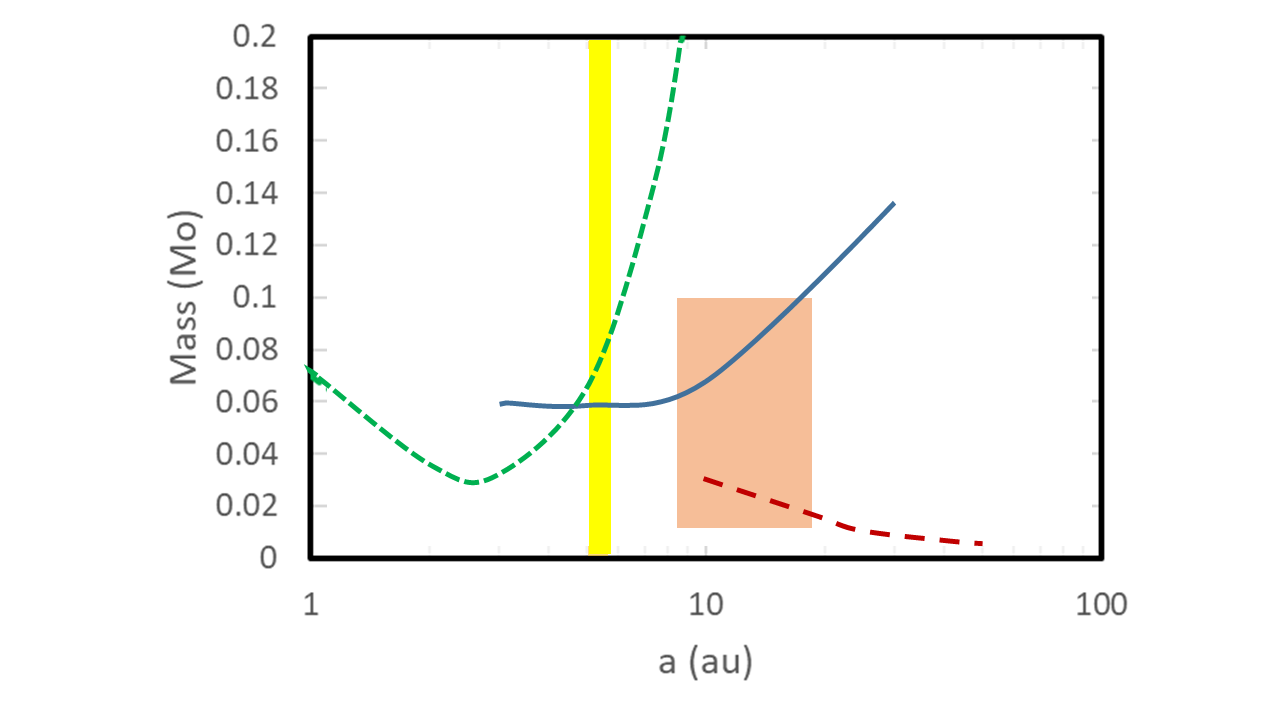}
    \caption{
    Upper panel: Comparison of the detection (upper) limit in mass as a function of separation for companions of HIP~62058 from data in the BEAST HCI (red long dashed line), the upper limit given by the lack of significant RUWE (green short dashed line) and the mass of the object responsible of the PMa as given by \citet{Kervella2022} (blue solid line). Bottom panel: the same for HIP~64053 from data in the BEAST HCI (red long dashed line), the upper limit given by the lack of significant RUWE (green short dashed line) and the mass of the object responsible of the PMa as given by \citet{Kervella2022}. The brown area in both panels is the region compatible with all constraints from our analysis. The orange shadowed area in the bottom panel marks the position of the warm debris disk \citep{Ballering2017}. }
    \label{fig:HIP62058}
\end{figure}

The values of the PMa \citep{Kervella2022} suggest the existence of potential substellar objects or small mass stars around HIP~62058 and HIP~64053 that were not previously detected in HCI surveys. We may derive constraints on the main characteristics of these objects by combining the mass-semi-major axis relation that \citet{Kervella2022} obtained from the measured PMa (that are rough values because they are computed in the assumption of circular orbits seen face-on), the limits set by non-detections in imaging, and those that are provided by the Gaia RUWE parameters. To estimate the mass limits corresponding to RUWE$<1.4$ \citep{Belokurov2020}, we computed the residuals around a best fit straight line fitting through astrometric points of a sequence simulating the Gaia observations but including the wobble of the primary due to the orbital motion of a companion of different mass. For low-mass objects, we may neglect the contribution due to the contribution of the secondary to the motion of the photocenter. For simplicity, we assumed circular orbit, but we assumed random phases and inclinations, respectively with uniform and $\sim \cos{i}$ distributions. For each mass of the secondary, we repeated the experiment with 1000 random trials, and derived the secondary mass where 90\% of the observation have quadratic sum of the residuals along RA and declination smaller than $\sigma \sim 0.3$~mas, that is the typical error of individual Gaia measurements \citep{Lindegren2018}. We notice that the astrometric signal is given by $(M_B/M_A) sep $, where $M_A$ and $M_B$ are the masses of the primary and secondary (in M$_\odot$), respectively, and $sep$ is the semi-major axis $a$ in au multiplied by the parallax $\pi$ (in mas). We found that the 90\% confidence upper limit of the mass of the secondary obtained through the Monte Carlo simulation is well reproduced as:
\begin{equation}
M_B<\sigma \frac{M_A}{a\,\pi} 
\end{equation}
if the time covered by the observations considered by Gaia DR3 $\Delta T=2.83$~yr is shorter than 0.6 times the orbital period $P$ (still in yr), and:
\begin{equation}
M_B<\sigma \frac{M_A}{a\,\pi} (\frac{P}{\sqrt{2}\,\Delta T})^2
\end{equation}
if $\Delta T> 0.6\,P$. With this in mind, we will then examine the individual cases.

{\bf HIP~62058}: This star is a B7/8V star with an estimated mass of 2.82~M$_\odot$; the large number of common proper motion companions allows to fix the age at $16.6\pm 0.3$~Myr \citep{Janson2021a}. The RV of the star is labelled as constant in \citet{Chini2012}, and the spread in Gaia RVs is not inconsistent with constant. The RUWE is also low, but the PMa has a $SNR=5.13$, indicating the presence of a companion. It was observed twice (at JD=58558.24 and JD=59310.25) in the BEAST survey, with no detection of companions. As shown in the upper panel of Figure~\ref{fig:HIP62058}, the object responsible for the PMa measured by \citet{Kervella2022} should be easily detectable on the BEAST images if its separation at the epoch of the BEAST observations were larger than 100 mas (projected separation of about 12 au). This essentially confines the semi-major axis to values lower than this limit, and the period to values lower than a few tens yr. At this small separation, the upper limit provided by the non-detection using the RUWE parameter is higher than the mass derived using the PMa, so that there is no further constraint on $a$. We then expect that the object responsible for the PMa is a BD with a mass of $\sim 0.05$~M$_\odot$ and semi-major axis $<12$~au. We notice that the expected radial velocity signal for such a companion is likely very low ($\sim 0.1$~km\,s$^{-1}$), not detectable on the existing data.

{\bf HIP~64053}: HIP~64053 is a B8/9V star in LCC for which our estimated mass is 2.58~M$_\odot$; the age should be about 15~Myr using the map by \citet{Pecaut2016}. The star hosts a warm debris disk, that should be located at a distance of $\sim 5.3$~ au \citep{Ballering2017}. The RV is considered constant by \citet{Chini2012}. The value of the Gaia RUWE$=1.365$ parameter is slightly below (though not far from) the threshold for indication of binarity. The PMa has a $SNR=5.82$, indicative of the presence of a companion, that is a BD if the semi-major axis is not much larger than 10~au, that is $\sim 100$~mas at the distance of the star. There are SPHERE HCI data published by \citet{Matthews2021}, who did not detect any physical companion to the star. They give an upper mass limit of $\sim 6$~M$_{\rm Jupiter}$ at 50 au, $\sim 9.5$~M$_{\rm Jupiter}$ at 25 au, and $\sim 30$~M$_{\rm Jupiter}$ at 10 au. The lower panel of Figure~\ref{fig:HIP62058} indicates that the object responsible for the PMa should have been detected with SPHERE if at a separation $>>100$~mas at the epoch of the observation by \citet{Matthews2021}, and it would cause a RUWE indicative of binarity if $a<5$~au. This suggests that the object is probably a BD  or a low-mass star, and it is probably farther from the star than the warm debris disk. The location of this low mass object is then quite constrained, because the disk may be stable only if the orbit is not highly eccentric and the semi-major axis is $>12$~au \citep{Holman1999}. This is reflected in the mean values of $M_B=0.058\pm 0.042$~M$_\odot$ and $a=13.3\pm 4.9$~au reported in Table~\ref{t:mass_dyn} (brown area in Figure~\ref{fig:HIP62058}). This suggests that the orbit has an apparent size similar to the field stop of the coronagraphs used in HCI, and that the object might be behind the coronagraph at the epoch of the observation by \citet{Matthews2021}, but it might be possibly detectable on new observations. We do not expect that such a low-mass companion with $q\sim 0.022$ would produce a signal detectable in RVs, given the difficulties of deriving high precision RVs for early type stars (the rotational velocity of the star is very high, $V \sin{i}=281$~km\,s$^{-1}$ \citealt{Glebocki2005}).


\FloatBarrier

\section{Long tables}
\label{sec:longtables}

Long tables containing the most relevant details for the programme stars are given in this Appendix. 

Table~\ref{t:targets} contains the target list. We give here the Hipparcos number, other designation, the HD number, right ascension (RA), declination (Dec), the distance d from \citet{Gaia2022a}, the probability of membership  to the Sco-Cen association according to \citet{Rizzuto2011}, the spectral type according to SIMBAD, the reddening $E(B-V)$ as estimated in Section~\ref{sec:reddening}, and a remark if the stars is either a Be or an $\alpha_2$~CVn variable. 

Table~\ref{t:info} gives a summary of information used to detect binary companions. The first column gives the Hipparcos number (if not indicated, it is the same as from the previous row); Col. 2 tells if the star has been observed either by TESS or K2; Col. 3 explains if the star is classified as as EB, an SB(2), has indication of variation in the RVs, or it has a constant RV from \citet{Pourbaix2004, Chini2012}; Col. 4 and 5 gives the time range and r.m.s. scatter of RV observations from \citet{Trifonov2020} ('no' means that the object is not in that catalogue); Col. 6-9 gives average RV, internal errors, the probability that the star has a constant RV, and a robust estimate of the amplitude of a potential RV curve from the GAIA DR3 catalogue \citep{Gaia2022a}; Col. 10 spells if the star was observed by \citet{Gullikson2013} and \citet{Gullikson2016a, Gullikson2016b} (an 'x') and if a companion was detected with that technique (a 'Y'); Col. 11 gives the Gaia RUWE parameter; Col. 12 contains the S/N of the eDR3 PMa obtained by \citet{Kervella2022}; Col. 13 gives the number of different studies containing HCI data for each star; Col. 14 tells if the star was observed in interferometry by \citet{Rizzuto2013} or \citet{Hutter2021}; finally, Col. 15 gives the separation of common proper motion companions detected by Gaia DR3 \citep{Gaia2022a} as separate sources.

Table~\ref{t:photometry} gives photometric data about the primaries and companions (if any). We give the Hipparcos number (if not indicated, it is the same as from the previous row), the projected separation of the companion in arcsec and in au, the apparent Gaia $G$\ (from \citealt{Gaia2022a}) and 2MASS $K$\ (from \citealt{Skrutskie2006}) magnitudes of the primary and secondary, the absolute magnitudes of both components (corrected for the contribution to the companions as spelled in Section~\ref{sec:massvisual}), the $G-K$ colour for the primaries, and the method used to detect the companion. 

Table~\ref{t:mass} gives data relevant to the masses of the star and the companions, derived as described in Section~\ref{sec:mass}. We give the Hipparcos number (if not indicated, it is the same as from the previous row), the masses of the primary and secondary (values derived for companions only detected from RVs, PMa and RUWE are given in a separate column), the mass ratio $q$, and a remark with references about the specific object.

\begin{table*}
  \caption[]{Target list (full table available through CDS). }
  \label{t:targets}
  \begin{tabular}{ccccccccccl}
  \hline
HIP     &       Other   &       HD      &       RA      &       Dec     &       d       &       Prob    &       ST      &       E(B-V)  & Age & Remark    \\
        &               &               &       degree  &       degree  &       pc      &               &               &       mag     &       Myr &       \\
 \hline
\\
\multicolumn{11}{l}{Stars in the BEAST sample}\\
\\
50847   &       L Car   &       90264   &       155.7422753     &       -66.90149731    &       123.2   &       80      &       B8      &       0.006   &       27.3 &       \\
52742   &       HR4221  &       93563   &       161.7394693     &       -56.75719167    &       161.3   &       0       &       B5      &       0.009   &       82.5 &Be     \\
54767   &       HR4355  &       95783   &       168.1883625     &       -64.16976977    &       97.7    &       4       &       B8      &       0.005   &       84.5 &       \\
58452   &               &       104080  &       179.7945014     &       -45.83222168    &       134.0   &       47      &       B8.5    &       0.007   &       20.0 &       \\
58901   &               &       104900  &       181.1885822     &       -59.25325091    &       117.0   &       99      &       B9.5    &       0.015   &       13.9 &       \\
 \hline
  \end{tabular}
\end{table*}

\begin{table*}
  \caption[]{Information about binarity (full table available through CDS). }
  \label{t:info}
  \begin{tabular}{ccccccccccccccc}
  \hline
HIP     &K2 or&SB orEB&\multicolumn{2}{c}{Trifonov2020}&\multicolumn{4}{c}{Gaia RV}&Gull.       &RUWE&PMa&      HCI     &       Int.    & GAIA  \\
      &TESS     &       &dt     &sigma&RV       &err    &prob.  &ampl.  &               &               &SNR&           &               & cpm     \\      
        &       &       &d      &km/s   &km/s   &km/s   &const. &km/s           &               &               &               &               &               &arcsec \\
\hline
\\
\multicolumn{15}{l}{B stars in BEAST}\\
\\
50847   &T      &SB2    &no     &       &no     &       &               &               &               &       1.038   &       1.39    &       3       &               &               \\
52742   &T      &       &no     &       &no     &       &               &               &               &       1.166   &       0.66    &       2       &               &               \\
54767   &T      &SB     &no     &       &14.4   &0.6    &       0.0903& 6.7     &               &       0.932   &       0.08    &       3       &               &               \\
58452   &K2     &C      &no     &       &no     &       &               &               &               &       0.973   &       1.59    &       2       &               &               \\
58901   &K2     &SB     &no     &       &no     &       &               &               &               &       2.750   &       35.28   &       1       &               &               \\
 \hline
  \end{tabular}
\end{table*}

\begin{table*}
  \caption[]{Photometry (full table available through CDS).}
  \label{t:photometry}
  \begin{tabular}{ccccccccccccl}
  \hline
HIP     &       sep     &       a       & $G_A$ & $G_B$ & $K_A$ & $K_B$ &       $M_{G~A~cor}$   & $M_{G~B~cor}$ & $M_{K~A~cor}$ & $M_{K~B~cor}$ & $G-K_A$ &       Method  \\
      & arcsec & au & mag & mag  & mag & mag & mag & mag & mag & mag & mag & \\
  \hline
\\
\multicolumn{13}{l}{Stars in the BEAST sample}\\
\\
50847   &       0.002   &       0.216   &       4.911   &       6.488   &       5.314   &               &       -0.320  &               &       0.175   &               &       -0.495  &       SB2     \\
        &       2.219   &       273.276 &       4.911   &               &       5.314   &       10.624  &       -0.320  &               &       0.175   &       5.170   &               &       VIS     \\
52742   &       1.092   &       176.0   &       5.080   &               &       5.225   &       11.965  &       -0.981  &               &       -0.815  &       5.925   &       -0.165  &       VIS     \\
54767   &       0.002   &       0.231   &       5.204   &               &       5.418   &               &       0.243   &               &       0.468   &               &       -0.225  &       SB      \\
58452   &               &               &       6.338   &               &       6.529   &               &       0.685   &               &       0.892   &               &       -0.207  &               \\
58901   &       0.043   &       4.987   &       6.201   &               &       6.278   &               &       0.824   &               &       0.934   &               &       -0.110  &       PMa     \\
 \hline
  \end{tabular}
\end{table*}

\begin{table*}
  \caption[]{Masses of primaries and companions (full table available through CDS).}
  \label{t:mass}
  \begin{tabular}{cccccl}
  \hline
HIP     &       $M_A$   &       $M_B$   &       $M_B$   &       q       & Remark  \\
    &           &           & PMa or RUWE or RV \\
      & M$_\odot$ & M$_\odot$ & M$_\odot$ & & \\
  \hline
\\
\multicolumn{6}{l}{Stars in the BEAST sample}\\
\\
50847   &       4.167   &       2.480   &               &       0.5952  &       \citet{Quiroga2010}     \\
        &       6.647   &       0.368   &               &       0.0553  &       \citet{Scholler2010}    \\
52742   &       4.69    &       0.510   &               &       0.1087  &       This paper, not member       \\
54767   &       3.062   &               &       0.335   &       0.1094  &       SB, not member                      \\
58452   &       2.619   &               &               &               &                                               \\
58901   &       2.499   &               &       0.307   &       0.1228  &       PMa     \citet{Makarov2005}     \\
 \hline
  \end{tabular}
\end{table*}

\end{appendix}

\end{document}